\documentclass[longauth]{aa} \usepackage{graphicx}
\usepackage{natbib}
\bibpunct{(}{)}{,}{a}{}{;}
\usepackage{amssymb}
\usepackage{amsmath}
\usepackage{url}
\usepackage{lscape}
\usepackage{rotating}
\usepackage{multirow}
\usepackage{txfonts}
\begin{document}
   \title{Water in star-forming regions with \textit{Herschel} (WISH)\thanks{\textit{Herschel} is an ESA space observatory with science instruments provided by European-led Principal Investigator consortia and with important participation from NASA.}}
\subtitle{II. Evolution of 557 GHz 1$_{10}$--1$_{01}$ emission in low-mass protostars}

\author{
	 L.E.~Kristensen\inst{1}
\and E.F.~van~Dishoeck\inst{1,2}
\and E.A.~Bergin\inst{3}
\and R.~Visser\inst{3}
\and U.A.~Y{\i}ld{\i}z\inst{1}
\and I.~San~Jose-Garcia\inst{1}
\and J.K.~J{\o}rgensen\inst{4}
\and G.J.~Herczeg\inst{2}
\and D.~Johnstone\inst{5,6}
\and S.F.~Wampfler\inst{7}
\and A.O.~Benz\inst{7}
\and S.~Bruderer\inst{2}
\and S.~Cabrit\inst{8}
\and P.~Caselli\inst{9,10}
\and S.D.~Doty\inst{11}
\and D.~Harsono\inst{1}
\and F.~Herpin\inst{12,13}
\and M.R.~Hogerheijde\inst{1}
\and A.~Karska\inst{2}
\and T.A.~van~Kempen\inst{1,14}
\and R.~Liseau\inst{15}
\and B.~Nisini\inst{16}
\and M.~Tafalla\inst{17}
\and F.~van~der~Tak\inst{18,19}
\and F.~Wyrowski\inst{20}
}

\institute{
Leiden Observatory, Leiden University, PO Box 9513, 2300 RA Leiden, The Netherlands
\and
Max Planck Institut f\"{u}r Extraterrestrische Physik, Giessenbachstrasse 1, 85748 Garching, Germany
\and
Department of Astronomy, The University of Michigan, 500 Church Street, Ann Arbor, MI 48109-1042, USA
\and
Niels Bohr Institute and Centre for Star and Planet Formation, University of Copenhagen, Juliane Maries Vej 30, DK-2100 Copenhagen {\O}., Denmark
\and
National Research Council Canada, Herzberg Institute of Astrophysics, 5071 West Saanich Road, Victoria, BC V9E 2E7, Canada
\and
Department of Physics and Astronomy, University of Victoria, Victoria, BC V8P 1A1, Canada
\and
Institute for Astronomy, ETH Zurich, 8093 Zurich, Switzerland
\and
LERMA, UMR 8112 du CNRS, Observatoire de Paris, 61 Av. de l'Observatoire, 75014 Paris, France
\and
School of Physics and Astronomy, University of Leeds, Leeds LS2 9JT, UK
\and
INAF - Osservatorio Astrofisico di Arcetri, Largo E. Fermi 5, 50125 Firenze, Italy
\and
Department of Physics and Astronomy, Denison University, Granville, OH, 43023, USA
\and
Universit\'{e} de Bordeaux, Observatoire Aquitain des Sciences de l'Univers, 2 rue de l'Observatoire, BP 89, F-33271 Floirac Cedex, France
\and
CNRS, UMR 5804, Laboratoire d'Astrophysique de Bordeaux, 2 rue de l'Observatoire, BP 89, F-33271 Floirac Cedex, France
\and
Joint ALMA Offices, Av. Alonso de Cordova 3107, Vitacura, Santiago, Chile
\and
Earth and Space Sciences, Chalmers University of Technology, SE-412 96 Gothenburg, Sweden
\and
INAF - Osservatorio Astronomico di Roma, 00040 Monte Porzio catone, Italy
\and
Observatorio Astron\'{o}mico Nacional (IGN), Calle Alfonso XII,3. 28014, Madrid, Spain
\and
SRON Netherlands Institute for Space Research, PO Box 800, 9700 AV, Groningen, The Netherlands
\and
Kapteyn Astronomical Institute, University of Groningen, PO Box 800, 9700 AV, Groningen, The Netherlands
\and
Max-Planck-Institut f\"{u}r Radioastronomie, Auf dem H\"{u}gel 69, 53121 Bonn, Germany
}

\date{Submitted: 26 Sept. 2011; Accepted: 19 March 2012}

\titlerunning{II. Evolution of 557 GHz 1$_{10}$--1$_{01}$ emission in low-mass protostars}

\abstract
{Water~is~a~key~tracer~of~dynamics~and~chemistry~in~low-mass~star-forming~regions, but~spectrally~resolved~observations~have~so~far been limited in sensitivity and angular~resolution, and~only~data~from~the~brightest~low-mass~protostars~have~been~published.}
{The first systematic survey of spectrally resolved water emission in 29 low-mass ($L<40~L_\odot$) protostellar objects is presented. The sources cover a range of luminosities and evolutionary states. The aim is to characterise the line profiles to distinguish physical components in the beam and examine how water emission changes with protostellar evolution.}
{H$_2$O was observed in the ground-state 1$_{10}$--1$_{01}$ transition at 557 GHz ($E_{\rm up}/k_{\rm B}$$\sim$60 K) as single-point observations with the Heterodyne Instrument for the Far-Infrared (HIFI) on \textit{Herschel} in 29 deeply embedded Class 0 and I low-mass protostars. Complementary far-IR and sub-mm continuum data (including PACS data from our programme) are used to constrain the spectral energy distribution (SED) of each source. H$_2$O intensities are compared to inferred envelope properties, e.g., mass and density, outflow properties and CO 3--2 emission.}
{H$_2$O~emission~is~detected~in all objects except one (TMC1A). The line profiles are complex and consist of several kinematic components tracing different physical regions in each system. In particular, the profiles are typically dominated by a broad Gaussian emission feature, indicating that the bulk of the water emission arises in outflows, not in the quiescent envelope. Several sources show multiple shock components appearing in either emission or absorption, thus constraining the internal geometry of the system. Furthermore, the components include inverse P-Cygni profiles in seven sources (six Class 0, one Class I) indicative of infalling envelopes, and regular P-Cygni profiles in four sources (three Class I, one Class 0) indicative of expanding envelopes. Molecular ``bullets'' moving at $\gtrsim$50 km s$^{-1}$ with respect to the source are detected in four Class 0 sources; three of these sources were not known to harbour bullets previously. In the outflow, the H$_2$O / CO abundance ratio as a function of velocity is nearly the same for all line wings, increasing from 10$^{-3}$ at low velocities ($<$5 km s$^{-1}$) to $\gtrsim$10$^{-1}$ at high velocities ($>$ 10 km s$^{-1}$). The water abundance in the outer cold envelope is low, $\gtrsim$10$^{-10}$. The different H$_2$O profile components show a clear evolutionary trend: in the younger Class 0 sources the emission is dominated by outflow components originating inside an infalling envelope. When large-scale infall diminishes during the Class I phase, the outflow weakens and H$_2$O emission all but disappears.}
{}


\maketitle

\section{Introduction}

Water is a unique probe of physical and chemical conditions in low-mass star-forming regions. In the molecular envelope from which the protostar accretes matter, water forms almost exclusively on the surfaces of dust grains \citep{tielens82, ioppolo08} and it is the dominant ice constituent \citep{whittet88, pontoppidan04}. As matter flows towards the protostar, it heats up and water desorbs when the dust temperature exceeds 100 K \citep{fraser01} leading to a jump in the abundance of several orders of magnitude \citep{ceccarelli96, boonman03}. At the same time, the protostar drives bipolar jets and winds into the envelope. It is unknown whether these are molecular or atomic in nature, regardless of their exact composition, water forms efficiently in the shocks involved, again enhancing the abundance by several orders of magnitude compared to the abundance in the quiescent part of the envelope \citep{kaufman96, franklin08, kristensen11}. 

The water molecule is an asymmetric rotor with many rotational transitions in the sub-millimetre (sub-mm) and far-infrared wavelength ranges. The combination of the abundance jumps and the high number of transitions makes water not only one of the most important molecular coolants in star-forming objects, but also one of the most effective diagnostics for understanding the chemical and dynamical evolution of star formation. Altogether, these effects make water a much better tracer of the various energetic components in a protostar than other commonly used molecules such as CO, which do not show large abundance variations.

Water has been observed previously in deeply embedded, low-mass protostars, primarily with space-based observatories such as ISO, SWAS, \textit{Odin} and \textit{Spitzer} \citep{kessler96, melnick00, nordh03, werner04}. ISO and \textit{Spitzer} both observed spectrally unresolved water lines in the wavelength range of 5 -- 180 $\mu$m \citep{ceccarelli98, boogert00, giannini01, larsson02, maret02, watson07}, where many higher-excited water lines ($E_{\rm up}/k_{\rm B}$$\sim$ few 100 K) were detected. With SWAS and \textit{Odin} it became possible to spectrally resolve the ortho-H$_2$O 1$_{10}$--1$_{01}$ ground-state transition at 557 GHz ($E_{\rm up}/k_{\rm B}$$\sim$60 K). The line profiles turned out to consist of a narrow absorption component ($\Delta\varv$$\sim$ a few km s$^{-1}$) superposed on a single outflow component ($\Delta\varv$$>$15--20 km s$^{-1}$) \citep{neufeld00, ashby00, bergin03, hjalmarson03, franklin08, bjerkeli09}. The large beams of \textit{Odin} and SWAS ($\sim$2\arcmin\ and 3$\times$4\arcmin, respectively) were naturally more sensitive to large-scale extended emission in the outflow.

While water emission turned out to be straightforward to detect in the deeply embedded Class 0 young stellar objects \citep[YSOs;][]{nisini02}, it was not detected in the more evolved Class I YSOs\footnote{In the following, Class 0 sources are charactised as YSOs with a bolometric temperature, $T_{\rm bol}$, $<$ 70 K. The Class 0 sources observed here are all Stage 0 sources, and the Class I sources are Stage I sources \citep[see][for a discussion of $T_{\rm bol}$ as an evolutionary parameter and a discussion of the difference between Stage and Class.]{evans09}.}. In particular, neither SWAS nor ISO detected water emission from any Class I objects. Two explanations were put forward \citep{nisini02}: either the non-detection is due to lower excitation in Class I objects, or it is due to a lower abundance. If the water is predominantly formed and excited in outflows, the Class I sources with their weaker outflows would go undetected, i.e., the water molecules would be less excited. On the other hand, the Class I envelopes are usually of lower mass and column density compared to their Class 0 siblings, allowing for the UV radiation from the accreting star to penetrate deeper into the envelope and outflow thereby photodissociating water more efficiently. The photodissociation naturally leads to a lower water abundance in the Class I objects rendering the emission difficult to detect.

With its much higher sensitivity and spatial resolution than any previous instrument capable of observing water, the Heterodyne Instrument for the Far-Infrared \cite[HIFI;][]{degraauw10} on \textit{Herschel} makes it possible to detect water emission from both Class 0 and I sources. Because of the improvement in sensitivity and resolution, early HIFI results already reveal water line profiles that are much more complex and show more kinematic components than those observed with SWAS and \textit{Odin}, and than those of other species, such as CO or SiO \citep{lefloch10, kristensen10b, kristensen11}. However, this analysis is still limited to only a handful of Class 0 sources associated with strong outflow activity, and is therefore not representative of how water emission and its profiles evolves in young low-mass YSOs, in particular the evolution from Class 0 to Class I objects.

As part of the ``Water in star-forming regions with \textit{Herschel}'' key programme \citep[WISH;][]{vandishoeck11}, 29 low-mass YSOs were observed with HIFI (see Table \ref{tab:source} for an overview). The selected sources span a range of luminosities (from $\sim$0.8 $L_\odot$ to 40 $L_\odot$), bolometric temperatures (from $\sim$ 30 K to $\sim$ 600 K) and envelope masses ($\sim$0.01 $M_\odot$ to 10 $M_\odot$), and are all closer than 450 pc. The Class 0 sources were selected from the list of \citet{andre00} and the Class I sources were selected from the lists of \citet{tamura91, andre94} and the \textit{Spitzer} ``Cores to Disks'' legacy programme \citep{evans09}. Furthermore, these objects have been the targets of numerous physical and chemical surveys \citep[e.g.,][]{hogerheijde99, shirley00, jorgensen02, jorgensen04, jorgensen07, emprechtinger09, kristensen10a, kristensen10b}. Some of these objects have also been observed with PACS (the Photodetector Array Camera and Spectrometer) on \textit{Herschel}, e.g., HH46-IRS and L1157 \citep{vankempen10, nisini10}. Eventually HIFI and PACS data will exist for all sources in several transitions of water, and in other species such as CO and OH \citep[for a summary and overview, see][]{vandishoeck11}. This paper focusses on observations of the ortho-H$_2$O ground-state 1$_{10}$--1$_{01}$ transition at 557 GHz in low-mass YSOs, and will thus greatly expand on previous SWAS and \textit{Odin} observations. 

The paper is organised as follows. The observations are described in Sect. \ref{sec:obs}. Section \ref{sec:results} contains the results, which are all discussed in Sect. \ref{sec:disc}. Concluding remarks are to be found in Sect. \ref{sec:conc}.

\section{Observations and complementary data}
\label{sec:obs}

\begin{figure*}
\begin{center}
\includegraphics[width = 17cm]{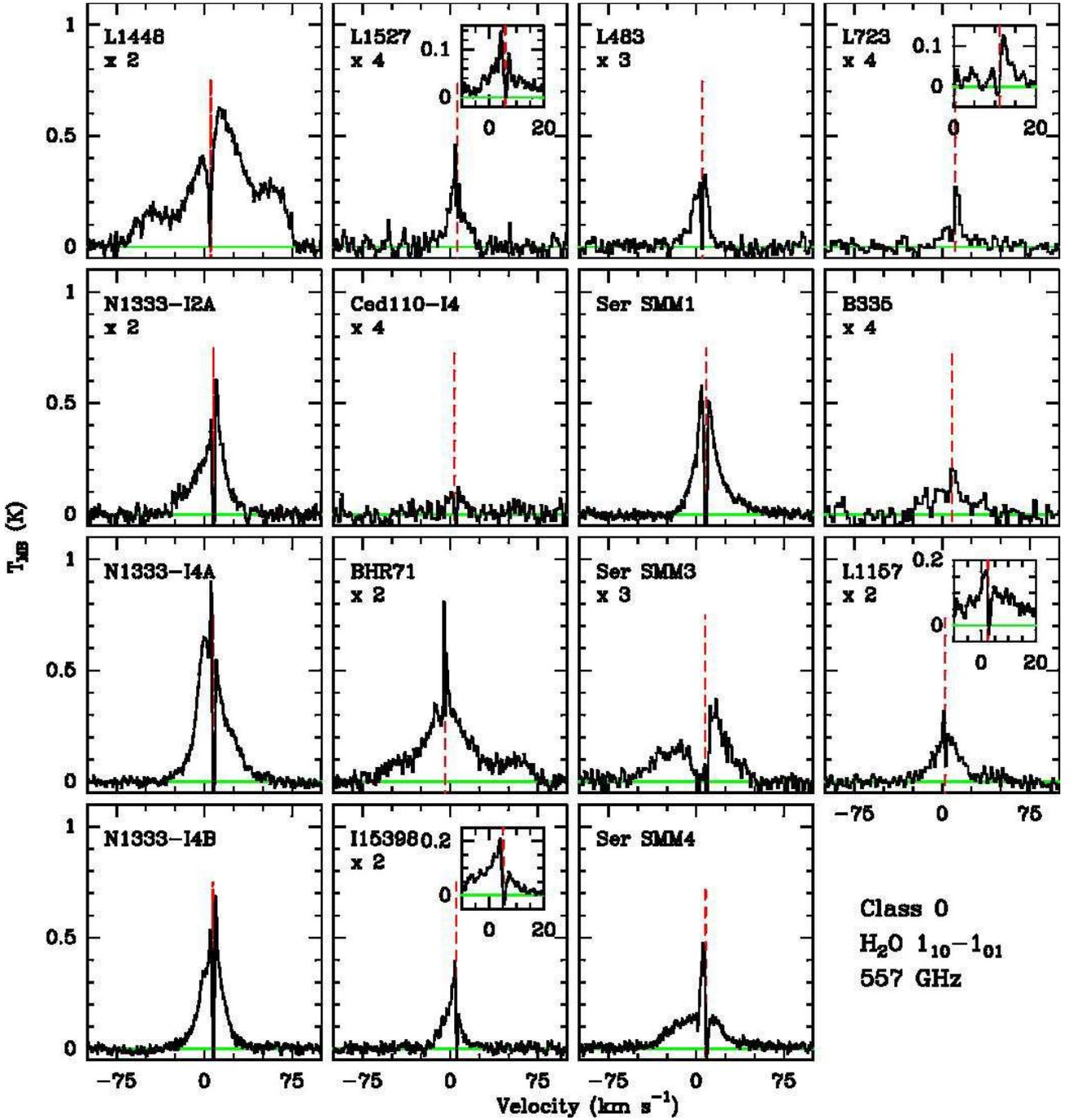}
\end{center}
\caption{Continuum-subtracted H$_2$O 1$_{10}$--1$_{01}$ spectra of the observed Class 0 sources. The red dashed line indicates the source velocity as derived from C$^{18}$O data \citep[][Y{\i}ld{\i}z et al. in prep.]{jorgensen02}. Some spectra have been rebinned for clarity, insets show unbinned spectra.}
\label{fig:class0}
\end{figure*}
\begin{figure*}
\begin{center}
\includegraphics[width = 17cm]{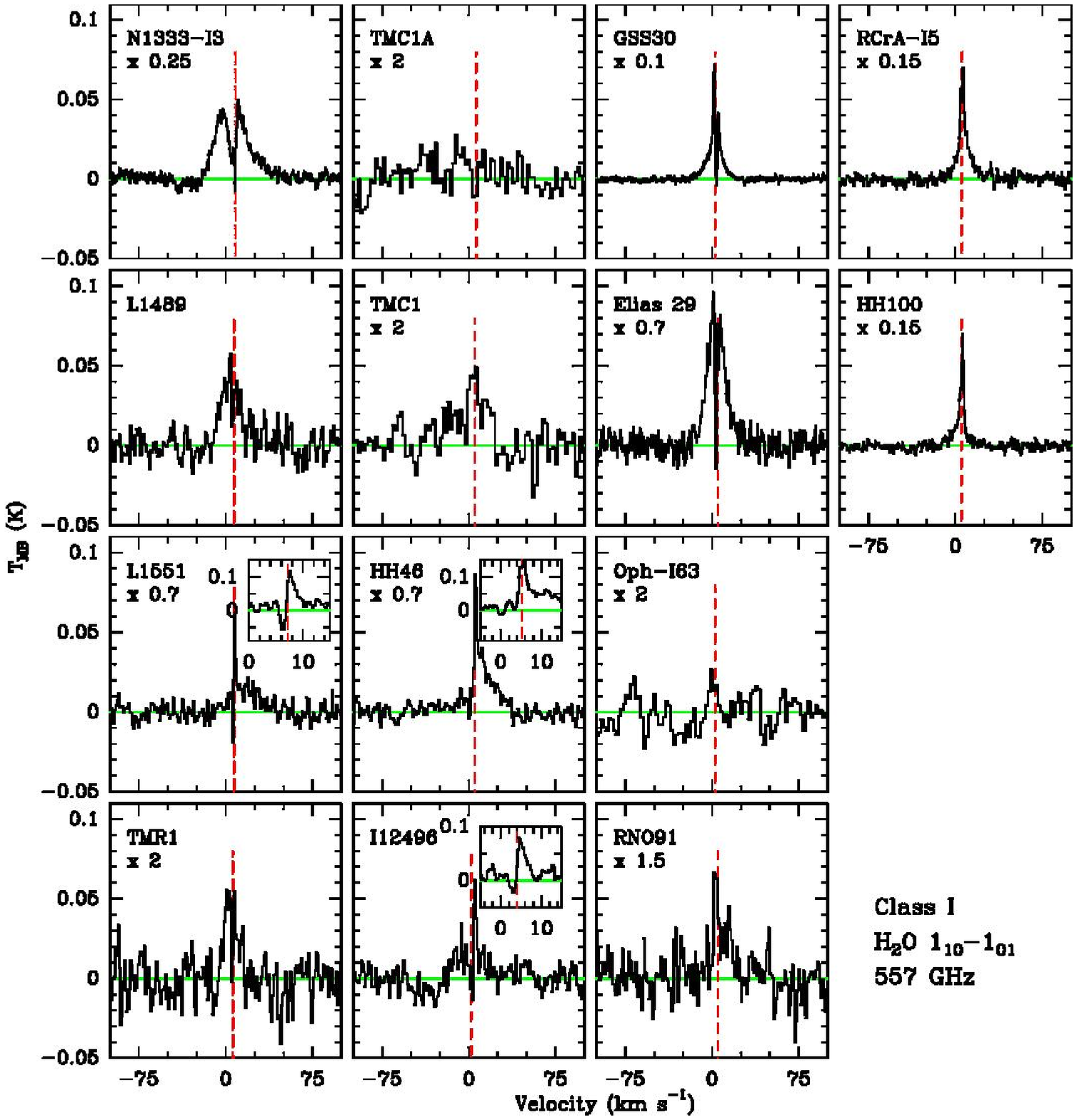}
\end{center}
\caption{Same as Fig. \ref{fig:class0} but for Class I sources.}
\label{fig:class1}
\end{figure*}

\subsection{\textit{Herschel}-HIFI observations}

A sample of 29 low-mass Class 0 and I YSOs was targeted with the Heterodyne Instrument for the Far-Infrared \citep[HIFI;][]{degraauw10} on the \textit{Herschel Space Observatory} \citep{pilbratt10}, see Table \ref{tab:source} for an overview of the sources. Observations were performed as single-pointing observations in either dual-beam-switch mode with a chop of 3\arcmin\ or position-switch mode with a throw large enough to ensure that the observations were not contaminated by cloud emission (reference positions are provided in Table \ref{tab:reference} in the Appendix). The latter mode was primarily used in crowded regions such as the NGC1333 region in Perseus, and the Ophiuchus star-forming region. The total telescope time (on + off + overhead) was 13 min for Class 0 sources and 20 min for Class I sources. HIFI was tuned to the 557 GHz H$_2$O 1$_{10}$--1$_{01}$ line in the lower sideband of Band 1b to avoid the strong spur in Band 1a; this setting does not include NH$_3$ 1$_0$--0$_0$ at 572 GHz in the upper sideband. The Wide Band Spectrometer and High Resolution Spectrometer backends were used for both horizontal and vertical polarizations. 

The beam size of HIFI is close to the diffraction limit at 557 GHz, i.e., 39\arcsec\ \citep{roelfsema12}. The pointing accuracy is better than $\sim$2\arcsec, with the caveat that the H- and V-polarization receivers are pointing towards slightly different positions in the sky (the separation is $\sim$ 6\farcs5). The calibration uncertainty is limited by the sideband gain ratio in Band 1b, but the overall uncertainty is likely $<$ 15\%. The velocity calibration is better than $\sim$100 kHz, or $\sim$0.05 km s$^{-1}$ at this frequency. 

Data were reduced with HIPE version 6.1 \citep{ott10} and exported to CLASS for further analysis. Typical system temperatures are $\sim$85 K. A main-beam efficiency of 0.76 is used to convert data from the antenna temperature scale ($T_{\rm A}^*$) to the main-beam temperature scale ($T_{\rm MB}$) \citep{roelfsema12}. H and V polarizations are averaged after individual inspection using a simple average. The rms noise in the V-polarization data is higher than in the H-polarization data by typically $\sim$20\%. The rms noise is typically 8--13 mK in 0.3 km s$^{-1}$ bins in the averaged spectra.

For the case of dual-beam-switch observations the baseline is always very stable and linear baselines are subtracted. The position-switched observations show more unstable baselines and higher-order polynomials are in some cases necessary to fit and subtract the baseline. The spectral ranges are 554.6 -- 558.6 GHz (lower sideband) and 566.6 -- 570.6 GHz (upper sideband). Detected lines besides H$_2$O include CH$_3$OH (568.566 GHz), CN (566.947 GHz) and SO$_2$ (555.666 and 567.593 GHz). 

The reduced spectra are all shown in Figs. \ref{fig:class0} (Class 0) and \ref{fig:class1} (Class I). A number of spectra show absorption against the continuum in addition to self-absorption. In the cases where these observations were done as dual-beam-switch (DBS) the data reduction was redone, and the level-1 reference-position data examined for emission. None of the DBS reference spectra shows emission, and thus the absorption is believed to be intrinsic to the source. Several of the absorption lines have flat bottoms at half the measured continuum level, i.e., they are saturated, thus corroborating that absorption below the continuum level is true. The observations of Serpens show a weak absorption feature at $\varv_{\rm LSR}$ = 1 km s$^{-1}$, a feature that also shows up in some observations of CO 3--2 \citep{dionatos10} but not all \citep{graves10}. This feature probably arises from emission in the reference position, and since it is easily identifiable and does not interfere with the science goals it is ignored.

\begin{table*}
\caption{Source parameters.}
\small
\begin{center}
\begin{tabular}{l c c c c c c c c}
\hline\hline
Source & RA & Dec & $D$ & $\varv_{\rm LSR}$\tablefootmark{a} & $L_{\rm bol}$\tablefootmark{b} & $T_{\rm bol}$\tablefootmark{b} & $M_{\rm env}$\tablefootmark{c} & References \\
       & (h m s) & (\degr\ \arcmin\ \arcsec) & (pc) & (km s$^{-1}$) & ($L_\odot$) & (K) & ($M_\odot$) \\ \hline
L1448-MM                  & 03 25 38.9 & $+$30 44 05.4 & 235 & \phantom{1}$+$4.7 & \phantom{1}9.0 & \phantom{1}46 &	\phantom{1}3.9 & 1,2 \\
NGC1333-IRAS2A            & 03 28 55.6 & $+$31 14 37.1 & 235 & \phantom{1}$+$7.7 &           35.7 & \phantom{1}50 &	\phantom{1}5.1 & 1,3 \\
NGC1333-IRAS4A            & 03 29 10.5 & $+$31 13 30.9 & 235 & \phantom{1}$+$7.2 & \phantom{1}9.1 & \phantom{1}33 &	\phantom{1}5.2 & 1,3 \\
NGC1333-IRAS4B            & 03 29 12.0 & $+$31 13 08.1 & 235 & \phantom{1}$+$7.4 & \phantom{1}4.4 & \phantom{1}28 &	\phantom{1}3.0 & 1,3 \\
L1527                     & 04 39 53.9 & $+$26 03 09.8 & 140 & \phantom{1}$+$5.9 & \phantom{1}1.9 & \phantom{1}44 &	\phantom{1}0.9 & 1,3 \\
Ced110-IRS4               & 11 06 47.0 & $-$77 22 32.4 & 125 & \phantom{1}$+$3.5 & \phantom{1}0.8 & \phantom{1}56 & \phantom{1}0.2 & 4,5 \\
BHR71                     & 12 01 36.3 & $-$65 08 53.0 & 200 & \phantom{1}$-$4.4 &           14.8 & \phantom{1}44 &	\phantom{1}3.1 & 6,7 \\
IRAS15398\tablefootmark{e}& 15 43 01.3 & $-$34 09 15.0 & 130 & \phantom{1}$+$5.1 & \phantom{1}1.6 & \phantom{1}52 &	\phantom{1}0.5 & 4,5 \\
L483                      & 18 17 29.9 & $-$04 39 39.5 & 200 & \phantom{1}$+$5.2 &           10.2 & \phantom{1}49 &	\phantom{1}4.4 & 1,8 \\
Ser SMM1 & 18 29 49.8 & $+$01 15 20.5 & 230\tablefootmark{d} & \phantom{1}$+$8.5 &           30.4 & \phantom{1}39 &	          16.1 & 9,10 \\
Ser SMM4 & 18 29 56.6 & $+$01 13 15.1 & 230\tablefootmark{d} & \phantom{1}$+$8.0 & \phantom{1}1.9 & \phantom{1}26 & \phantom{1}1.9 & 9,10 \\
Ser SMM3 & 18 29 59.2 & $+$01 14 00.3 & 230\tablefootmark{d} & \phantom{1}$+$7.6 & \phantom{1}5.1 & \phantom{1}38 &	\phantom{1}3.2 & 9,10 \\
L723                      & 19 17 53.7 & $+$19 12 20.0 & 300 &           $+$11.2 & \phantom{1}3.6 & \phantom{1}39 &	\phantom{1}1.3 & 11,12 \\
B335                      & 19 37 00.9 & $+$07 34 09.6 & 250 & \phantom{1}$+$8.4 & \phantom{1}3.3 & \phantom{1}36 &	\phantom{1}1.2 & 1,13 \\
L1157                     & 20 39 06.3 & $+$68 02 15.8 & 325 & \phantom{1}$+$2.6 & \phantom{1}4.7 & \phantom{1}46 &	\phantom{1}1.5 & 1,14 \\ \hline
NGC1333-IRAS3             & 03 29 03.8 & $+$31 16 04.0 & 235 & \phantom{1}$+$8.5 &           41.8 &           149 &	\phantom{1}8.6 & 15,2 \\
L1489                     & 04 04 43.0 & $+$26 18 57.0 & 140 & \phantom{1}$+$7.2 & \phantom{1}3.8 &           200 & \phantom{1}0.2 & 16,17 \\
L1551-IRS5                & 04 31 34.1 & $+$18 08 05.0 & 140 & \phantom{1}$+$7.2 &           22.1 & \phantom{1}94 & \phantom{1}2.3 & 11,17 \\
TMR1\tablefootmark{e}     & 04 39 13.7 & $+$25 53 21.0 & 140 & \phantom{1}$+$6.3 & \phantom{1}3.8 &           133 &	\phantom{1}0.2 & 18,17 \\
TMC1A\tablefootmark{e}    & 04 39 34.9 & $+$25 41 45.0 & 140 & \phantom{1}$+$6.6 & \phantom{1}2.7 &           118 &	\phantom{1}0.3 & 18,17 \\
TMC1                      & 04 41 12.4 & $+$25 46 36.0 & 140 & \phantom{1}$+$5.2 & \phantom{1}0.9 &           101 &	\phantom{1}0.2 & 18,17 \\
HH46-IRS                  & 08 25 43.9 & $-$51 00 36.0 & 450 & \phantom{1}$+$5.2 &           27.9 &           104 &	\phantom{1}4.4 & 19,20 \\
IRAS12496                 & 12 53 17.2 & $-$77 07 10.6 & 178 & \phantom{1}$+$2.3 &           35.4 &           569 &	\phantom{1}0.8 & 4,21 \\
GSS30-IRS1                & 16 26 21.4 & $-$24 23 04.0 & 125 & \phantom{1}$+$2.8 &           13.9 &           142 & \phantom{1}0.6 & 22,23 \\
Elias 29                  & 16 27 09.4 & $-$24 37 19.6 & 125 & \phantom{1}$+$5.0 &           14.1 &           299 & \phantom{1}0.3 & 24,23 \\
Oph-IRS63                 & 16 31 35.6 & $-$24 01 29.6 & 125 & \phantom{1}$+$2.8 & \phantom{1}1.0 &           327 & \phantom{1}0.3 & 24,23 \\
RNO91                     & 16 34 29.3 & $-$15 47 01.4 & 125 & \phantom{1}$+$5.0 & \phantom{1}2.6 &           340 & \phantom{1}0.5 & 4,23 \\
RCrA-IRS5A                & 19 01 48.0 & $-$36 57 21.6 & 130 & \phantom{1}$+$5.7 & \phantom{1}7.1 &           126 & \phantom{1}2.0 & 4,5 \\
HH100-IRS                 & 19 01 49.1 & $-$36 58 16.0 & 130 & \phantom{1}$+$5.6 &           17.7 &           256 & \phantom{1}8.1 & 4,5 \\
\hline
\end{tabular}
\tablefoot{Sources above the horizontal line are Class 0, sources below are Class I.
	\tablefoottext{a}{Obtained from ground-based C$^{18}$O or C$^{17}$O observations.}
	\tablefoottext{b}{Measured using \textit{Herschel}-PACS data from the WISH and DIGIT key programmes (Karska et al. in prep., Green et al. in prep.).}
	\tablefoottext{c}{Determined from DUSTY modelling of the sources; see Appendix \ref{app:dusty}.}
	\tablefoottext{d}{Using VLBA observations of a star thought to be associated with the Serpens cluster, \citet{dzib10} derive a distance of 415 pc.}
	\tablefoottext{e}{The coordinates used in WISH; more accurate SMA coordinates of the sources are 15$^{\rm h}$43$^{\rm m}$02\fs2, $-$34\degr09\arcmin06\farcs8 (IRAS15398), 04$^{\rm h}$39$^{\rm m}$13\fs9, $+$25\degr53\arcmin20\farcs6 (TMR1) and 04$^{\rm h}$39$^{\rm m}$35\fs2, $+$25\degr41\arcmin44\farcs4 \citep[TMC1A;][]{jorgensen09}.}
	}\tablebib{
	(1) \citet{jorgensen07}; (2) \citet{hirota11}; (3) \citet{hirota08}; (4) \citet{evans09}; (5) \citet{knude98}; (6) \citet{chen08}; (7) \citet{seidensticker89}; (8) \citet{fuller95}; (9) \citet{hogerheijde99}; (10) \citet{eiroa08}; (11) \citet{jorgensen02}; (12) \citet{goldsmith84}; (13) \citet{frerking87}; (14) \citet{straizys92}; (15) \citet{bachiller98}; (16) \citet{brinch07}; (17) \citet{kenyon08}; (18) \citet{hogerheijde98}; (19) \citet{velusamy07}; (20) \citet{noriega-crespo04}; (21) \citet{whittet97}; (22) \citet{jorgensen09}; (23) \citet{degeus89}; (24) \citet{lommen08}.}
\end{center}
\label{tab:source}
\end{table*}

\subsection{Complementary data}

\subsubsection{CO emission and $F_{\rm CO}$}

In addition to the \textit{Herschel} data, CO 3--2 data at 345 GHz have been obtained at the JCMT and APEX \citep[][Y{\i}ld{\i}z et al. in prep.]{vankempen06, vankempen09a, vankempen09b, parise06} in the form of Nyquist-sampled maps that have subsequently been convolved with a 39\arcsec\ beam to be directly comparable to the H$_2$O data presented here. The CO data are presented in the Appendix (Fig. \ref{fig:co3-2} and Table \ref{tab:intensity_co}) and will be further discussed in Y{\i}ld{\i}z et al. (in prep.).

Carbon monoxide is clearly detected in all sources. The profiles have higher $T_{\rm MB}^{\rm peak}$ than H$_2$O by more than an order of magnitude (up to 25 K) and consist typically of a single Gaussian component with $\Delta\varv$ $\sim$ 10 km s$^{-1}$, with a self-absorption feature located at the source velocity\footnote{$\Delta\varv$ is used throughout to refer to full width at half maximum and is always obtained from Gaussian fitting; $\Delta\varv_{\rm max}$ refers to the full width at 4-$\sigma$ rms in 2 km s$^{-1}$ channels (representative of the full width at zero intensity) and is always measured from the data, i.e., without fitting a profile function.}. Two sources, L1448-MM and BHR71, also show bullet emission in the CO 3--2 line; this was known for L1448-MM \citep{bachiller90} but not reported for BHR71 \citep{parise06}.

Values for the CO outflow force, $F_{\rm CO}$, have been assembled from the literature (Table \ref{tab:intensity_co} in the Appendix). Values of $F_{\rm CO}$ are chosen such that they are all measured using the same method and corrections. The corrections consist of taking the opacity of the emission in the outflow wings into account and correcting for inclination effects using the correction curves of \citet{cabrit92}. When more than one value exist in the literature, the average is used here. Eventually these values will be derived in a coherent manner for all sources from our own data (Y{\i}ld{\i}z et al. in prep.); the literature data are used to obtain preliminary results.

\subsubsection{Bolometric luminosity and temperature}

The bolometric luminosity, $L_{\rm bol}$, and temperature, $T_{\rm bol}$, are measured for each source using recent \textit{Spitzer}-IRAC and MIPS fluxes \citep{evans09} and 2MASS fluxes when available \citep{skrutskie06}. Furthermore, \textit{Herschel}-PACS fluxes obtained within the WISH \citep{vandishoeck11} and DIGIT (Dust, Ice and Gas In Time; PI: N. Evans) key programmes are used, when available, as well as sub-mm and mm fluxes obtained from the literature. The data reduction and analysis of the PACS data will be presented in two forthcoming papers (Karska et al. in prep.; Green et al. in prep.). The newer values of $L_{\rm bol}$ do not change significantly ($<$20\%) for most sources compared to literature values. The bolometric temperature, $T_{\rm bol}$, is also measured for each source using the same data; the availability of the PACS data covering the peak of the spectral energy distribution in most sources has improved the value significantly \citep{evans09}. Using the bolometric temperature to classify sources as either Class 0 ($T_{\rm bol}$$<$70 K) or Class I ($T_{\rm bol}$$>$70 K) implies that Ced110-IRS4 and IRAS15398 are now Class 0 sources and NGC1333-IRAS3 is a Class I source.

\subsubsection{DUSTY modelling}

To quantify the large-scale physical structure of the envelopes and to compare the water emission to envelope parameters, each source is modelled using the 1D dust radiative transfer code DUSTY \citep{ivezic97} following the prescriptions outlined in \citet{schoier02} and \citet{jorgensen02}. Details of the procedure are provided in Appendix \ref{app:dusty}. The density is assumed to have a power-law structure, and the temperature is obtained as a function of position by solving for the dust radiative transfer through the (assumed) spherical envelopes, given the luminosity. The free parameters are fit to the SED and the spatial extent of the sub-mm continuum. Furthermore, the mass of each envelope is obtained. The mass is calculated at the smallest radius where either the dust temperature reaches 10 K or where the density drops below 10$^4$ cm$^{-3}$, in both cases corresponding to where the envelope merges with the ambient cloud.

\section{Results}\label{sec:results}

\begin{figure}
\begin{center}
\includegraphics[width = 0.8\columnwidth]{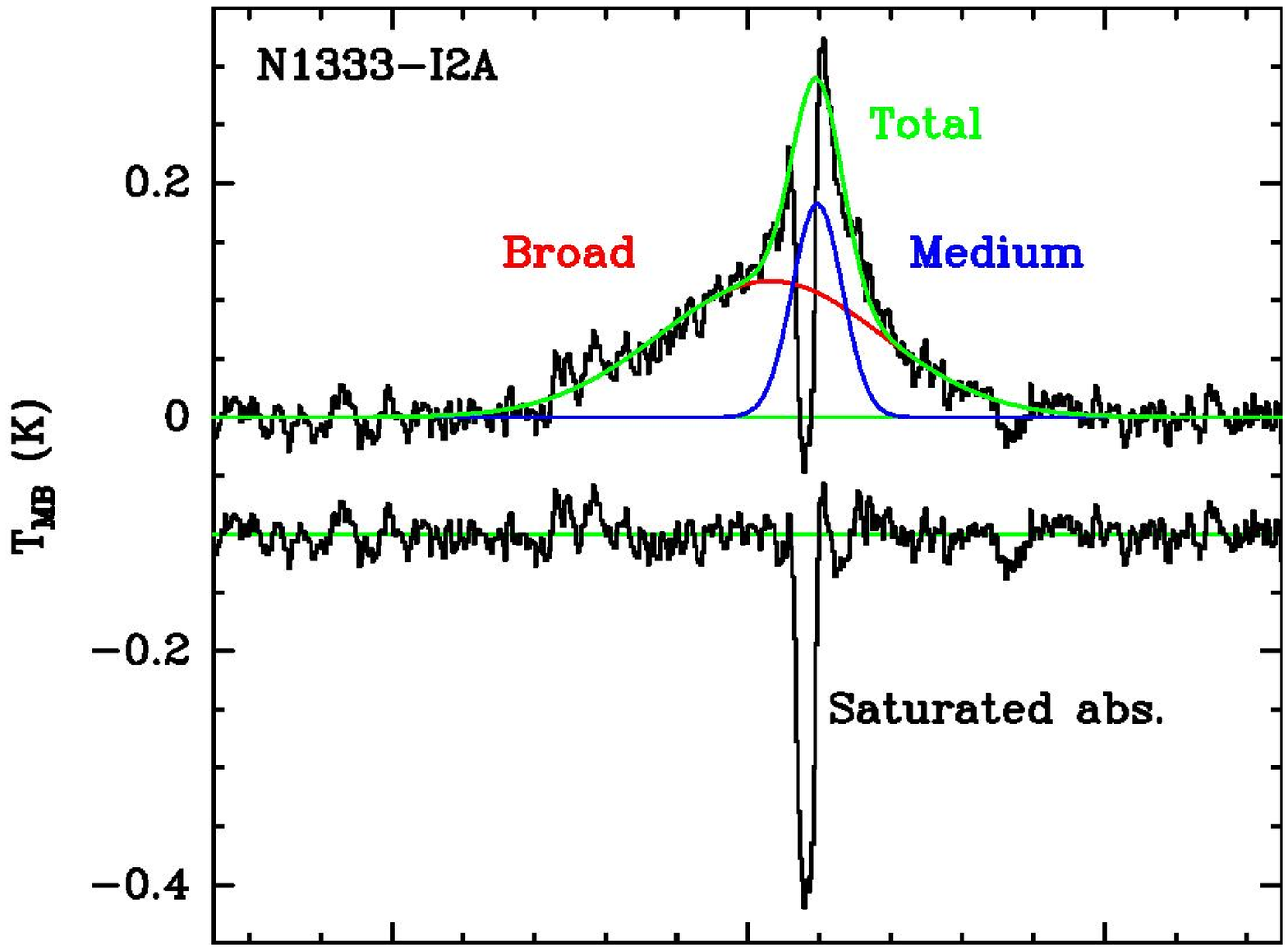}
\includegraphics[width = 0.8\columnwidth]{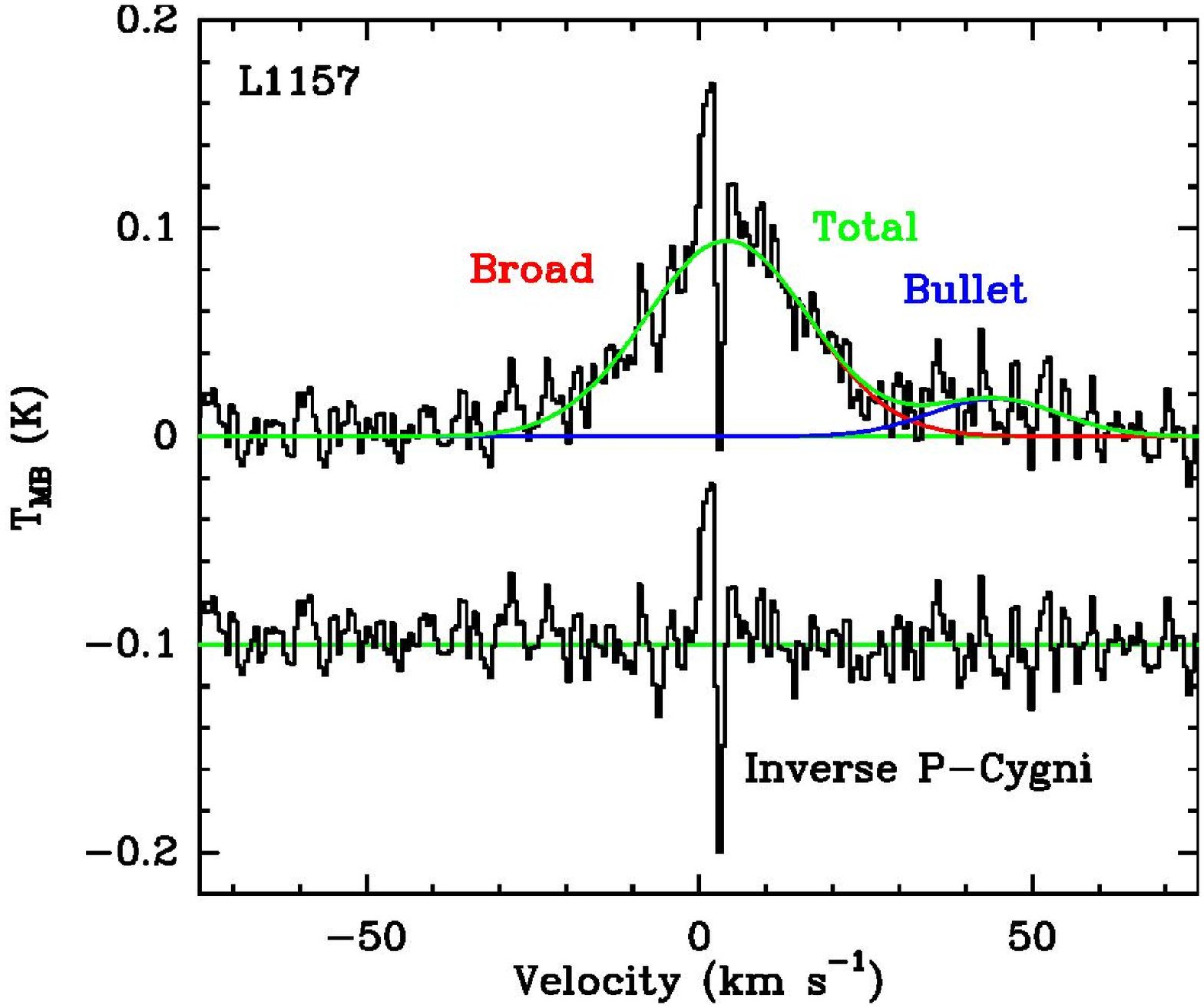}
\end{center}
\caption{Gaussian decomposition of the H$_2$O 1$_{10}$--1${01}$ spectra towards the two sources NGC1333-IRAS2A and L1157. The individual Gaussian components are shown overlaid on the spectra and the residual is plotted below. The saturated absorption and inverse P-Cygni profile are not fitted by Gaussians.}
\label{fig:gauss_decomp}
\end{figure}

\begin{table}
\caption{Observed properties of the H$_2$O line profiles.}
\centering
\tiny
\begin{tabular}{l c c c r}
\hline \hline
Source & rms\tablefootmark{a}  & $T_{\rm MB}^{\rm peak}$ & $\int T_{\rm MB}^{\rm total}~{\rm d}\varv$\tablefootmark{b} & $\Delta\varv_{\rm max}$\tablefootmark{c} \\
       & (mK) & (K) & (K\,km\,s$^{-1}$) & (km\,s$^{-1}$) \\
\hline
L1448-MM				&				12	&	0.32	&			18.7	&	138	\\
NGC1333-IRAS2A			&				12	&	0.31	&	\phantom{1}5.2	&	58	\\
NGC1333-IRAS4A			&				12	&	0.90	&			17.4	&	84	\\
NGC1333-IRAS4B			&				12	&	0.69	&			10.6	&	54	\\
L1527					&				10	&	0.14	&	\phantom{1}1.2	&	22	\\
Ced110-IRS4				&	\phantom{1}8	&	0.03	&	\phantom{1}0.3	&	11	\\
BHR71					&				12	&	0.44	&			10.2	&	131	\\
IRAS15398				&	\phantom{1}8	&	0.21	&	\phantom{1}2.0	&	28	\\
L483					&				12	&	0.14	&	\phantom{1}3.2	&	22	\\
Ser SMM1				&				12	&	0.58	&			10.8	&	69	\\
Ser SMM4				&				12	&	0.47	&	\phantom{1}8.8	&	80	\\
Ser SMM3				&				12	&	0.13	&	\phantom{1}3.6	&	80	\\
L723					&				11	&	0.13	&	\phantom{1}0.7	&	15	\\
B335					&				13	&	0.09	&	\phantom{1}1.1	&	26	\\
L1157					&				13	&	0.17	&	\phantom{1}3.0	&	67	\\ \hline
NGC1333-IRAS3			&				11	&	0.20	&	\phantom{1}5.2	&	54	\\
L1489					&				11	&	0.08	&	\phantom{1}0.8	&	41	\\
L1551-IRS5				&	\phantom{1}9	&	0.12	&	\phantom{1}0.7	&	19	\\
TMR1					&	\phantom{1}8	&	0.04	&	\phantom{1}0.3	&	9	\\
TMC1A					&	\phantom{1}8	&	0.04	&	$<$0.1 & \ldots	\\
TMC1					&				11	&	0.05	&	\phantom{1}0.5	&	19	\\
HH46-IRS				&	\phantom{1}8	&	0.15	&	\phantom{1}1.2	&	24	\\
IRAS12496				&	\phantom{1}8	&	0.08	&	\phantom{1}0.6	&	24	\\
GSS30-IRS1\tablefootmark{d}	&			10	&	0.72	&	\phantom{1}4.4	&	35	\\
Elias29					&				10	&	0.14	&	\phantom{1}2.0	&	28	\\
Oph-IRS63				&				10	&	0.02	&	\phantom{1}0.1	&	5	\\
RNO91					&				10	&	0.16	&	\phantom{1}0.5	&	13	\\
RCrA-IRS5A				&				17	&	0.50	&	\phantom{1}3.8	&	34	\\
HH100-IRS				&				16	&	0.63	&	\phantom{1}5.3	&	28	\\
\hline\\
\end{tabular}
\tablefoot{
	\tablefoottext{a}{Measured in 0.3 km\,s$^{-1}$ bins.}
	\tablefoottext{b}{Integrated over the entire line, including bullet emission.}
	\tablefoottext{c}{Spectra smoothed to 2 km s$^{-1}$ resolution and line width measured at the 4-$\sigma$ level.}
	\tablefoottext{d}{Possible contamination from the VLA1623 outflow.}
		}
\label{tab:intensity_sum}
\end{table}

\begin{figure}
\begin{center}
\includegraphics[width = 0.9\columnwidth]{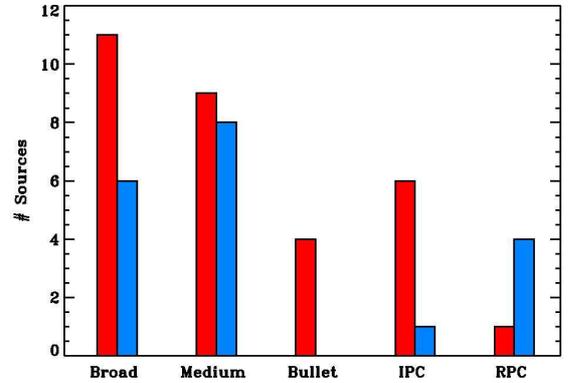}
\end{center}
\caption{Histogram over detected features in Class 0 (red) and Class I (blue) sources; the features include the broad and medium components, bullets and inverse and regular P-Cygni profiles (IPC and RPC, respectively).}
\label{fig:feature}
\end{figure}

\begin{figure}
\begin{center}
\includegraphics[width = 0.9\columnwidth]{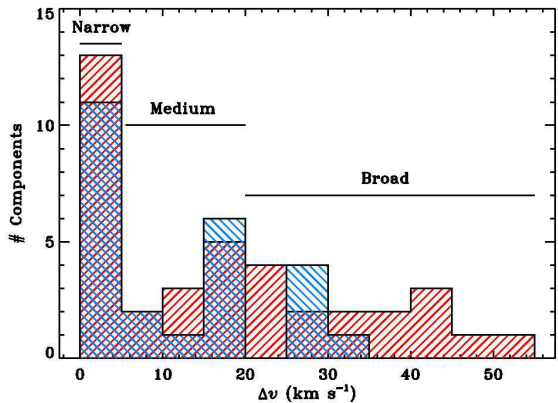}
\end{center}
\caption{Distribution of $\Delta\varv$ of the different Gaussian components. Class 0 sources are marked in red, and Class I sources in blue. The horizontal lines show the range of velocities adopted for the narrow, medium and broad components.}
\label{fig:histo}
\end{figure}

\subsection{H$_2$O at 557 GHz}\label{sec:h2o_results}

The H$_2$O 557 GHz line is detected in all Class 0 sources and all but one Class I source, TMC1A, and the full data set is presented in Figs. \ref{fig:class0} (Class 0) and \ref{fig:class1} (Class I). These observations represent the first detection of cold H$_2$O emission from Class I YSOs, i.e., emission from the ground-state 557 GHz transition. Integrated line intensities are provided in Table \ref{tab:intensity_sum}. The dynamical range of the integrated intensities is more than two orders of magnitude, from $\sim$0.1 K km s$^{-1}$ (Oph-IRS63) to $\sim$20 K km s$^{-1}$ (L1448-MM and NGC1333-IRAS4A). On average, the integrated intensity of the 1$_{10}$--1$_{01}$ line is almost an order of magnitude weaker in Class I sources than in Class 0 sources, explaining why the emission was not detected with \textit{Odin} and SWAS. Peak intensities span more than an order of magnitude from $\sim$0.03 K to 0.9 K. While the Class 0 sources typically have higher peak intensities than the Class I sources, there are notable exceptions in both categories (Class 0 YSOs with low peak intensity: Ced110-IRS4, Ser SMM3, L723, L483 and L1527; Class I YSOs with high peak intensity: R CrA-IRS5, HH100-IRS and GSS30-IRS1). Both R CrA-IRS5 and HH100-IRS are embedded in the dense photodissociating region (PDR) created by the Herbig Ae star, R CrA \citep{neuhauser08}, which may be causing the higher H$_2$O intensity in these two sources. Indeed, both profiles show a strong narrow component emission component ($\Delta\varv$ $\sim$ 3--5 km s$^{-1}$).

The main difference between the Class 0 and Class I YSOs leading to the dichotomy in integrated intensity is the linewidth. The linewidth at zero intensity, here defined as the 4-$\sigma$ level, is also listed in Table \ref{tab:intensity_sum}, and the range extends over more than one order of magnitude. The maximum linewidth, $\Delta\varv_{\rm max}$, includes contributions from bullets. The line profiles are very different from envelope-only model predictions \citep{poelman07, vankempen08}, indicating that very little emission is coming directly from the envelope, but that the bulk arises from outflows. In general, very little water emission originates in cold, quiescent regions \citep[e.g.,][]{bergin10, caselli10, hogerheijde11}.

\begin{table*}
\caption{Observed features in the H$_2$O 1$_{10}$--1$_{01}$ 557 GHz line profiles\tablefootmark{a}.}
\small
\begin{center}
\begin{tabular}{l c c c c c c c c}
\hline\hline
Source & Broad\tablefootmark{b} & Medium\tablefootmark{c} & Narrow\tablefootmark{d} & Bullets & Saturated abs. & Multiple abs. & Inverse P-Cygni & P-Cygni \\ \hline
L1448-MM        & E &                 & A\phantom{$+E$} & X   &       \\
NGC1333-IRAS2A  & E & \phantom{$A+$}E & A\phantom{$+E$} &     & X &   \\
NGC1333-IRAS4A  & E & \phantom{$A+$}E & A+E             &     & X &   & X \\
NGC1333-IRAS4B  & E & \phantom{$A+$}E & A & X\tablefootmark{e} & X \\
L1527           & E &                 & A+E             &  &  &  & X \\
Ced110-IRS4       &   & \phantom{$A+$}E & A\phantom{$+E$} &     & X \\
BHR71           & E &	              & A+E             & X   &   &  & X \\
IRAS15398       &   & \phantom{$A+$}E & A+E             &     & X &   & X \\	
L483            &   & \phantom{$A+$}E & A\phantom{$+E$} &     & X \\
Ser SMM1        & E & \phantom{$A+$}E & A\phantom{$+E$} &     & X & X \\
Ser SMM4        & E &                 & A+E             &     & X & X & X \\
Ser SMM3        & E & A\phantom{$+E$} & A\phantom{$+E$} &     & X & X \\
L723            &   & \phantom{$A+$}E & A+E             &     & X &   &   & X \\
B335            & E &                 & \phantom{$A+$}E &     &   &   &   & \\
L1157           & E &                 & A+E             & X   &   &   & X \\ \hline
NGC1333-IRAS3   & E & A\phantom{$+E$} & A\phantom{$+E$} &     &   & X \\
L1489           & E &                 & A\phantom{$+E$} & \\
L1551-IRS5      & E &                 & A+E             &     & X &   &   & X \\
TMR1            &   & \phantom{$A+$}E & A\phantom{$+E$} & \\
TMC1A           &  \\
TMC1            & E &                 &                 & \\
HH46-IRS        & E &                 & A+E             &     &   &   &   & X \\
IRAS12496       & E &                 & A+E             &     & X &   &   & X \\
GSS30-IRS1      &   & \phantom{$A+$}E & A+E             &     & X & & X \\
Elias 29        &   & \phantom{$A+$}E & A+E             &     & X \\
Oph-IRS63       &   & \phantom{$A+$}E &                 & \\
RNO91           &   & \phantom{$A+$}E &  & \\
RCrA-IRS5A      &   & \phantom{$A+$}E & A+E             & \\
HH100-IRS       &   & \phantom{$A+$}E & \phantom{$A+$}E & \\
\hline
\end{tabular}
\tablefoot{
	\tablefoottext{a}{Detected components marked with E (emission) or A (absorption). Detected features are marked with X.}
	\tablefoottext{b}{$\Delta\varv$ $>$ 20 km s$^{-1}$.}
	\tablefoottext{c}{5 $<$ $\Delta\varv$ $<$ 20 km s$^{-1}$.}
	\tablefoottext{d}{$\Delta\varv$ $<$ 5 km s$^{-1}$.}
	\tablefoottext{e}{The bullet emission is only detected in the excited 2$_{02}$--1$_{11}$ line at 988 GHz.}}
\end{center}
\label{tab:features}
\end{table*}

\begin{figure*}
\begin{center}
\begin{minipage}{8.5cm}
\includegraphics[width=7.8cm]{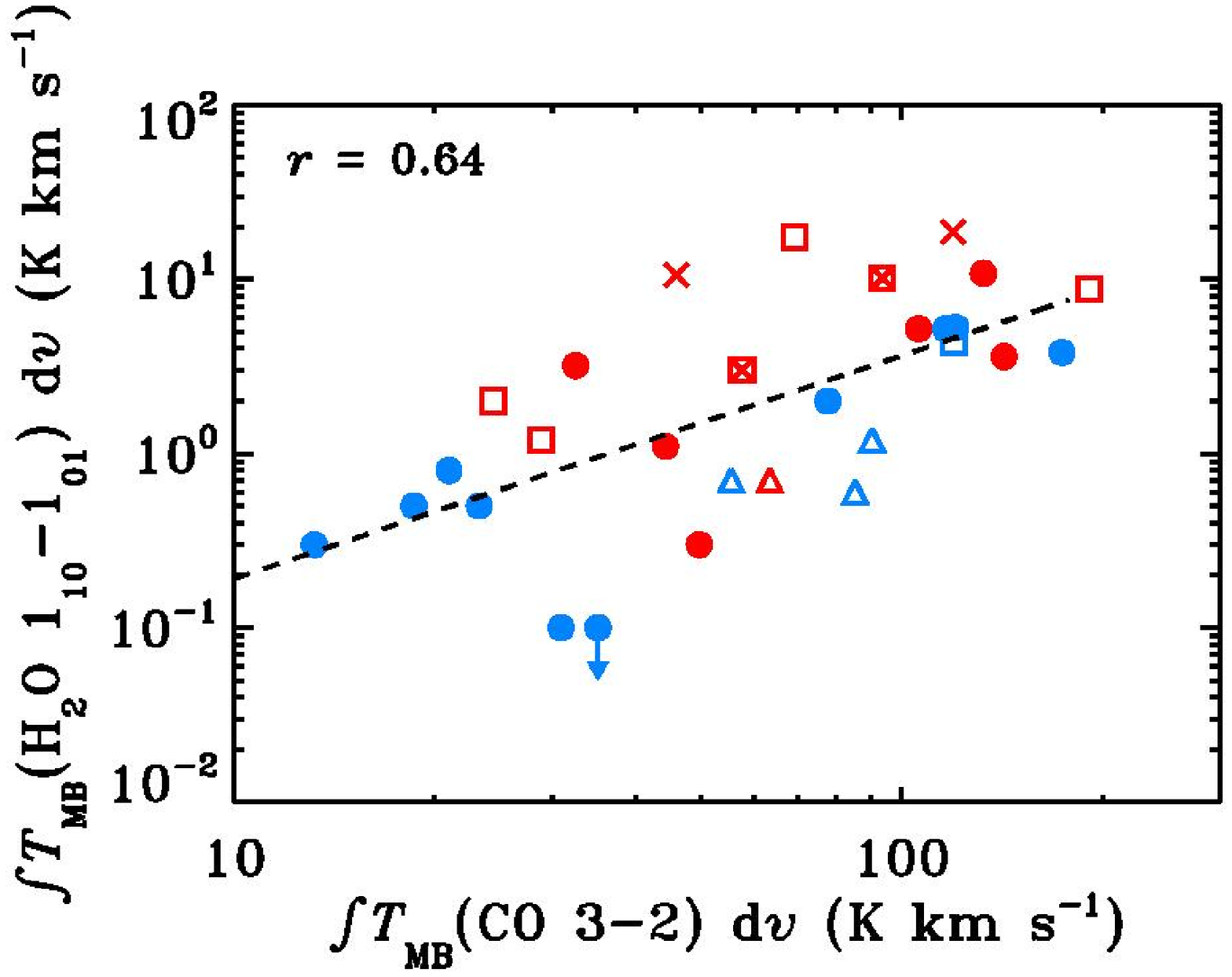}
\includegraphics[width=7.8cm]{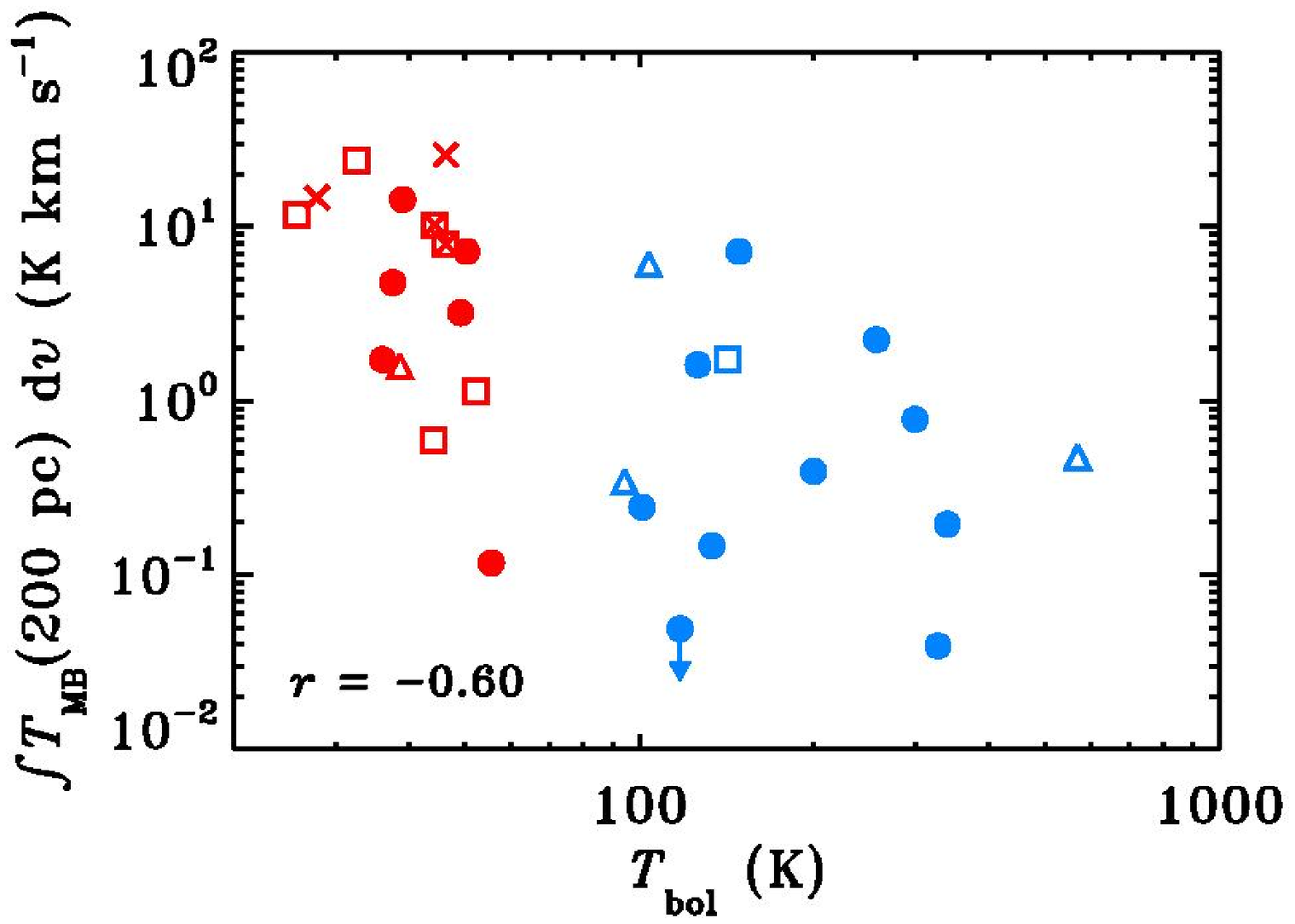}
\includegraphics[width=7.8cm]{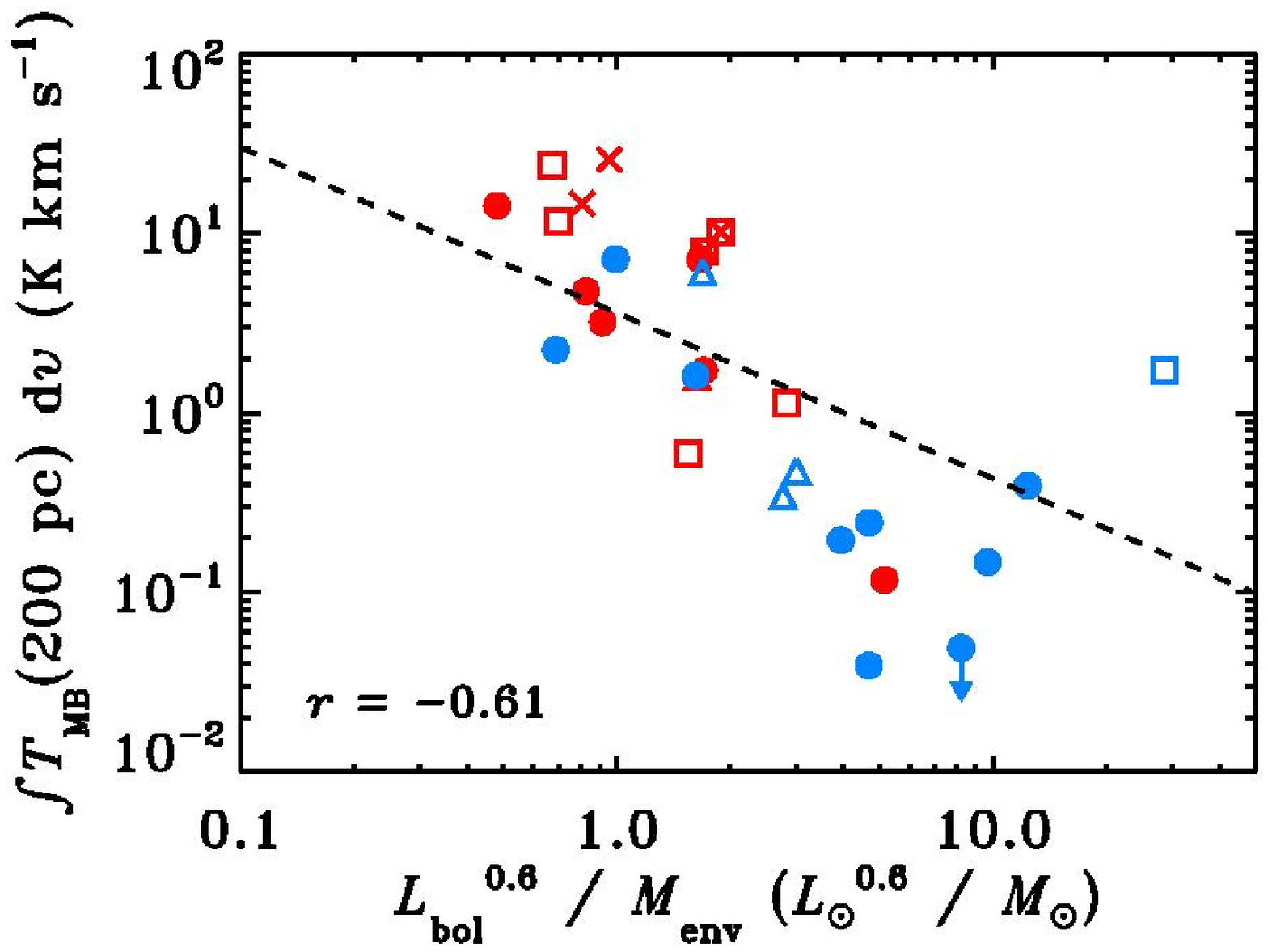}
\includegraphics[width=7.8cm]{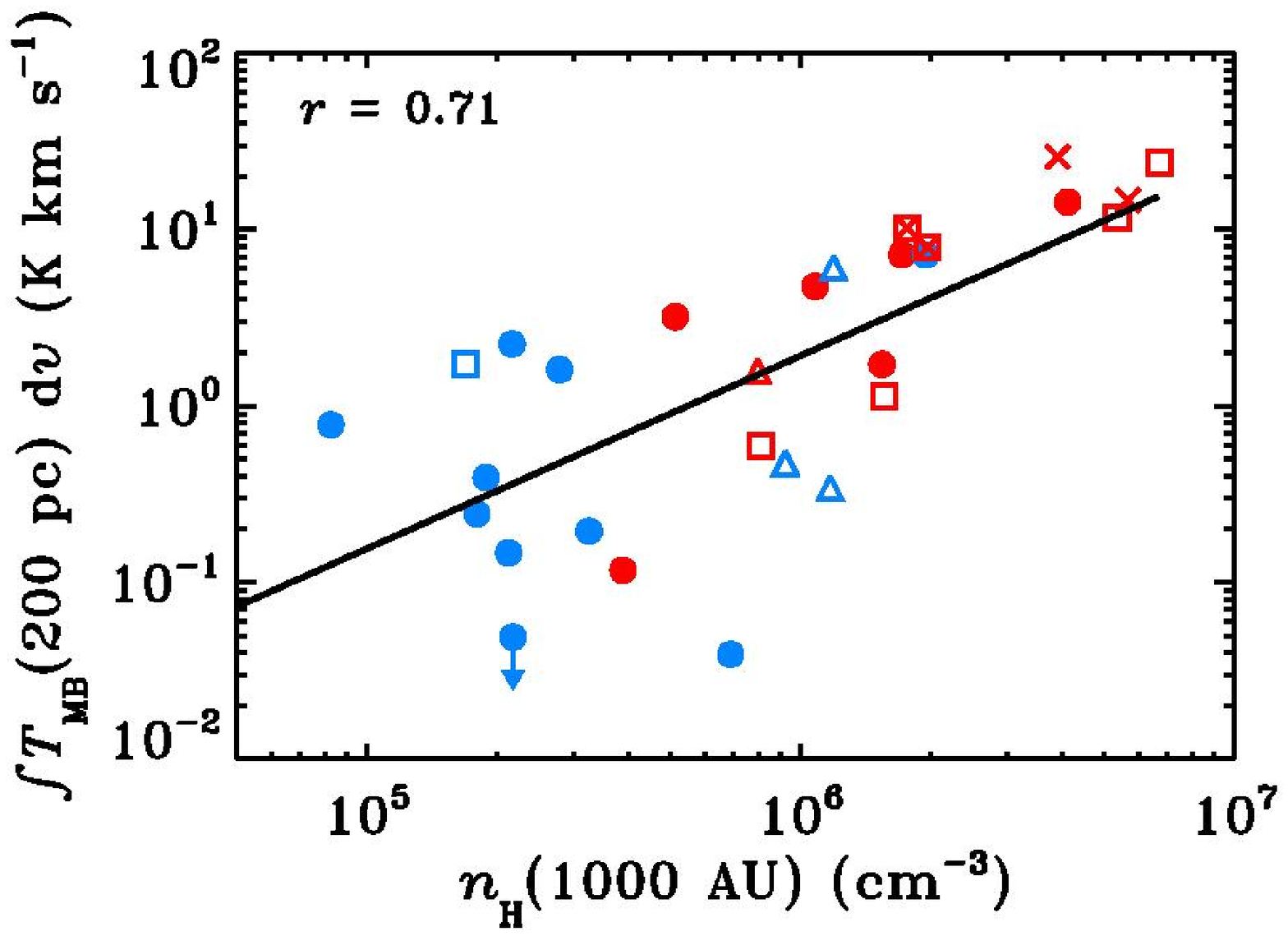}
\end{minipage}
\begin{minipage}{8.5cm}
\includegraphics[width=7.8cm]{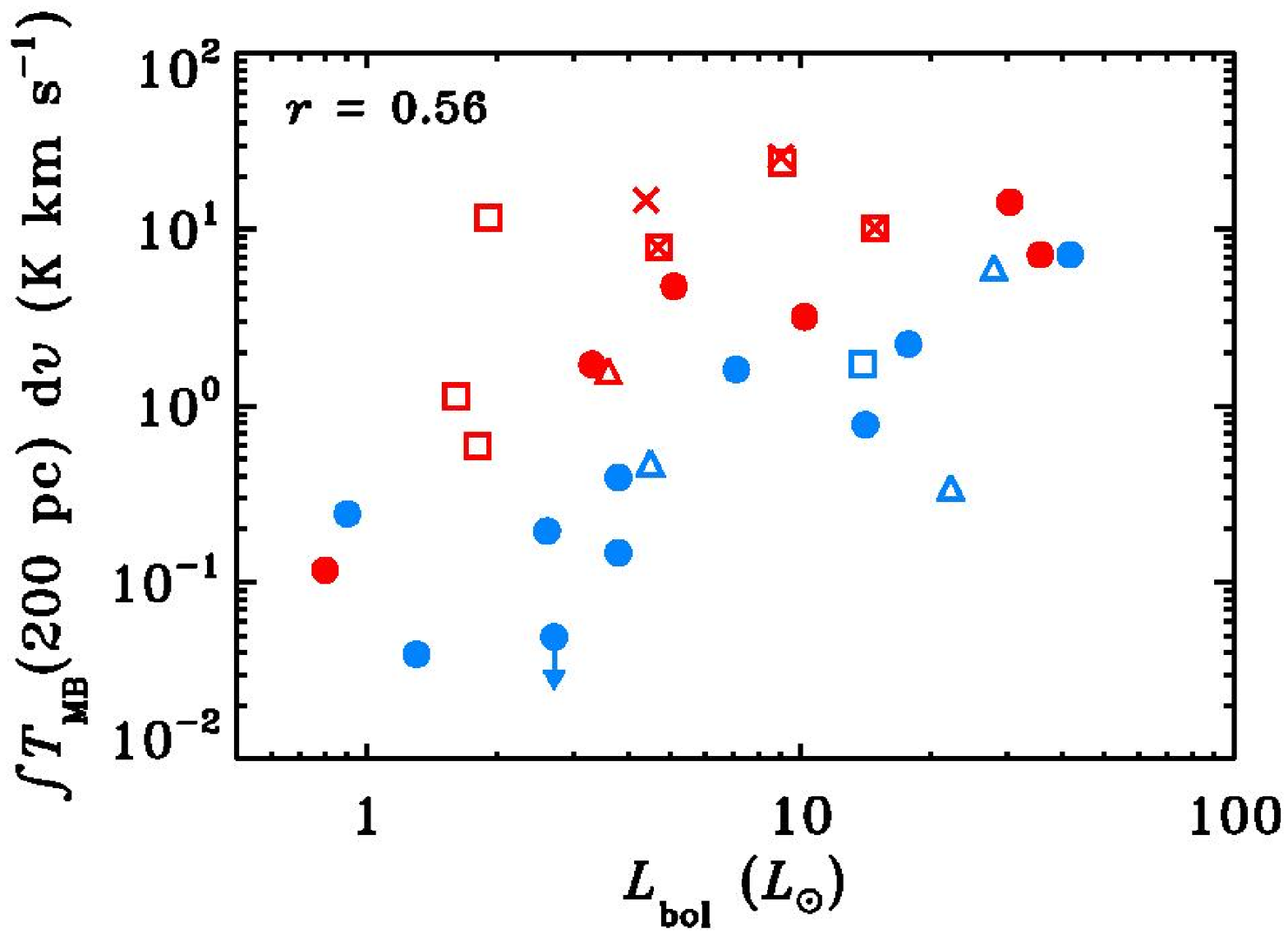}
\includegraphics[width=7.8cm]{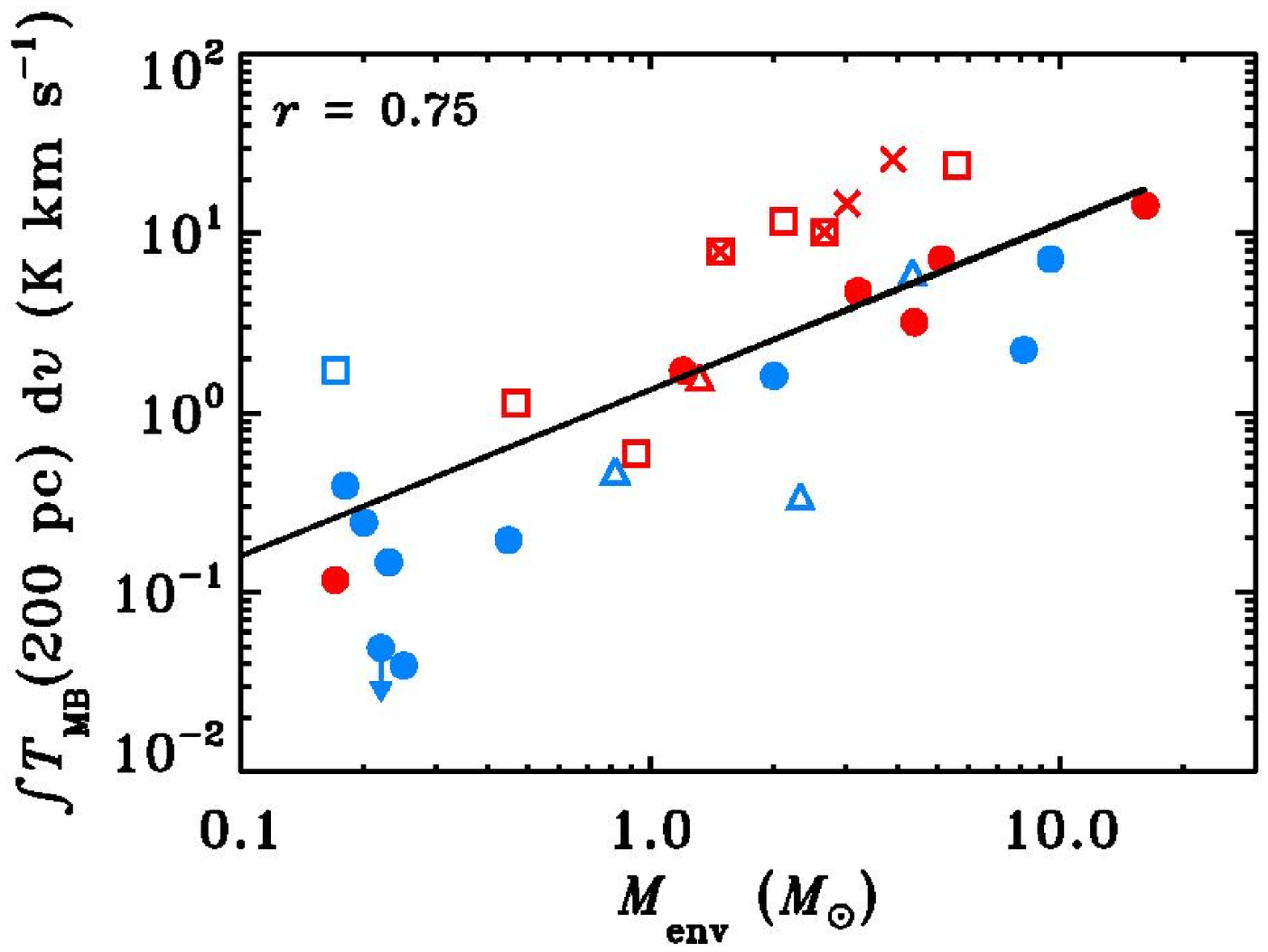}
\includegraphics[width=7.8cm]{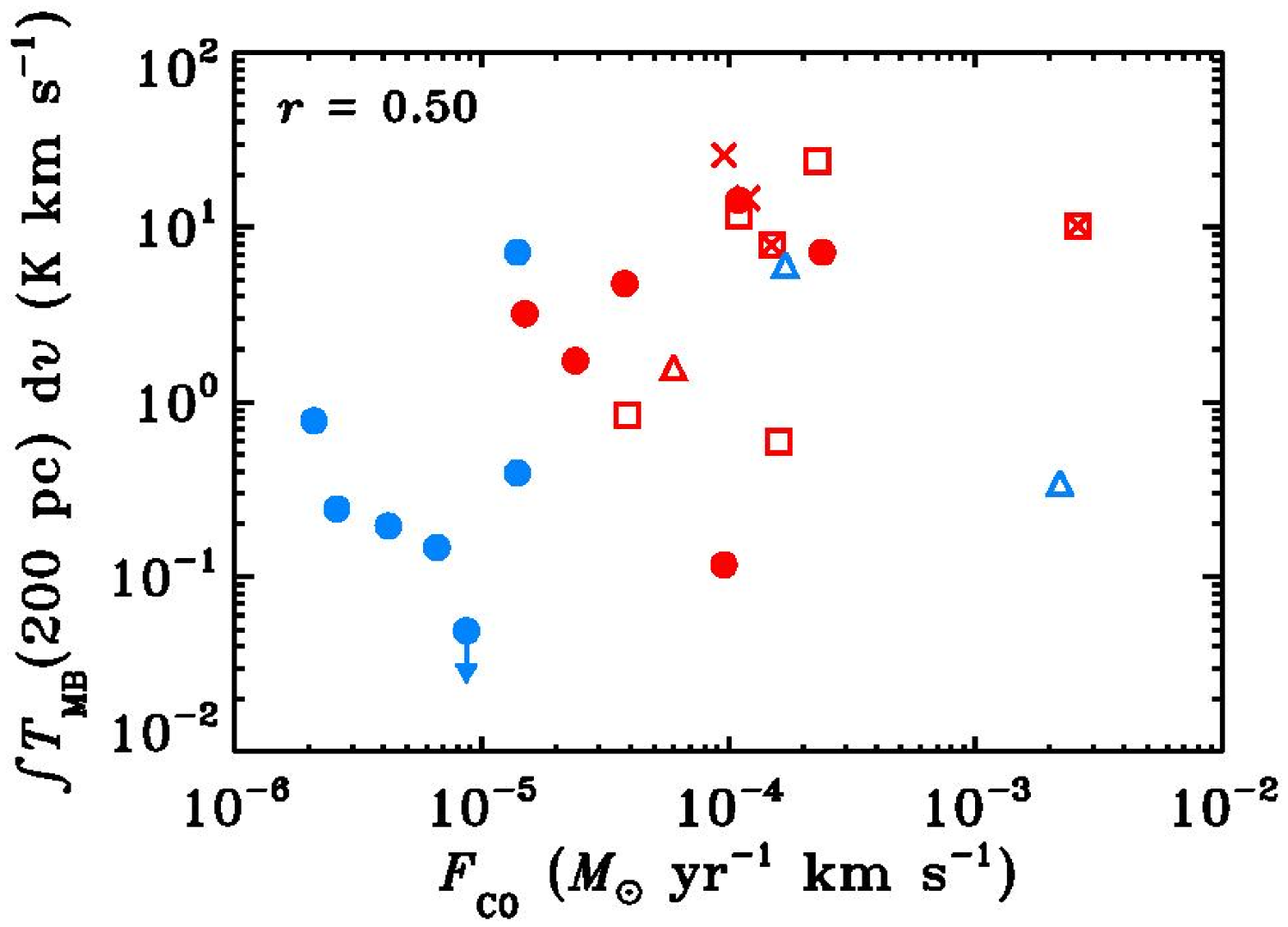}
\includegraphics[width=7.8cm]{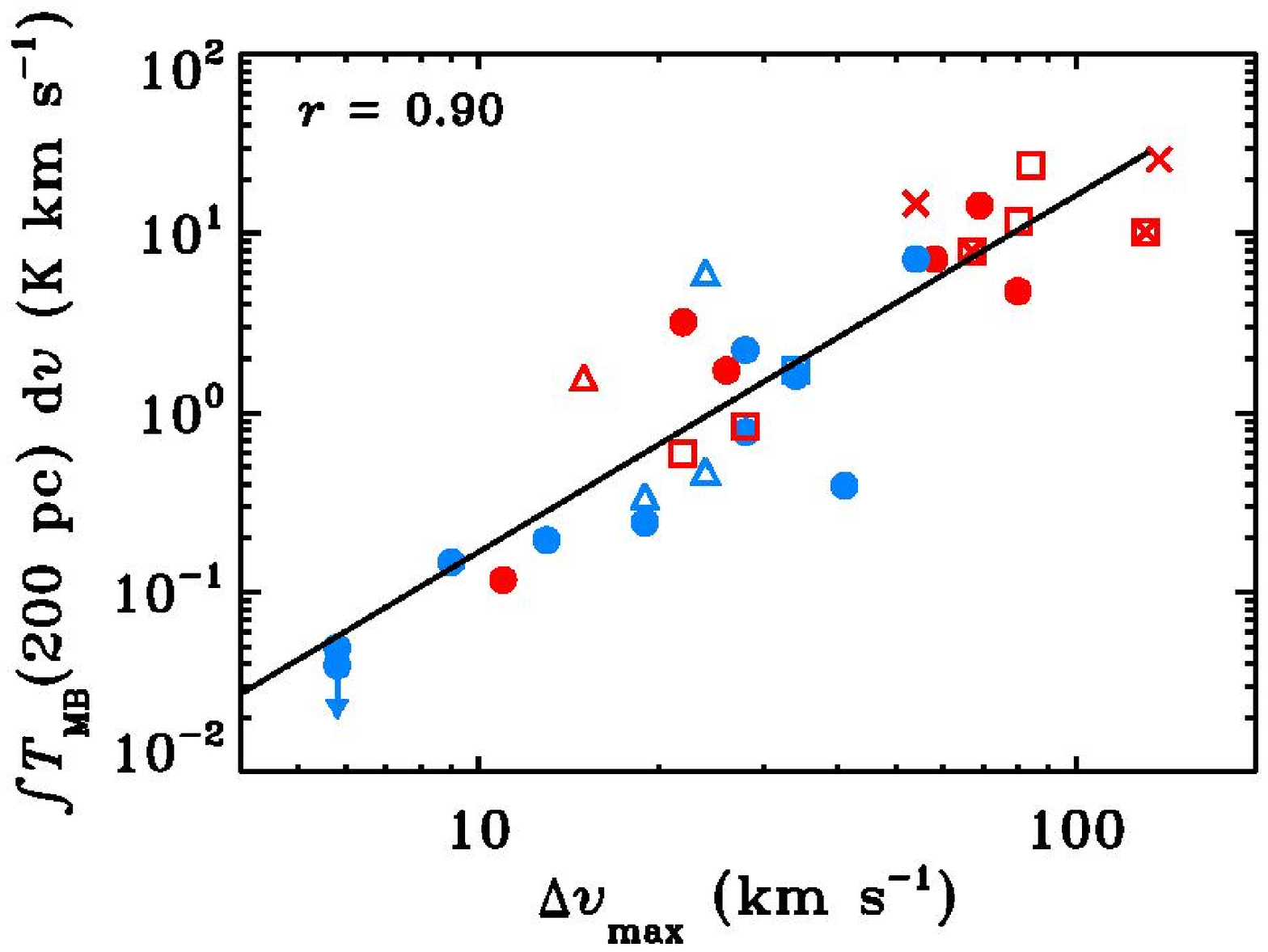}
\end{minipage}
\end{center}
\caption{Correlation plots showing the integrated H$_2$O 1$_{10}$--1$_{01}$ 557 GHz line intensity scaled to a distance of 200 pc as a function of various envelope parameters; the H$_2$O intensity is not scaled when comparing to CO 3--2 emission. Class 0 sources are marked in red, Class I sources in blue; sources with bullet emission are marked with crosses, inverse P-Cygni profiles with squares and regular P-Cygni profiles with triangles; the upper limit for TMC1A is marked with an arrow. Strong correlations ($|r|$ $>$ 0.7) are marked with full lines, tentative correlations (0.6 $<$ $|r|$ $<$ 0.7) with dashed lines. The parameters are (from left to right, top to bottom): CO 3--2 emission, $L_{\rm bol}$, $T_{\rm bol}$, $M_{\rm env}$, the evolutionary parameter $L_{\rm bol}^{0.6}/M_{\rm env}$, $F_{\rm CO}$, $n$(1000 AU) and $\Delta\varv_{\rm max}$.}
\label{fig:correlation}
\end{figure*}

The most prominent features of the line profiles are their complexity and width, with no two profiles being identical. Nevertheless, it is possible to identify a set of common traits based on Gaussian decomposition of the line profiles (see Fig. \ref{fig:gauss_decomp} for examples and Table \ref{tab:features} and Fig. \ref{fig:feature} for an overview):
\begin{itemize}
\item a broad component with $\Delta\varv$ $>$ 20 km s$^{-1}$; this component is present in most Class 0 sources and several Class I sources and when it is detected it is always in emission. The center of this component is sometimes shifted by up to 10 km s$^{-1}$ with respect to the source velocity, as determined from C$^{18}$O lines (e.g., NGC1333-IRAS2A);
\item a medium component with a $\Delta\varv$ of 5--20 km s$^{-1}$; in NGC1333-IRAS3 and Ser SMM3 this component is seen in absorption, otherwise it is in emission. The component is sometimes blue-shifted by up to 10 km s$^{-1}$ with respect to the source velocity (e.g., NGC1333-IRAS4A);
\item a narrow component with a $\Delta\varv$ of $\sim$2--3 km s$^{-1}$; this component is mostly seen in absorption, but not always. When the component is in absorption, it is often saturated, i.e., all of the continuum photons are absorbed as well. A few sources show multiple absorption components at velocities shifted by up to a few km s$^{-1}$ from that of the source (e.g., the Serpens sources);
\item extremely-high-velocity (EHV) components or ``bullets'' appearing as near-Gaussian profiles separated from the broad or medium components as well as from the source velocity by several tens of km s$^{-1}$; only three Class 0 sources show these components in the 557 GHz line (L1448-MM, BHR71 and L1157);
\item regular or inverse P-Cygni-type profiles where narrow emission and absorption are seen next to each other and always centred at the source velocity.
\end{itemize}
The results of the Gaussian decomposition are provided in Table \ref{tab:intensity_all} in the Appendix, and all features are discussed in Sect. \ref{sec:prof}. Many of these features have never been identified in these sources in other species before, such as the bullet emission from BHR71 and L1157. The broad component is observed in several low-mass YSOs, but it never appears so prominently in, e.g., CO rotational lines. For the sources showing both medium and broad components, the decomposition was checked against data for higher-excited lines where the absorption feature is not interfering \citep[see][for examples]{kristensen10a}. Inverse P-Cygni profiles have so far only been detected in one low-mass YSO, NGC1333-IRAS4A, in H$_2$CO, $^{13}$CO, HCN and H$_2$O emission \citep[][]{difrancesco01, jorgensen07, attard09, kristensen10b}. Our H$_2$O data increase the sample by almost an order of magnitude.

\begin{table*}
\caption{Pearson's correlation coefficient for H$_2$O and CO emission, envelope and outflow parameters.}
\centering
\begin{tabular}{l c c c c c c c c c}
\hline \hline
   				& H$_2$O & CO & $L_{\rm bol}$ & $T_{\rm bol}$ & $M$ & $L^{0.6}/M$ & $F_{\rm CO}$ & $n$(1000 AU) & $\Delta\varv_{\rm max}$ \\
\hline
H$_2$O 			        & \ldots &   0.64 &   0.56 & $-$0.60 & \phantom{$-$}0.75 &           $-$0.61 & \phantom{$-$}0.50 & \phantom{$-$}0.71 & \phantom{$-$}0.90 \\
CO 				        & \ldots & \ldots &   0.56 & $-$0.44 & \phantom{$-$}0.71 &           $-$0.56 & \phantom{$-$}0.46 & \phantom{$-$}0.63 & \phantom{$-$}0.66 \\
$L_{\rm bol}$ 	        & \ldots & \ldots & \ldots & \phantom{$-$}0.03 & \phantom{$-$}0.59 & $-$0.18 & \phantom{$-$}0.28 & \phantom{$-$}0.19 & \phantom{$-$}0.46 \\
$T_{\rm bol}$ 	        & \ldots & \ldots & \ldots &  \ldots &           $-$0.45 & \phantom{$-$}0.56 &           $-$0.70 &           $-$0.66 & $-$0.49 \\
$M$ 			        & \ldots & \ldots & \ldots &  \ldots &            \ldots &           $-$0.90 & \phantom{$-$}0.43 & \phantom{$-$}0.70 & \phantom{$-$}0.58 \\
$L^{0.6}/M$             & \ldots & \ldots & \ldots &  \ldots &            \ldots &            \ldots &           $-$0.38 &           $-$0.75 & $-$0.46	\\
$F_{\rm CO}$ 	        & \ldots & \ldots & \ldots &  \ldots &            \ldots &            \ldots &            \ldots & \phantom{$-$}0.65 & \phantom{$-$}0.51 \\
$n$(1000 AU) 	        & \ldots & \ldots & \ldots &  \ldots &            \ldots &            \ldots &            \ldots & \ldots & \phantom{$-$}0.61 \\
$\Delta\varv_{\rm max}$ & \ldots & \ldots & \ldots &  \ldots &            \ldots &            \ldots &            \ldots & \ldots & \ldots \\
\hline\\
\end{tabular}
\label{tab:corr}
\end{table*}

Figure \ref{fig:histo} shows a histogram of $\Delta\varv$ of each component. The broad H$_2$O component is present in almost all Class 0 sources (except Ced110-IRS4, IRAS15398, L483 and L723) but less than half of the Class I sources (NGC1333-IRAS3, L1489, L1551-IRS5, TMC1, HH46-IRS and IRAS12496). The medium component is found in $\sim$2/3 of all the sources, independent of evolutionary stage. The narrow component, however, is present in nearly all sources. With the exception of two sources (B335 and HH100-IRS), narrow emission always appears together with absorption either in a regular or inverse P-Cygni profile. Two sources show absorption broader than 5 km s$^{-1}$ (NGC1333-IRAS3 and Ser SMM3). NGC1333-IRAS3 shares a reference position with the three other NGC1333 sources, and since the absorption is not seen in any of them, it likely originates in the source. Similarly,  Ser SMM3 shares a reference position with SMM1 and SMM4, where the absorption is not seen. These reference positions were carefully chosen to be free of $^{12}$CO emission and thus should be free of H$_2$O emission given the difficulty in detecting any H$_2$O emission from cold clouds \citep{caselli10}.

\subsection{Correlations between H$_2$O and outflow/envelope parameters}

To quantify the origin of H$_2$O emission further and to use the full sample of 29 sources, the total H$_2$O integrated line intensity (including bullet contributions) is compared to different envelope and outflow parameters. The parameters are the envelope mass ($M_{\rm env}$), the bolometric luminosity and temperature ($L_{\rm bol}$ and $T_{\rm bol}$), the envelope density at a distance of 1000 AU from the source ($n$(1000 AU)), integrated CO 3--2 emission, linewidth of the 557 GHz line ($\Delta\varv_{\rm max}$), CO outflow force ($F_{\rm CO}$) and the alternative evolutionary parameter $L_{\rm bol}^{0.6}/M_{\rm env}$ \citep{saraceno96}.

For these comparisons, the observed line intensities are scaled by the square of the source distance to a common distance of 200 pc, the average source distance. For the comparison with CO 3--2 emission, neither H$_2$O nor CO 3--2 emission is scaled.  The comparisons are quantitatively examined using Pearson's correlation coefficient, $r$. For values of $r$ close to 1, the correlation is good, for values of $r$ close to 0 it is non-existing, and for values close to $-$1 there is a strong anti-correlation. In the following, a strong correlation is defined as $|r|$$>$0.7 and weak correlations are defined to have 0.6$<$$|r|$$<$0.7; these values correspond to $>$3.5$\sigma$ and $\sim$3$\sigma$, respectively \citep{marseille10}. The results of all the correlations are shown in Table \ref{tab:corr} and sample correlations are shown in Fig. \ref{fig:correlation}, where the sources with special features, such as inverse P-Cygni profiles, are highlighted separately.

\begin{figure}
\begin{center}
\includegraphics[width=\columnwidth]{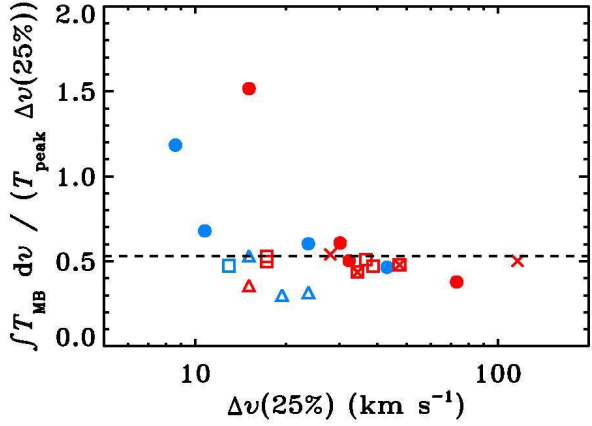}
\end{center}
\caption{Correlation between the width measured at the 25\% level of the peak intensity, $T_{\rm peak}$ and the integrated intensity ($\int T_{\rm MB}$ d$\varv$) divided by $T_{\rm peak}$. Symbols are as in Fig. \ref{fig:correlation}. The thick dashed line is for a proportionality constant of 0.53, corresponding to a Gaussian curve.}
\label{fig:corr_special}
\end{figure}

H$_2$O emission at 557 GHz correlates the strongest with $\Delta\varv_{\rm max}$ ($r$ = 0.90) followed by envelope mass ($r$ = 0.75), and $n$(1000 AU) ($r$ = 0.71). Emission correlates weakly with CO 3--2 emission ($r$ = 0.64), $L_{\rm bol}^{0.6}/M_{\rm env}$ ($r$ = $-$0.61) and $T_{\rm bol}$ ($r$ = $-$0.60), but not with $L_{\rm bol}$ and $F_{\rm CO}$. The strong correlation between the H$_2$O integrated intensity and $\Delta\varv_{\rm max}$ is not a surprise, considering that most of the emission of each profile is located in a single medium or broad Gaussian component, i.e., the integrated area of the profile is directly proportional to the peak intensity ($T_{\rm peak}$) times the width. To demonstrate that the profiles are dominated by a single Gaussian component, the width is measured at 25\% of $T_{\rm peak}$; sources with $T_{\rm peak}$/4 $<$ 4$\sigma$ are not included. For a Gaussian function, the area is equal to 0.53 $\times$ $T_{\rm peak}$ $\times$ $\Delta\varv$(25\%). Figure \ref{fig:corr_special} shows the correlation between $\int T_{\rm MB}$ d$\varv$ / ($T_{\rm peak}$ $\times$ $\Delta\varv$(25\%)) and $\Delta\varv$(25\%). The correlation itself is not strong, the scatter being dominated by profiles with a width of $<$ 20 km s$^{-1}$ (Fig. \ref{fig:corr_special}). However, with the exception of two sources, all points have values close to the expected, 0.53, and we conclude that the profiles are indeed dominated by a single Gaussian component.

\begin{figure}
\begin{center}
\includegraphics[width=0.8\columnwidth]{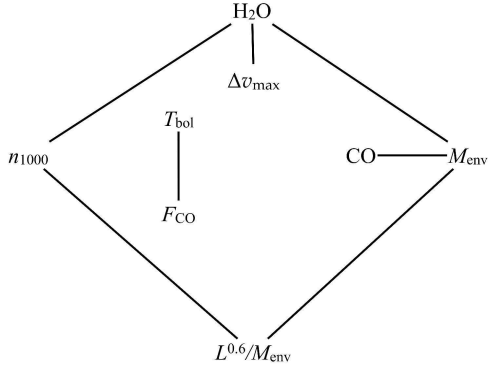}
\end{center}
\caption{Schematic diagram showing which parameters correlate with each other. Only strong correlations with $|r|$ $>$ 0.70, for which the correlation is confirmed, are shown.}
\label{fig:corr_1}
\end{figure}

The correlations are displayed in Fig. \ref{fig:corr_1} and discussed in Sect. \ref{sec:disc}. To further confirm that the correlations are real, the partial correlation factors are calculated for all combinations of three parameters \citep[e.g.,][]{marseille10}. At the 2$\sigma$ level, all strong correlations are confirmed except that between density and envelope mass, i.e., there is a 95\% likelihood that the correlations are real.

\section{Discussion}\label{sec:disc}

\subsection{H$_2$O and dynamics: profile components}\label{sec:prof}

\subsubsection{Outflow components}\label{sec:outflow}

\begin{figure}
\begin{center}
\includegraphics[width = 0.8\columnwidth]{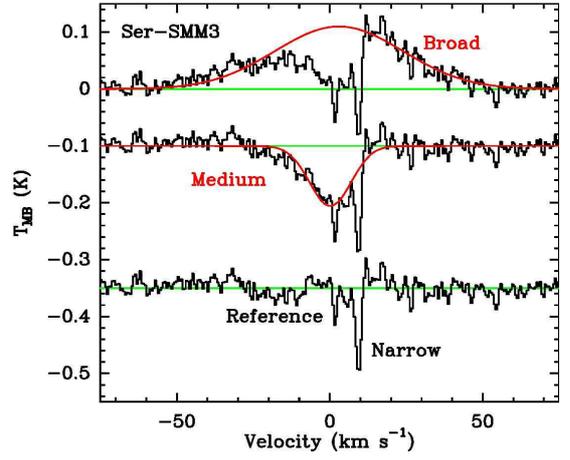}
\end{center}
\caption{Gaussian decomposition of the source Ser-SMM3. This source shows a broad emission feature ($\Delta\varv\sim$50 km s$^{-1}$). A medium absorption feature ($\Delta\varv\sim$15 km s$^{-1}$) is superposed on the emission along with two narrow absorption features at $\varv_{\rm LSR}$$\sim$1 km s$^{-1}$ and $\sim$8 km s$^{-1}$. The former of the narrow absorption features is likely due to emission in the reference position (see text) and the latter is intrinsic to the source and is saturated.}
\label{fig:smm3-abs}
\end{figure}

The broad and medium line components have widths exceeding the line width typically seen originating from the quiescent envelope \citep[$\sim$ 1 km s$^{-1}$; e.g.,][]{jorgensen02} and are thus likely related to the protostellar outflow. The justification for distinguishing between two  components comes from NGC1333-IRAS2A, NGC1333-IRAS3, NGC1333-IRAS4A, Ser SMM1 and Ser SMM3, where both components are seen, and they are offset from each other by up to 10 km s$^{-1}$. The offset results in very asymmetric line profiles, as also illustrated in \citet{kristensen10b} (see Fig. \ref{fig:gauss_decomp} and \ref{fig:smm3-abs}). The statistics presented in Fig. \ref{fig:histo} show that the broad component is more closely associated to the earlier evolutionary stage, whereas the medium component is equally present at both stages. 

\paragraph{Viewing angle.} 

The viewing angle may play a role in whether a source has a broad component associated with it, or both a broad and medium component, but because of the relatively large sample size such an effect should be statistically small. For example, the two Class 0 sources L483 and L1527 both have outflows moving close to the plane of the sky and none of them show broad emission components. Yet, the outflow in NGC1333-IRAS2A is likely moving close to the plane of the sky, but the profile shows a broad component. Because the Class I sources show weaker emission it is also conceivable that broad emission components are hiding in the noise in these more evolved sources. For the weak Class I sources (TMR1, Oph-IRS63, RNO91) this may be the case, but the stronger Class I sources (GSS30-IRS1, Elias29, RCrA-IRS5 and HH100-IRS) should have detectable broad components associated with them if they exist. Therefore, the disappearance of the broad component at later evolutionary stages is likely real, whereas the medium component remains through the Class I stage, but decreases in intensity. The disappearance implies an evolutionary trend in the broad component as it weakens in both width and intensity from the Class 0 phase to the Class I phase.

\paragraph{Broad or medium.} 

In sources where only the broad or the medium component is observed, the profile typically corresponds to what has previously been observed in low-$J$ CO emission from ground-based telescopes. This component is spatially extended over $>$ 1000 AU and associated with the large-scale outflow. A map of H$_2$O emission in L1157 from PACS \citep{nisini10} shows that this component is physically closer to the region which is currently being shocked (as traced by H$_2$ emission) than to the swept-up gas in the molecular outflow (as traced by low-$J$ CO emission). With the data presented here, it is not possible to infer whether H$_2$O emission is always associated with shocked gas rather than with swept-up gas. No matter the exact excitation conditions, when only the broad or medium component is present it traces the outflow.

\paragraph{Broad and medium.} 

\begin{figure*}
\sidecaption
\includegraphics[width = 12cm]{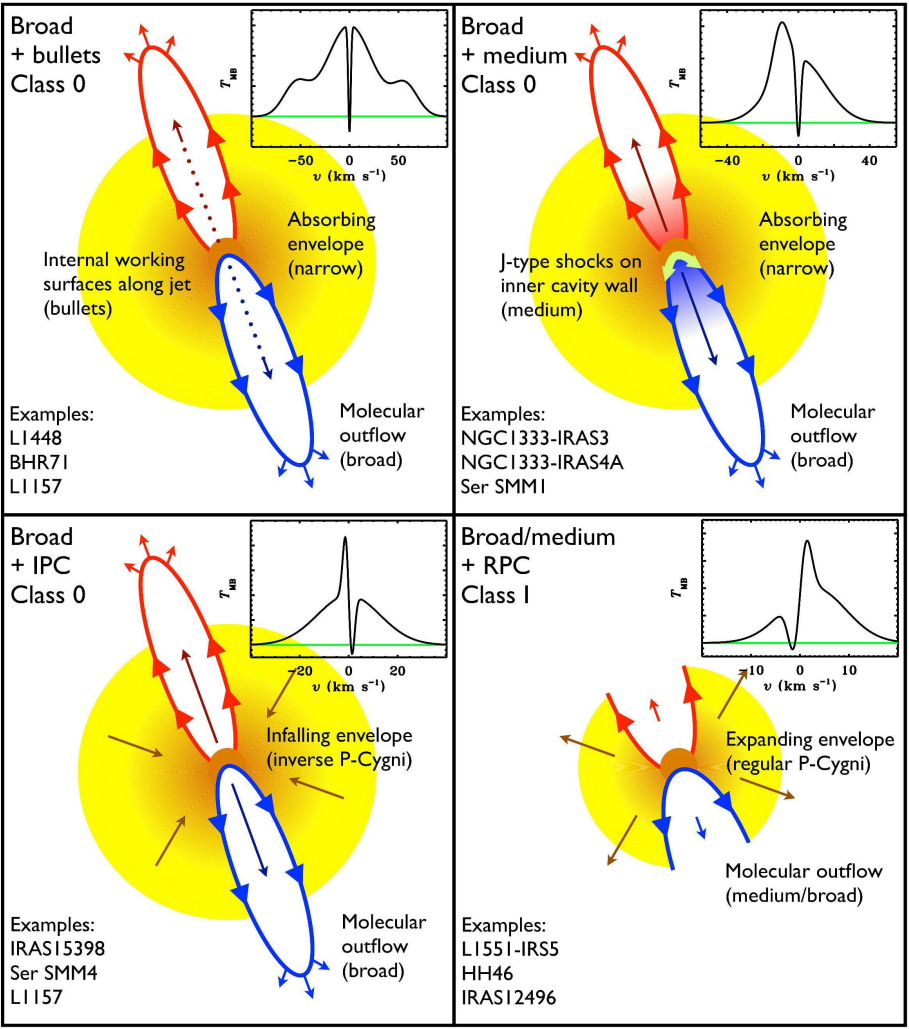}
\caption{Cartoon showing the different dynamical and energetic components in protostars directly traced by water emission. Each panel shows a cartoon of a protostellar system with a sample spectrum and a number of specific examples of objects in this category. Note, the velocity scale of the inserted spectra change from the top left panel to bottom right. Inverse P-Cygni and regular P-Cygni profiles have been abbreviated IPC and RPC, respectively. The labels Class 0 and I refer to the Class where this type of profile is typically found.}
\label{fig:schematic}
\end{figure*}

In the sources where both a broad and medium component is present (NGC1333-IRAS2A, NGC1333-IRAS3, NGC1333-IRAS4A, NGC1333-IRAS4B, Ser SMM1 and Ser SMM3), the medium component is either centred at the source velocity (offset $<$ 1 km s$^{-1}; $NGC1333-IRAS2A) or blue-shifted by 2--10 km s$^{-1}$. The medium component is not seen in CO 3--2 or any other species, except for the case of NGC1333-IRAS2A. Furthermore, the medium component is seen in absorption against the broad component in NGC1333-IRAS3 and Ser SMM3 (see Fig. \ref{fig:smm3-abs} for the example of Ser SMM3). The absorption implies that the medium component is located in front of the blue-shifted part of the broad component, and that the medium component is less extended than the broad (if it were more extended, it would also show emission; see Fig. \ref{fig:schematic} for a cartoon illustration). The broad component is not observed in absorption, neither against the continuum nor the medium component.

If there is a corresponding red-shifted medium component (not seen in any of the objects) it must be located behind the red-shifted part of the broad component, and be less extended than the broad component. Furthermore, the broad component must be optically thick over the extent of the red-shifted medium component, if it exists, or the corresponding red-shifted medium component would have been detected in emission. The medium component must be located inside the envelope; the narrow absorption components are saturated in Ser SMM3 and NGC1333-IRAS3 showing that the absorbing envelope is between us and the medium absorption component. These geometrical considerations along with the non-detection of the component in CO point to an origin in the inner part of the envelope on scales $<$ 1000 AU. 

Viewing angle appears not to play a role in determining whether a source shows both a broad and medium component. The three NGC1333 sources IRAS2A, IRAS4A and IRAS4B have outflows moving close to the plane of the sky, close to an angle of 45\degr\ and close to the line-of-sight, respectively \citep[][Y{\i}ld{\i}z et al. in press]{jorgensen04b}.

\paragraph{Association with masers.}

A Gaussian peak offset from the source velocity is a signature of a Jump-type shock, as opposed to a Continuous-type shock which shows up as a triangular profile peaking at the source velocity \citep{hollenbach97}. This is not the first time H$_2$O emission is linked with J-type shocks on small spatial scales in protostellar envelopes; H$_2$O masers at 22 GHz are also thought to be associated with J-type shocks in dense gas \citep[see review by][]{elitzur92}. Many low-mass protostars have H$_2$O masers associated with them \citep{furuya03}, and VLBI observations show them to originate typically within the inner few 100 AU of the source. The sources showing both broad and medium components all have H$_2$O masers associated with them \citep[e.g.,][]{furuya03, moscadelli06}. Most of the masers are variable in terms of velocity over time-scales of years or less. It is therefore no surprise that maser spots are found matching the observed velocity of the medium components seen here. It is interesting to note that in some of the sources with little maser variability \citep[e.g., NGC1333-IRAS3;][]{furuya03}, the velocity of the maser spot coincides with the velocity of the medium component (4.6 km s$^{-1}$ versus 4.9 km s$^{-1}$). The same is seen in Ser SMM1 in the 183 GHz H$_2$O maser line \citep{vankempen09_h2o}, where the velocity of their maser spot 2 is 7.4 km s$^{-1}$, identical to the velocity of the medium component in Ser SMM1. It is important to note here that we do not propose that the maser causes the medium component. However, the phenomenon producing the masers likely has a thermal component associated with it, which produces the medium component. This is similar to what has previously been observed in high-mass star-forming regions \citep{marseille10}. Indeed, the physical conditions for maser emission, $n$ $>$ 10$^9$ cm$^{-3}$ and $T$ $\sim$ 400 K \citep{elitzur92}, are such that a thermal component must be present. The only limiting factor is the column density of the thermal component: if the beam-averaged column density is low, the component is not detected.

Five sources have H$_2$O maser activity associated with them, but do not show both broad and medium components (Ser SMM4, L483, L723, L1157 and GSS30-IRS1). The water masers are not constant over the observed epochs in these five sources \citep{furuya03}, but seem to switch on and off. When the masers are on in these five sources, the intensity is weaker than in the sources showing both broad and medium components. Furthermore, L483, L723 and L1157 are weak in 557 GHz emission, and it cannot be ruled out that the non-detection of the medium component is a question of $S/N$. 

While the exact origin of the masers and the J-type shock causing the medium component is, is unknown, however there are several candidate mechanisms. If the protostellar wind directly impacts on the cavity wall, this could drive a J-type shock into the cavity wall, a mechanism which has been proposed for masers in high-mass protostars \citep{goddi06}. Another possibility is that the masers originate in the jet itself, or at least that they are connected to the jet \citep{furuya03}, which is almost certainly true for the high-velocity masers ($\varv$$>$20 km s$^{-1}$). Because the medium component is located physically between the inner outflow close to the base and the outer absorbing envelope, these observations suggest that the lower-velocity masers are more directly linked to the outflow cavity walls rather than the jet itself, if the physical mechanism behind the masers and the medium component is the same.

\subsubsection{Molecular bullets}\label{sect:bullets}

Molecular ``bullets'' or extremely high-velocity (EHV) gas has been detected in four Class 0 sources, L1448, BHR71, NGC1333-IRAS4B and L1157, with the bullet emission in NGC1333-IRAS4B only seen in the higher-excited H$_2$O 2$_{02}$--1$_{11}$ line at 988 GHz \citep{kristensen10b}. The three previously unknown bullet features highlight H$_2$O as a good dynamical probe, thereby complementing previous SiO observations of bullets \citep[e.g.,][]{girart01, nisini02, nisini07, cabrit07}. \citet{kristensen11} analysed the H$_2$O bullet emission in L1448, and came to the conclusion that the bullets are rich in H$_2$O and H$_2$, the former requiring the presence of the latter. In particular, H$_2$O abundances of 10$^{-5}$--10$^{-4}$ are derived with respect to H$_2$. 

High spatial resolution images of, e.g., CO and SiO, show that the bullets consist of sub-arcsecond-sized clumps typically separated by 1--2\arcsec\ in the plane of the sky \citep[e.g.,][for the case of L1448]{maury10}. The bullets are located along the jet axis, where they are likely internal working surfaces along the jet \citep[e.g.,][]{santiago-garcia09}. 

The bullets appear in more sources than were previously known to harbour them, indicating that the phenomenon is more common than previously known. Whether all protostars show bullet emission, whether it is a specific phase that protostars go through or whether it is limited to specific sources will need to be tested by observing more sources at higher $S/N$ and in other species. To date, bullets have not been detected in Class I sources, and it is not known whether this is due to an absence of the mechanism creating the bullets, or if it is a lack of $S/N$. For the two sources showing the brightest bullet emission, L1448-MM and BHR71, the bullet emission contributes $\sim$25--30\% of the total integrated intensity of the 557 GHz emission (for L1157 the bullet contribution is $\sim$15\%). Furthermore, the bullets are only seen in Class 0 sources with a broad component. Of the Class I sources with a broad component, only NGC1333-IRAS3 is bright enough that bullets would be detected, were they present. Therefore it is not possible to conclude whether the phenomenon is limited to Class 0 sources; deeper integrations on Class I sources are required to determine whether it is particular to Class 0 sources.

\subsubsection{Absorption features}\label{sec:abs}

The narrow component ($\Delta\varv<5$ km s$^{-1}$) is seen in absorption in all but one Class 0 source and a few Class I sources (B335, TMC1, Oph-IRS63, RNO91 and HH100-IRS). This component is interpreted as being caused by absorption by the outer envelope and ambient cloud, consistent with previous interpretations \citep{kristensen10b} and originates in the same region as low-J C$^{18}$O and HCO$^+$ emission \citep[e.g.,][]{jorgensen02}. Because the absorption is seen against both the outflow and continuum emission, the absorbing layer must be located in front of both the emitting layers, i.e., the outflows are also embedded. If the absorption layer were located behind the outflow, flat-bottomed absorption features not extending down to the zero-level should have been observed; this is not the case.

The non-LTE radiative transfer code RADEX \citep{vandertak07} is used to estimate the H$_2$O column density required to cause the saturated absorption. For physical conditions such as those found in the outer envelope ($n_{\rm H}$=10$^5$ cm$^{-3}$, $T$=15 K, $\Delta\varv$=1 km s$^{-1}$) an o-H$_2$O column density of $>$ 10$^{13}$ cm$^{-2}$ is required to reproduce the saturated absorption, i.e., for those conditions, the line optical depth, $\tau$, is more than 3. The result is not sensitive to the precise physical conditions since the bulk of the population is in the 1$_{01}$ ground state. Flat-bottomed profiles typically require $\tau$ $\gtrsim$ 7, which is obtained for a column density of $\sim$3--4$\times$10$^{13}$ cm$^{-2}$. Thus, deep non-saturated profiles are found in a very narrow range of column densities. Limits obtained from deep H$_2^{18}$O observations will be used to further constrain the column density of H$_2$O in the outer envelope.

The H$_2$ column density calculated from spherical envelope models ranges from a few times 10$^{22}$ cm$^{-2}$ (e.g., L483, L1489, TMC1, IRAS12496) to $\sim$10$^{24}$ cm$^{-2}$ (NGC1333-IRAS4A, NGC1333-IRAS4B) in a pencil beam towards these sources (see Appendix \ref{app:dusty}). Therefore, the o-H$_2$O abundances range from $>$ 10$^{-11}$ for the highly embedded sources, to $>$10$^{-9}$ for the less embedded sources. To determine the abundance of the absorbing layer more accurately, envelope models are required taking the range of densities and temperatures found in the envelope into account. NGC1333-IRAS2A was modelled in some detail \citep{kristensen10a}, where the abundance is of the order of 10$^{-8}$. The abundance is higher than what is found for other quiescent regions such as prestellar cores or protoplanetary disks \citep{caselli10, bergin10, hogerheijde11}. Because these values are all lower limits, there is no reason to believe that there is a real trend with evolution or column density. There is, however, a trend for more Class 0 sources showing saturated absorption (10/15) than Class I sources (4/14). If the abundance in the outer envelope is the same in Class 0 and I sources, then this difference can arise from the Class 0 envelopes having a higher H$_2$ column density than the Class I sources. In this scenario, the absorption is less likely to be saturated in the more dilute Class I envelopes.

\subsubsection{Inverse P-Cygni profiles}

\begin{figure}
\begin{center}
\includegraphics[width=0.91\columnwidth]{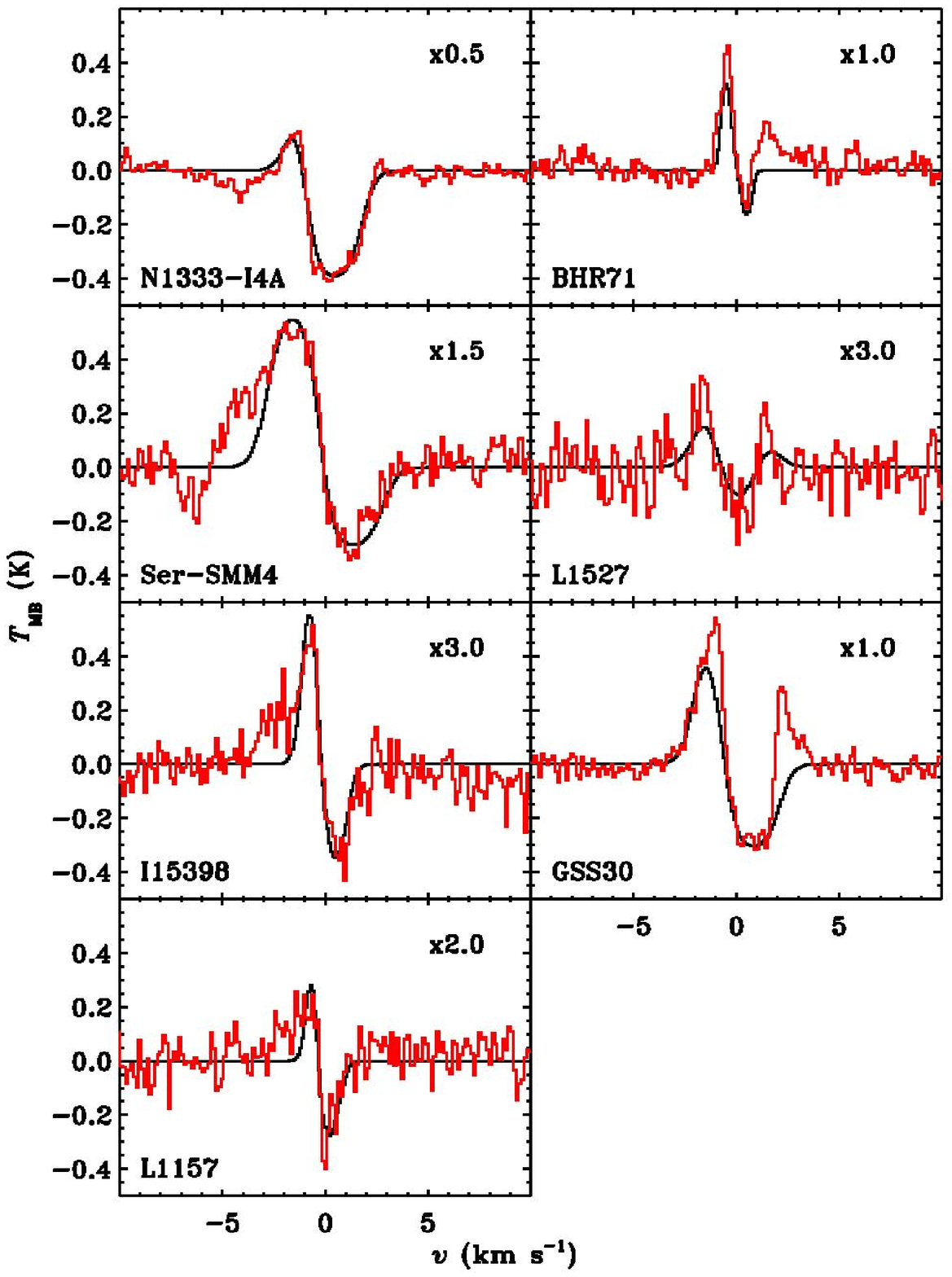}
\includegraphics[width=0.91\columnwidth]{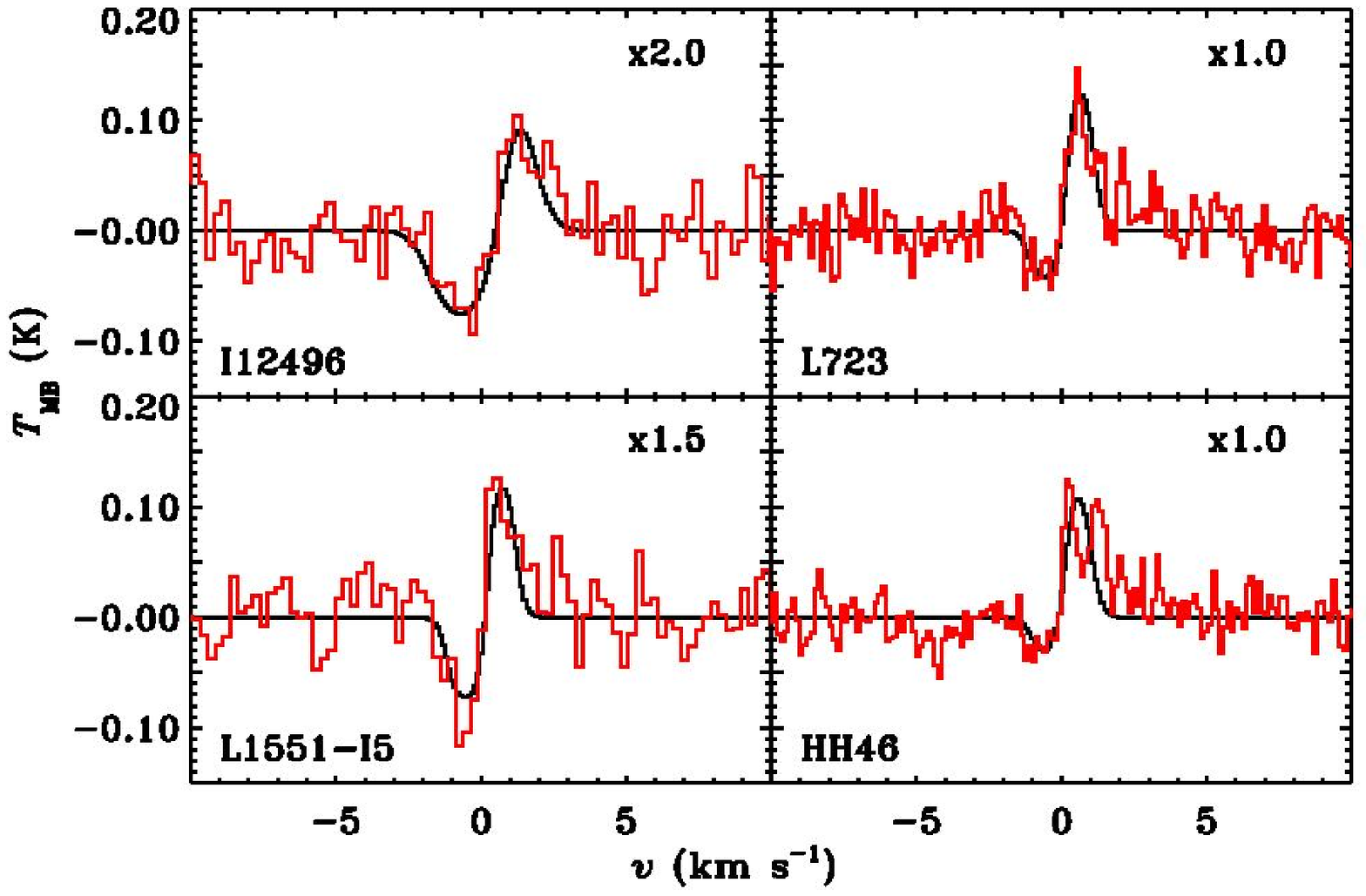}
\end{center}
\caption{Inverse (top) and regular (bottom) P-Cygni profiles detected in the eleven sources (red) with the infall or expansion model shown on top (black). The broad outflow component has been subtracted from each spectrum. Spectra have been centred at 0 km s$^{-1}$.}
\label{fig:infall}
\end{figure}

Inverse P-Cygni profiles are detected towards seven sources: NGC1333-IRAS4A, Ser SMM4, L1157, IRAS15398, L1527, BHR71 (all Class 0) and GSS30-IRS1 (Class I). An inverse P-Cygni profile is characterised by a narrow absorption feature red-shifted with respect to a narrow emission feature and is typically caused by a velocity and excitation gradient in gas falling in towards a central object, in this case the protostar. The profiles are located on top of the broader outflow component, and the absorption takes place against both the outflow component and the continuum. This implies that the absorbing layer is located between the outflow component and the edge of the envelope. Inverse P-Cygni profiles have previously been detected for NGC1333-IRAS4A, in both interferometric and single-dish data \citep{difrancesco01, jorgensen07, attard09, kristensen10b}. These observations constitute the first detections of seven inverse P-Cygni profiles in a 40\arcsec\ beam of the envelopes surrounding low-mass protostars. As noted above (Sect. \ref{sec:obs}), none of the observations show signs of contamination from the reference position or foreground material. 

To quantify the infall in a simplistic manner, \citet{difrancesco01} used a toy model consisting of two slabs moving towards each other with a continuum source between them by modifying the model of \citet{myers96}. The two slabs are considered to be isothermal and of uniform density. There are two limitations to the model: the simplicity and the large number of free parameters. First, an infalling envelope has temperature, velocity and density gradients along the line of sight which are not taken into account. Second, the large number of free parameters for such a toy model includes the excitation temperature of the front layer, the rear layer and the continuum source ($T_{\rm f}$, $T_{\rm r}$, $T_{\rm c}$), the filling factor of the absorbing source with respect to the telescope beam ($\Phi$), the peak optical depth of the front and rear layers ($\tau_0$), the turbulent velocity dispersion in each layer ($\Delta\varv_{\rm tur}$) and the infall velocity ($\varv_{\rm in}$). Furthermore, the absorption also takes place against the outflow component in Ser SMM4, NGC1333-IRAS4A and GSS30-IRS1, where the absorption is saturated. Absorption against multiple layers is taken into account in the modified 4-slab model of \citet{attard09}. Introducing four slabs with different velocities and excitation temperatures increases the number of free parameters significantly, making any solution even more degenerate. Instead of adopting more parameters to reproduce the entire profile, only the outflow-subtracted inverse P-Cygni profiles are analysed in the following. The outflow contribution only plays a part in the value of $\Phi$, which is larger than if just the continuum emission is considered, and the value of $T_{\rm c}$, which includes both a contribution from the source and outflow. The main focus will be on the dynamics rather than the excitation.

Any infall velocity found here should be considered as a ``characteristic'' infall velocity as also noted in \citet{myers96}. To estimate the excitation conditions, H$_2$O is assumed to be sub-thermally excited because the density is likely lower than the critical density for this transition, $\sim$ 10$^8$ cm$^{-3}$ in the optically thin limit \citep{dubernet06}. For densities in the range of 10$^4$ -- 10$^6$ cm$^{-3}$, the excitation temperature in the front layer is 3--4 K for a column density of 10$^{13}$ cm$^{-2}$ (Sect. \ref{sec:abs}) and a kinetic temperature of 10--20 K. Under the same assumptions, the excitation temperature of the rear layer is closer to the kinetic temperature \citep[see Fig. 1 of][for details]{myers96}, which in this case is expected to be 10--20 K for the outer parts of the envelope. The excitation temperature of the continuum layer is assumed fixed at 15 K; this parameter is degenerate with $T_{\rm r}$ and $\tau_0$, and it is possible to find other solutions if the continuum temperature is changed to, say, 10 or 20 K. However, since the goal is not to find the best-fit solution for the excitation conditions, this is ignored at present.

The shape of the profile is dominated by $\varv_{\rm inf}/\Delta\varv_{\rm turb}$, and it is therefore possible to estimate the infall velocity of the two slabs along with the turbulent width, $\Delta\varv_{\rm turb}$. The fitting is done by eye, as the goal is to get a qualitative estimate of the infall velocity. The numbers are given in Table \ref{tab:infall} along with the other parameters used to describe the profiles, and the resulting profiles are compared to observations in Fig. \ref{fig:infall}. Characteristic infall velocities range from 0.1 km s$^{-1}$ (L1157) to 1.2 km s$^{-1}$ (Ser SMM4) for $\Delta\varv_{\rm turb}$ ranging from 0.2 km s$^{-1}$ (BHR71) to 1.0 km s$^{-1}$ (Ser SMM4, L1527). The turbulent velocities of $\Delta\varv$(C$^{18}$O 3--2) agree to within 0.1--0.3 km s$^{-1}$ with the values found here \citep[][; Y{\i}ld{\i}z et al. in prep.]{jorgensen02}. The uncertainty on both the infall velocity and the velocity dispersion is estimated to be 0.2 km s$^{-1}$ by comparing different model profiles to the observations. \citet{difrancesco01} and \citet{attard09} both modelled the infall in NGC1333-IRAS4A using the same method but for different molecules (H$_2$CO and HCN), and both found a typical infall velocity of 0.6 km s$^{-1}$ in good agreement with 0.5 km s$^{-1}$ inferred here.

If the envelopes are in free fall throughout, the velocity at any point through the envelope is $\varv=\sqrt{2GM_{\rm enc}/r}$, where $M_{\rm enc}$ is the mass enclosed within radius $r$. In the classic inside-out Shu collapse \citep{shu77}, the infall wave travels through the envelope at the sound speed, and only the inner parts of the envelope are free-falling towards the centre at early times. If the inferred enclosed mass is less than the total envelope mass obtained from DUSTY modelling, the infall wave has likely not reached the outer edges of the envelope, and the assumption that the entire envelope is in free fall is no longer valid. The infall velocity at the outer edge of the envelope will therefore range between 0 km s$^{-1}$ and the free-fall velocity at the outer radius. Moreover, if the envelope is rotating, the infall velocity may be systematically lower than the free-fall velocity, depending on the viewing angle \citep{terebey84}.

\begin{table*}
\tiny
\caption{Best-fit parameters of the infall modelling.}
\centering
\begin{tabular}{l c c c c c c c}
\hline \hline
Parameter & NGC1333-IRAS4A & Ser SMM4 & IRAS15398 & L1157\tablefootmark{a} & BHR71 & L1527\tablefootmark{a} & GSS30-IRS1 \\
\hline
$\varv_{\rm in}$ (km s$^{-1}$)	&	0.5	&	1.2	&	0.3	&	0.1 & 0.5 & 0.08 & 0.7	\\
$\Delta\varv_{\rm tur}$  (km s$^{-1}$)			&	0.8	&	1.0	&	0.5	&	0.3 & 0.2 & 1.0 & 0.8	\\
$\Delta\varv$(C$^{18}$O) (km s$^{-1}$)\tablefootmark{b}	&	0.6	&	0.9	&	0.2	&	0.3	\\
$\tau_0$						&	3.0	&	1.3	&	0.9	&	2.0 & 1.5 & 1.5 & 3.0	\\
$\Phi$							&	0.3	&	0.2	&	0.2	&	0.2 & 0.2 & 0.15 & 0.25	\\
$T_{\rm r}$ (K)					&	\multicolumn{7}{c}{11.0} \\
$T_{\rm f}$ (K)\tablefootmark{c}&	\multicolumn{7}{c}{4.0}	\\
$T_{\rm c}$ (K)\tablefootmark{c}&	\multicolumn{7}{c}{15}	\\
$R_{\rm inf}$ (AU)				&	3500	&	600	&	9900	&	88000 & 3500 & 140000 & 1800	\\
$R$(10 K) (AU)				&	6900	&  2800	&	2700	&	5400  & 9900 &   4600 & 16000	\\
$n_{\rm inf}$ (cm$^{-3}$)		& 7$\times$10$^5$ & 9$\times$10$^6$ & 4$\times$10$^5$ & 1$\times$10$^5$ & 2$\times$10$^5$ & 2$\times$10$^5$ & 7$\times$10$^4$ \\
$\dot{M}_{\rm in}$ ($M_\odot$ yr$^{-1}$) & 9$\times$10$^{-5}$ & 8$\times$10$^{-5}$ & $<$2$\times$10$^{-4}$\tablefootmark{d} & $<$2$\times$10$^{-3}$\tablefootmark{d} & 3$\times$10$^{-5}$ & $<$7$\times$10$^{-3}$\tablefootmark{d} & 3$\times$10$^{-6}$\\
\hline\\
\end{tabular}
\tablefoot{
	\tablefoottext{a}{A free-fall collapse model does not describe these envelopes, and no values are derived for the characteristic infall radius, or mass infall rate.}
	\tablefoottext{b}{From Y{\i}ld{\i}z et al. (in prep.).}
	\tablefoottext{c}{Fixed parameter, see text.}
	\tablefoottext{d}{Mass infall rate derived through a sphere bounded by the 10-K radius. The infall rate is therefore a strict upper limit to the actual infall rate (see text).}
		}
\label{tab:infall}
\end{table*}

The characteristic distance of the infall zone for the envelopes that may be undergoing free-fall collapse can be estimated. For infall onto a 0.5 $M_\odot$ central object the velocity is 6 km s$^{-1}$ at 25 AU. This stellar mass is a typical upper limit given that Class 0 sources are thought to have accreted less than half of their final mass \citep{andre00}, and the peak of the initial mass function is at 0.3--0.5 $M_\odot$ \citep[e.g.,][]{chabrier03}. If the mass is decreased to 0.1 $M_\odot$ any derived velocity decreases by a factor of $\sim$2 only and the characteristic distance increases by a factor of $\sim$2. For a mass of 0.5 $M_\odot$, the inferred characteristic infall distances are between 600 and 10$^5$ AU, with Ser SMM4 and L1527 representing the extreme cases. Three sources, IRAS15398, L1157 and L1527, have such low infall velocities that the inferred infall radius is larger than the outer radius of the envelope as derived from the DUSTY modelling. Therefore, the entire envelope of these three sources is not in free fall, and the 10-K radius is adopted as the characteristic infall radius. Thus, the profiles trace large-scale infall ranging from several 100 AU to scales of $>$ 1000 AU.

Assuming spherical infall symmetry, the mass infall rates can be estimated through \mbox{$\dot{M}_{\rm inf}\,=4\pi\, R_{\rm inf}^2\,\mu\, m_{\rm H}\, n_{\rm inf}\, \varv_{\rm inf}$} where $\mu$ is the mean molecular weight \citep[2.8 amu;][]{kauffmann08} and $n_{\rm inf}$ the density at the infall radius. For a free-falling envelope, this may be reformulated as
\begin{eqnarray}
\dot{M} &=& 4\pi\, \mu\, m_{\rm H}\, n_{\rm inf}\, \left(2\,GM_{\rm enc}\right)^2\, \varv_{\rm inf}^{-3} \nonumber \\ 
&=& 1.3\times10^{-5}\ M_\odot\, {\rm yr}^{-1} \left(\frac{n_{\rm inf}}{10^5\ {\rm cm}^{-3}} \right) \left(\frac{\varv_{\rm inf}}{0.5\ {\rm km\ s}^{-1}} \right)^{-3}
\end{eqnarray}
for a 0.5 $M_\odot$ enclosed mass ($M_{\rm enc} \neq M_\star$). The density is taken from the relevant DUSTY model at the infall radius, $R_{\rm inf}$, and ranges from 7$\times$10$^4$ cm$^{-3}$ (GSS30-IRS1) to 9$\times$10$^6$ cm$^{-3}$ (Ser SMM4). Mass infall rates range from 3$\times$10$^{-6}$ $M_\odot$ yr$^{-1}$ (GSS30-IRS1) to 2$\times$10$^{-3}$ $M_\odot$ yr$^{-1}$ (L1157). The very high mass infall rate inferred for L1157 is a result of the very large infall radius (and therefore large surface area over which the infall takes place) and the very low velocity. As argued above, it is unlikely that IRAS15398, L1157 and L1527 are in free fall, and the inferred properties are thus upper limits. We emphasize again, that the model used to derive these infall parameters is a toy model; to derive more accurate infall velocities and conditions it is necessary to run full radiative-transfer models of infalling envelopes which is beyond the scope of this paper. The values derived here serve primarily as order-of-magnitude estimates.

The mass infall rate derived for NGC1333-IRAS4A (9$\times$10$^{-5}$ $M_\odot$ yr$^{-1}$) is very similar to what \citet{difrancesco01} derived for the same source (1.1$\times$10$^{-4}$ $M_\odot$ yr$^{-1}$). The infall radius is found to be nearly the same (2700 AU versus 3500 AU), and the density at that radius differs by a factor of two (6.9$\times$10$^5$ cm$^{-3}$ versus 3.0$\times$10$^5$ cm$^{-3}$), thus accounting for the small difference. For the case of NGC1333-IRAS4A, the interferometric observations of H$_2$CO and the H$_2$O data presented here thus trace infall on the same physical scale, not limited by the optical depth of the H$_2$O line.

The location of the sources are marked in the correlation plots shown in Fig. \ref{fig:correlation} (with a square), but the sources do not stand out in terms of envelope or outflow parameters. Thus, the envelope parameters alone do not appear to determine whether a source shows an inverse P-Cygni profile or not, nor is it the particularly luminous or massive sources which show inverse P-Cygni profiles. Moreover, the geometry of the system does not seem to play a significant role in determining whether a source has an inverse P-Cygni profile; the L1527 outflow is located close to the plane of the sky, that of NGC1333-IRAS4A is likely at an inclination of $\sim$ 45\degr\ whereas the outflow from IRAS15398 is seen nearly pole-on. Further modelling of the sources showing inverse P-Cygni profiles is under way to properly constrain the infall parameters of the envelopes (Mottram et al. in prep.).

\subsubsection{Regular P-Cygni profiles}

Two sources have regular P-Cygni profiles superposed on the outflow component (IRAS12496 and L1551-IRS5; both Class I), with two more sources showing signs of P-Cygni profiles (L723 and HH46-IRS; the former being Class 0\footnote{Although L723 is formally classified as a Class 0 object, it is also the Class 0 object with the highest $T_{\rm bol}$; therefore it may be close to transitioning to Class I.}, the latter Class I; Fig. \ref{fig:infall}). A regular P-Cygni profile is a signature of expansion rather than infall. For the same reasons as above, the signatures are thought to be real and not the effect of chopping into nearby emission or absorption or emission caused by a foreground cloud.

A similar analysis as above can be performed to examine characteristic expansion velocities; the only change in the model is that the infall velocity is negative. All profiles can be reproduced by an expansion velocity of 0.5 km s$^{-1}$ (Fig. \ref{fig:infall}), with the same uncertainty of $\sim$ 0.2 km s$^{-1}$ as found above for the infall velocities. A reversal of infall has been observed in the high-mass star-forming region SgrB2(M) close to the Galactic Centre, where it is proposed that the powerful outflows push out the surrounding envelope, while infall continues along certain preferred directions \citep{rolffs10}. The high-mass star-forming region W3-IRS5 shows the same phenomenon \citep{chavarria10} and it is thus possible that this mechanism is at play on a smaller scale in these low-mass objects, assuming that the protostellar wind is strong enough to reverse infall in large parts of the envelope. In this respect, it is worth noting that the two sources showing the strongest P-Cygni profiles are both Class I sources where the envelope is less dense and easier to push aside. With a constant expansion velocity of 0.5 km s$^{-1}$ the envelope will fully disperse on timescales of 10$^4$ years, which is not consistent with inferred Class I lifetimes of $>$10$^5$ years \citep{evans09}, indicating that the expansion is either a local phenomenon and does not involve the entire envelope, or that the phenomenon is transient and turns on at a certain stage of the evolution when the outflow overcomes the infall motion of the envelope and then disperses the envelope in a short period.

\subsection{H$_2$O versus CO}\label{sec:h2oco}

\begin{figure}
\begin{center}
\includegraphics[width = 0.9\columnwidth]{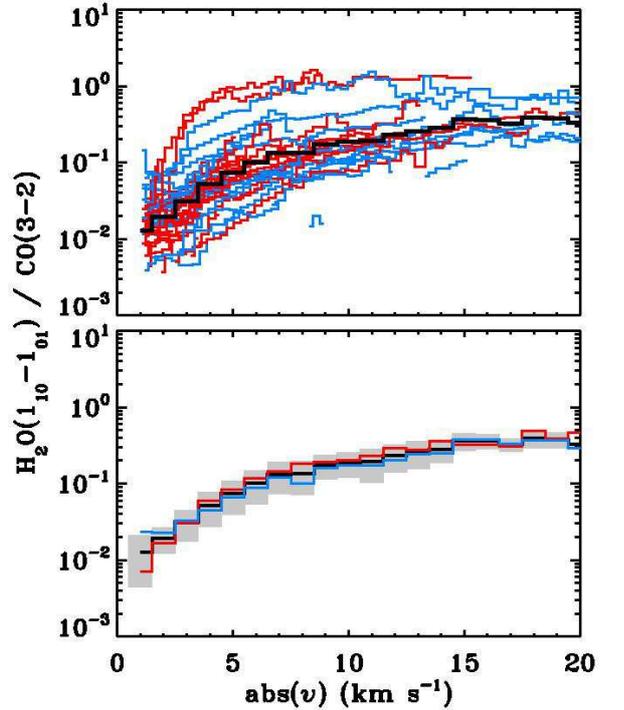}
\end{center}
\caption{Ratio of H$_2$O 1$_{10}$--1$_{01}$ / CO 3--2 line intensities as a function of absolute velocity for all sources. All spectra have been centred at 0 km s$^{-1}$. The top panel shows the ratio for all line wings where both H$_2$O and CO emission exceeds 4 $\sigma$. Red-shifted line wings are shown in red and blue-shifted wings in blue. The average value is shown in black. The bottom panel shows the average value with the standard deviation shown in grey. The average values for the red- and blue-shifted line wings are shown in red and blue, respectively. The average ratios are taken in 1 km s$^{-1}$ bins.}
\label{fig:h2o_co}
\end{figure}

\begin{figure}
\begin{center}
\includegraphics[width = 0.95\columnwidth]{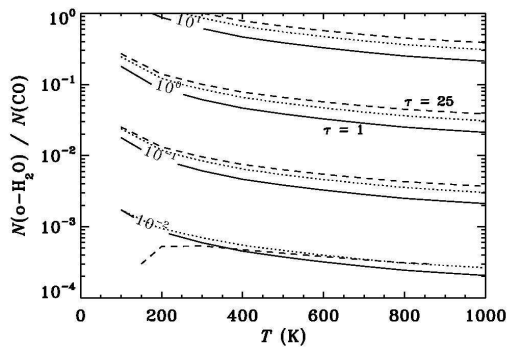}
\includegraphics[width = 0.95\columnwidth]{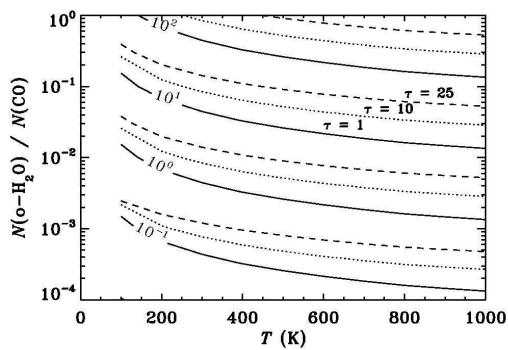}
\end{center}
\caption{H$_2$O / CO column density ratio as a function of kinetic gas temperature for three H$_2$O optical depths, $\tau$=1, 10 and 25 (CO is assumed to be optically thin in all cases). Contours are for intensity ratios of H$_2$O 1$_{10}$--1$_{01}$ / CO 3--2. The top panel shows the ratio for a density of $n_{\rm H}$ = 10$^5$ cm$^{-3}$ while the bottom panel is for a density of 10$^6$ cm$^{-3}$.}
\label{fig:h2o_co2}
\end{figure}

One of the results from SWAS observations is that the H$_2$O / CO abundance ratio increases with velocity, as shown for the case of Orion \citep{franklin08}. In our sample of 29 YSOs, the H$_2$O 1$_{10}$--1$_{01}$ / CO 3--2 line ratio increases with velocity, as shown in Fig. \ref{fig:h2o_co}, from an initial value of $\sim$0.01 to $\sim$0.2. There is no significant difference between the line ratio measured in red-shifted line wings compared to those in blue-shifted wings. For L1448, where both H$_2$O and CO emission are detected to $\varv$$>$40 km s$^{-1}$, the line ratio remains constant at $\sim$0.3 for $\varv$$>$20 km s$^{-1}$. 

The non-LTE statistical equilibrium radiative transfer code RADEX \citep{vandertak07} is used to estimate the abundance of H$_2$O with respect to CO in the optically thin limit. The kinetic gas temperature of the H$_2$O and CO emitting regions are assumed to be identical, and the derived abundance ratio is shown for two different densities, $n_{\rm H}$=10$^5$ and 10$^6$ cm$^{-3}$ (Fig. \ref{fig:h2o_co2}). Recent independent studies suggest that this assumption may not be valid \citep{bjerkeli11, santangelo12, vasta12}, but we use it here to illustrate the typical order-of-magnitude abundance ratios of H$_2$O/CO. If the density is low, the observed range of line ratios corresponds to abundance ratios of the order of 10$^{-3}$ -- 10$^{-1}$ in the H$_2$O and CO optically thin limit. For a standard abundance of CO of 10$^{-4}$ with respect to H$_2$, this translates to H$_2$O abundances of 10$^{-7}$--10$^{-5}$ relative to H$_2$. For the high-density case, these numbers are a factor of 10 lower. 

The resulting H$_2$O abundance is consistent with what is inferred elsewhere \citep[$x$(H$_2$O) $\sim$ 10$^{-7}$--10$^{-5}$;][]{franklin08, lefloch10, kristensen10b}. \citet{kristensen11} studied the L1448-MM outflow and analysed the H$_2$O abundance in the line wings reaching the conclusion that the integrated water emission is very optically thick ($\tau$$>$10) based on a tentative detection of broad H$_2^{18}$O emission. If the same exercise as above is performed but now for the optically thick case, higher H$_2$O / CO abundance ratios are inferred for the high-velocity and high-density gas leading to H$_2$O abundances of $\sim$10$^{-5}$ with respect to H$_2$. Figure \ref{fig:h2o_co2} shows the abundance ratio for three optical depths of the H$_2$O line, $\tau$ = 1, 10 and 25. The CO 3--2 emission remains optically thin for all the considered physical conditions ($\tau_{\rm CO}$$\ll$1). Thus, to obtain even an approximate value for the H$_2$O / CO abundance ratio, the optical depth of the H$_2$O emission should always be taken into account.

The weak H$_2$O-CO intensity correlation (Fig. \ref{fig:correlation}) is likely due to differences in the density of the emitting material and the abundance ratio of H$_2$O / CO. Because H$_2$O is sub-thermally excited (the critical density for the 1$_{10}$--1$_{01}$ line is $\sim$10$^8$ cm$^{-3}$), the intensity scales as $n$(H$_2$)$N$(H$_2$O) whereas the intensity of the thermally excited CO scales directly with $N$(CO) (see also Fig. \ref{fig:h2o_co2}). Both of these proportionalities are only valid in the optically thin limit. The CO emission in the line wings is usually close to optically thin \citep[$\tau$$<$1; e.g.,][Y{\i}ld{\i}z et al. subm.]{curtis10}, whereas water is effectively optically thin \citep[Fig. \ref{fig:h2o_co2};][]{snell00}. The H$_2$O / CO abundance ratio is the same for all sources to within an order of magnitude (see above), and therefore the weak correlation is likely dominated by differences in the density. The density is estimated using the results from the DUSTY modelling (Appendix \ref{app:dusty}). The correlation between H$_2$O emission and the density at 1000 AU is strong ($r$ = 0.71; Fig. \ref{fig:correlation}) and does not degrade if the density is taken at 500 or 1500 AU. This further corroborates the result that variations in the H$_2$O-CO correlation are dominated by density effects.

\subsection{H$_2$O and evolution: Class 0 versus Class I}

\begin{figure}
\begin{center}
\includegraphics[width = 0.9\columnwidth]{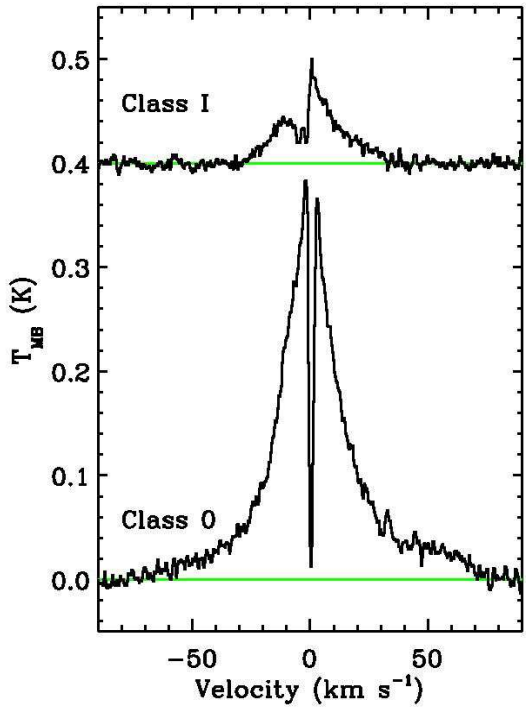}
\end{center}
\caption{H$_2$O spectra of Class 0 and Class I sources averaged independently and compared. All spectra were shifted to 0 km s$^{-1}$ before averaging, rebinned to 0.3 km s$^{-1}$ channels and shifted to a common distance of 200 pc. GSS30-IRS1, R CrA-IRS5 and HH100-IRS are not included, see text.}
\label{fig:evolution}
\end{figure}

A clear trend found in this study is the lower integrated 557 GHz intensity of Class I sources compared to Class 0 sources. Figure \ref{fig:evolution} compares the average Class 0 spectrum with the average Class I spectrum and there is a difference of a factor of 8 in integrated intensity between the two. For this comparison all spectra have been scaled to a common distance of 200 pc, and GSS30-IRS1, R CrA-IRS5 and HH100-IRS have not been included, since their excitation may not be representative of Class I sources (see Sect. \ref{sec:h2o_results}). Two evolutionary parameters have previously been used in the literature: $T_{\rm bol}$ and $L_{\rm bol}^{0.6}$ / $M_{\rm env}$ \citep{chen95, saraceno96}, both parameters increasing in value with evolution. There is indeed an anti-correlation of H$_2$O emission with both evolutionary tracers (albeit weak, $r$ = $-$0.60 and $-$0.61, respectively; Fig. \ref{fig:correlation}), thus showing that water emission decreases with evolutionary stage.

Class I envelopes are generally characterised by lower masses and lower densities throughout, conclusions that are confirmed by the envelope modelling presented here. As the protostellar system evolves, the outflow cavity opens, and the outflow becomes less collimated \citep[e.g.,][]{arce06}. Furthermore, the material swept up by the outflow is less dense, because the envelope is less dense, and therefore the H$_2$O intensity decreases. This has the effect of decreasing the outflow force, as there is less material to interact with \citep[e.g.,][]{bontemps96, hogerheijde97}. Therefore the decrease in H$_2$O emission in Class I systems compared to Class 0 systems may be directly linked to the weaker outflows, which in turn is linked to lower mass infall rates and envelope dispersal.

It has previously been proposed that UV irradiation of the envelope may lead to photodissociation of water, thereby lowering the abundance \citep{nisini02}. Also, X-rays have been found to effectively destroy water \citep{stauber06}. These mechanisms cannot be ruled out with the observations presented here, however no decrease in the H$_2$O / CO ratio is observed in the Class I objects compared to the Class 0 ones (see Fig. \ref{fig:h2o_co_class01} in the Appendix). Because the bulk of H$_2$O emission in the Class I sources is associated with the medium component, it is only possible to analyse the H$_2$O / CO ratio over a few km s$^{-1}$ from the source velocity for most of the Class I outflows. To fully test this scenario, high $S/N$ velocity-resolved observations of the photodissociation product OH are required, such as has recently been done for the high-mass object W3-IRS5 \citep{wampfler11}.

\section{Conclusions}\label{sec:conc}

Emission from the o-H$_2$O 1$_{10}$--1$_{01}$ 557 GHz transition is detected in 28 low-mass Class 0 and I YSOs from a sample of 29 sources within the WISH key programme on \textit{Herschel}. Line profiles are complex and rich in different dynamical components. In particular, the maximum width of the Class 0 profiles extends beyond 100 km s$^{-1}$. Several dynamical components that have not been seen before in other molecular tracers show up in the H$_2$O line profiles. These components include molecular ``bullets'' and regular and inverse P-Cygni profiles, indicative of expansions and infall, respectively. In particular, the occurence of water bullets in more sources than was previously known indicates that this is a common phenomenon, and that water is one of the best tracers of these objects.

Class 0 and I sources appear to be quantitatively different in terms of water line profiles and their intensity. Class 0 sources are associated with broad and medium components, both Gaussian in nature, whereas the Class I sources tend to be associated more with the medium component. In sources where both a medium and broad component are detected, the medium component is spatially less extended than the broad component and located between us and the region causing the broad component. We speculate that the medium component is the thermal component of the 22 GHz H$_2$O maser emission observed in these sources. If that is the case, the excitation conditions ($T$, $n$) are higher in the medium component than the broad. Infall is observed directly in six Class 0 sources and one Class I, whereas expansion is observed primarily in Class I sources. Bullets appear exclusively in Class 0 sources, but not in Class I sources. Absorption features are predominantly saturated in Class 0 sources, but not in Class I sources, directly deomonstrating that the water abundance in the outer envelope is low, $\gtrsim$10$^{-11}$--10$^{-9}$. This confirms earlier ideas of the Class 0 phase being dominated by prominent outflows, while at the same time being dominated by infall. By the time the YSO enters the Class I phase, outflow activity has decreased significantly, and infall has in some cases turned to large-scale expansion of the envelope. 

Modelling of the inverse P-Cygni profiles in Class 0 objects leads to inferred infall velocities of $\sim$ 0.1 -- 1.0 km s$^{-1}$. The profiles are tracing large-scale infall ($\gtrsim$1000 AU). Corresponding mass accretion rates are in the range of $\sim$10$^{-6}$ -- 10$^{-4}$ $M_\odot$ yr$^{-1}$ depending on source. The expanding Class I envelopes are modelled using the same prescription, leading to characteristic expansion velocities of $\sim$0.5 km s$^{-1}$. Expansion profiles have previously been observed in high-mass star-forming regions, but this is the first time they are observed in several low-mass objects. The dispersion timescale is short ($\sim$10$^4$ years) indicating that the expansion is either localized to small regions in the envelope or that it is a phenomenon taking place at a certain stage of protostellar evolution when the outflow overcomes the infall motion of the envelope and then disperses the envelope in a short period.

Water emission from the outflow is found to be controlled by the envelope density, which is used as a proxy for the density of the outflowing gas. Only small chemical differences between the outflows are found, and any difference is dominated by different excitation conditions. There is a tendency for more evolved Class I objects to appear weaker in H$_2$O emission, an observational fact that is linked to the lower density in the envelopes and the weaker outflow force.

The water abundance with respect to CO is found to increase with velocity in all sources from 10$^{-3}$ close to the source velocity to 0.1 at velocities greater than 10 km s$^{-1}$ from the source, assuming that the water and CO excitation conditions are similar. The increase is uniform in both the red- and blue-shifted line wings. There is no systematic difference between the H$_2$O / CO abundance ratio in Class 0 and I sources, although the Class I line wings are weaker and it is difficult to make this conclusion robust.

The results presented here are one of the first steps towards an understanding of the role water plays in star formation. Thanks to its high spatial resolution and sensitivity, HIFI is delivering velocity-resolved water data for a much larger sample of low-mass protostars than has ever been done before on scales where the relevant physical processes occur. For the first time it is possible to study the evolution of water in star-forming regions, something that will continue to be done within the framework of the WISH programme.

\begin{acknowledgements}
We would like to thank both the WISH and DIGIT teams for many stimulating discussions, in particular Carolyn M$^{\rm c}$Coey and Joseph Mottram. Astrochemistry in Leiden is supported by the Netherlands Research School for Astronomy (NOVA), by a Spinoza grant and grant 614.001.008 from the Netherlands Organisation for Scientific Research (NWO), and by the European Community's Seventh Framework Programme FP7/2007-2013 under grant agreement 238258 (LASSIE). HIFI has been designed and built by a consortium of institutes and university departments from across Europe, Canada and the US under the leadership of SRON Netherlands Institute for Space Research, Groningen, The Netherlands with major contributions from Germany, France and the US. Consortium members are: Canada: CSA, U.Waterloo; France: CESR, LAB, LERMA, IRAM; Germany: KOSMA, MPIfR, MPS; Ireland, NUI Maynooth; Italy: ASI, IFSI-INAF, Arcetri-INAF; Netherlands: SRON, TUD; Poland: CAMK, CBK; Spain: Observatorio Astronomico Nacional (IGN), Centro de Astrobiolog{\'i}a (CSIC-INTA); Sweden: Chalmers University of Technology - MC2, RSS \& GARD, Onsala Space Observatory, Swedish National Space Board, Stockholm University - Stockholm Observatory; Switzerland: ETH Z{\"u}rich, FHNW; USA: Caltech, JPL, NHSC. HIPE is a joint development by the Herschel Science Ground Segment Consortium, consisting of ESA, the NASA Herschel Science Center, and the HIFI, PACS and SPIRE consortia.
\end{acknowledgements}

\bibliographystyle{aa}
\bibliography{bibliography}

\Online
\appendix

\section{H$_2$O supplementary material}\label{app:h2o}

\begin{figure}
\begin{center}
\includegraphics[width = 0.9\columnwidth]{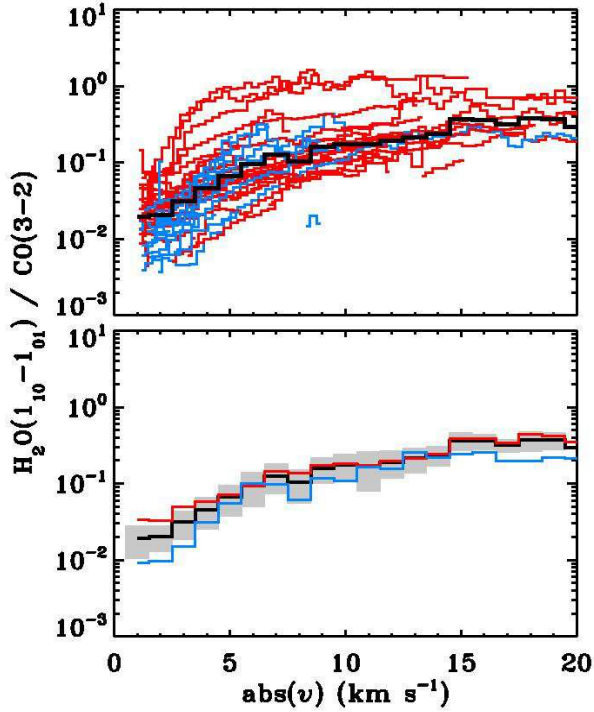}
\end{center}
\caption{Emission ratio of H$_2$O 1$_{10}$--1$_{01}$ / CO 3--2 as a function of absolute velocity. All spectra have been centred at 0 km s$^{-1}$. The top panel shows the ratio for all line wings where both H$_2$O and CO emission exceeds 4 $\sigma$. Line wings from Class 0 sources are shown in red, wings from Class I sources in blue. The average value is shown in black. The bottom panel shows the average value with the standard deviation shown in grey. The average values for the Class 0 and Class I sources in red and blue, respectively. The average ratios are taken in 1 km s$^{-1}$ bins.}
\label{fig:h2o_co_class01}
\end{figure}

Table \ref{tab:reference} provides an overview of the specific observing mode used for each object, including (when applicable) the reference position used. Reference positions were primarily chosen from SWAS observations showing these regions to be emission free, and regions were in general checked against ground-based CO 1--0 observations.

\begin{table*}
\caption{Reference positions chosen for each source and obs ids.}
\tiny
\begin{center}
\begin{tabular}{l c c c c c c}
\hline\hline
Source & Obs. mode\tablefootmark{a} & RA\tablefootmark{b} & Dec\tablefootmark{b} & $\Delta$RA\tablefootmark{c} & $\Delta$Dec\tablefootmark{c} & Obs id \\ \hline
L1448-MM   &	PSSW	&		&		&	10\arcmin		&	5\arcmin	&	1342203202 \\
NGC1333-IRAS2A  &	PSSW	&	52\fdg3525	&	31\fdg1665		&		&		&	1342202067 \\
NGC1333-IRAS4A  &	PSSW	&	52\fdg3525	&	31\fdg1665		&		&		&	1342202065 \\
NGC1333-IRAS4B  &	PSSW	&	52\fdg3525	&	31\fdg1665		&		&		&	1342202064 \\
L1527      &	DBS		&				&					&		&		&	1342192524 \\
Ced110-IRS4  &	DBS		&				&					&		&		&	1342201525 \\
BHR71      &	PSSW	&		&		&	3\arcmin		&	0\arcmin 	&	1342201677 \\
IRAS15398  &	DBS		&				&					&		&		&	1342213732 \\
L483       &	PSSW	&		&		&	3\arcmin		&	3\arcmin	&	1342217691 \\
Ser SMM1   &	PSSW	&	277\fdg4011	&	1\fdg3112		&		&		&	1342208580 \\
Ser SMM4   &	PSSW	&	277\fdg4011	&	1\fdg3112		&		&		&	1342208579 \\
Ser SMM3   &	PSSW	&	277\fdg4011	&	1\fdg3112		&		&		&	1342208577 \\
L723       &	DBS		&				&					&		&		&	1342210077 \\
B335       &	PSSW	&	294\fdg2271	&	7\fdg4356		&		&		&	1342196409 \\
L1157      &	PSSW	&		&		&	10\arcmin		&	2\arcmin 	&	1342196407 \\ \hline
NGC1333-IRAS3   &	PSSW	&	52\fdg3525	&	31\fdg1665		&		&		&	1342202066 \\
L1489      &	PSSW	&	59\fdg5100	&	26\fdg6428		&		&		&	1342203197 \\
L1551-IRS5   &	PSSW	&	68\fdg0388	&	17\fdg9564		&		&		&	1342203194 \\
TMR1       &	DBS		&				&					&		&		&	1342192525 \\
TMC1A      &	DBS		&				&					&		&		&	1342192527 \\
TMC1       &	DBS		&				&					&		&		&	1342192526 \\
HH46-IRS       &	DBS		&				&					&		&		&	1342196410 \\
IRAS12496  &	DBS		&				&					&		&		&	1342201526 \\
GSS30-IRS1      &	PSSW	&		&		&	60\arcmin		&	0\arcmin	&	1342205302 \\
Elias 29   &	PSSW	&		&		&	60\arcmin		&	0\arcmin	&	1342204011 \\
Oph-IRS63    &	PSSW	&		&		&	60\arcmin		&	0\arcmin	&	1342205304 \\
RNO91      &	PSSW	&	249\fdg5017	&	$-$16\fdg2703	&		&		&	1342205297 \\
RCrA-IRS5A   &	PSSW	&	284\fdg5913	&	$-$36\fdg5986	&		&		&	1342215840 \\
HH100-IRS      &	PSSW	&	284\fdg5913	&	$-$36\fdg5986	&		&		&	1342215841 \\
\hline
\end{tabular}
\tablefoot{
	\tablefoottext{a}{Observations were done in position switch mode (PSSW) or dual-beam-switch mode (DBS). In the latter case, a throw of 3\arcmin\ was always used.}
	\tablefoottext{b}{Absolute coordinates (J2000) for the reference position.}
	\tablefoottext{c}{Relative coordinates with respect to the source position.}
		}
\end{center}
\label{tab:reference}
\end{table*}

Table \ref{tab:intensity_all} provides the decomposition of each profile into Gaussians. Only saturated absorption features are not listed here.

\begin{table*}
\caption{Gaussian decomposition of the H$_2$O line profiles.}
\centering
\begin{tabular}{l l c c c c c}
\hline \hline
Source & Component & $T^{\rm peak}_{\rm MB}$ & $\int$ $T_{\rm MB}$ d$\varv$ & $\Delta\varv$\tablefootmark{a} & $\varv_{\rm LSR}$ & $\varv_{\rm source}$ \\
       &           & (K)                     & (K\,km\,s$^{-1}$)            & (km s$^{-1}$) & (km s$^{-1}$) & (km s$^{-1}$) \\
\hline
L1448-MM					    &  Broad & \phantom{$-$}0.28 &             15.68 &			52.8	& \phantom{$-$}14.5 & \phantom{1}$+$4.7 \\
							    &  EVH-R & \phantom{$-$}0.11 & \phantom{$-$}2.17 &			19.2	& \phantom{$-$}61.9 \\
							    &  EHV-B & \phantom{$-$}0.07 & \phantom{$-$}2.21 &			30.8	&           $-$47.0 \\
							    & Narrow &           $-$0.19 &           $-$0.99 &	\phantom{1}4.9	& \phantom{$-$1}4.8 \\
NGC1333-IRAS2A				    &  Broad & \phantom{$-$}0.12 & \phantom{$-$}4.49 &			36.4	& \phantom{$-$1}3.4 & \phantom{1}$+$7.7 \\
							    & Medium & \phantom{$-$}0.19 & \phantom{$-$}1.61 &	\phantom{1}7.9	& \phantom{$-$1}9.6 \\
NGC1333-IRAS4A\tablefootmark{b}	&  Broad & \phantom{$-$}0.43 &             15.02 &			33.0	& \phantom{$-$1}7.8 & \phantom{1}$+$7.2 \\
							    & Medium & \phantom{$-$}0.28 & \phantom{$-$}3.00 &			10.1	& \phantom{$-$1}0.0 \\
NGC1333-IRAS4B				    &  Broad & \phantom{$-$}0.41 &             10.07 &			23.6	& \phantom{$-$1}6.2 & \phantom{1}$+$7.4 \\
							    & Medium & \phantom{$-$}0.27 & \phantom{$-$}1.43 &	\phantom{1}8.8	& \phantom{$-$1}5.0 \\
L1527						    & Medium & \phantom{$-$}0.05 & \phantom{$-$}0.99 &			20.3	& \phantom{$-$1}5.6 & \phantom{1}$+$5.9 \\
Ced110-IRS4					    & Medium & \phantom{$-$}0.02 & \phantom{$-$}0.35 &			14.3	& \phantom{$-$1}4.9 & \phantom{1}$+$3.5 \\
BHR71\tablefootmark{b}		    &  Broad & \phantom{$-$}0.15 & \phantom{$-$}7.21 &			44.6	& \phantom{1}$-$2.7 & \phantom{1}$-$4.4 \\
							    &  EVH-R & \phantom{$-$}0.05 & \phantom{$-$}1.50 &			29.0	& \phantom{$-$}55.4 \\
							    &  EHV-B & \phantom{$-$}0.03 & \phantom{$-$}0.82 &			23.0	&           $-$48.7 \\
							    & Narrow & \phantom{$-$}0.04 & \phantom{$-$}0.09 &	\phantom{1}1.8	&           $-$12.3 \\
IRAS15398\tablefootmark{b}	    & Medium & \phantom{$-$}0.09 & \phantom{$-$}1.78 &			18.9	& \phantom{$-$1}0.6 & \phantom{1}$+$5.1 \\
L483						    & Medium & \phantom{$-$}0.09 & \phantom{$-$}1.62 &			15.3	& \phantom{$-$1}3.5 & \phantom{1}$+$5.2 \\
Ser SMM1					    &  Broad & \phantom{$-$}0.16 & \phantom{$-$}6.62 &			39.0	& \phantom{$-$}14.2 & \phantom{1}$+$8.5 \\
							    & Medium & \phantom{$-$}0.45 & \phantom{$-$}6.46 &			13.6	& \phantom{$-$1}7.4 \\
Ser SMM4\tablefootmark{b}	    &  Broad & \phantom{$-$}0.14 & \phantom{$-$}7.72 &			40.2	& \phantom{$-$1}0.7 & \phantom{1}$+$8.0 \\
Ser SMM3					    &  Broad & \phantom{$-$}0.12 & \phantom{$-$}5.92 &			46.8	& \phantom{$-$1}3.2 & \phantom{1}$+$7.6 \\
							    & Medium &           $-$0.11 &           $-$2.03 &			17.3	& \phantom{$-$1}0.0 \\
L723\tablefootmark{c}		    &  Broad & \phantom{$-$}0.04 & \phantom{$-$}0.68 &			17.4	& \phantom{$-$1}9.9 &           $+$11.2 \\
B335						    &  Broad & \phantom{$-$}0.03 & \phantom{$-$}1.15 &			40.5	& \phantom{$-$1}6.7 & \phantom{1}$+$8.4 \\
							    & Narrow & \phantom{$-$}0.07 & \phantom{$-$}0.08 &	\phantom{1}1.1	& \phantom{$-$1}9.3 \\
L1157\tablefootmark{b}		    &  Broad & \phantom{$-$}0.09 & \phantom{$-$}2.64 &			26.5	& \phantom{$-$1}4.2 & \phantom{1}$+$2.6 \\
							    &  EVH-R & \phantom{$-$}0.01 & \phantom{$-$}0.30 &			20.5	& \phantom{$-$}42.6 \\ \hline
NGC1333-IRAS3				    &  Broad & \phantom{$-$}0.19 & \phantom{$-$}6.22 &			30.1	& \phantom{$-$1}5.2 & \phantom{1}$+$8.5 \\
							    & Medium &           $-$0.14 &           $-$0.90 &	\phantom{1}5.9	& \phantom{$-$1}4.9 \\
L1489						    &  Broad & \phantom{$-$}0.05 & \phantom{$-$}0.87 &			17.6	& \phantom{$-$1}4.1 & \phantom{1}$+$7.2 \\
							    & Narrow &			 $-$0.06 &           $-$0.07 &	\phantom{1}1.2	& \phantom{$-$1}5.8 \\
L1551-IRS5\tablefootmark{c}	    &  Broad & \phantom{$-$}0.02 & \phantom{$-$}0.66 &			28.1	& \phantom{$-$}16.1 & \phantom{1}$+$7.2 \\
TMR1						    & Medium & \phantom{$-$}0.03 & \phantom{$-$}0.36 &			11.4	& \phantom{$-$1}4.4 & \phantom{1}$+$6.3 \\
							    & Narrow &           $-$0.04 &           $-$0.06 &	\phantom{1}1.6	& \phantom{$-$1}5.7 \\
TMC1A						    &		 &           $<$0.02 &           $<$0.04 &			\ldots	& \ldots	                            \\
TMC1						    &  Broad & \phantom{$-$}0.02 & \phantom{$-$}0.49 &			28.4	& \phantom{$-$1}3.4 & \phantom{1}$+$5.2 \\
HH46-IRS\tablefootmark{c}	    &  Broad & \phantom{$-$}0.04 & \phantom{$-$}1.23 &			26.8	& \phantom{$-$1}9.8 & \phantom{1}$+$5.2 \\
IRAS12496\tablefootmark{c}	    &  Broad & \phantom{$-$}0.02 & \phantom{$-$}0.58 &			25.4	& \phantom{$-$1}1.8 & \phantom{1}$+$2.3 \\
GSS30-IRS1					    & Medium & \phantom{$-$}0.18 & \phantom{$-$}3.32 &			17.8	& \phantom{$-$1}1.9 & \phantom{1}$+$2.8 \\
Elias29						    & Medium & \phantom{$-$}0.13 & \phantom{$-$}2.25 &			16.7	& \phantom{$-$1}3.4 &                   \\
							    & Narrow &           $-$0.06 &           $-$0.07 &	\phantom{1}1.1	& \phantom{$-$1}5.5   \phantom{1}$+$5.0 \\
Oph-IRS63					    & Medium & \phantom{$-$}0.01 & \phantom{$-$}0.09 &	\phantom{1}6.6	& \phantom{$-$1}0.5 & \phantom{1}$+$2.8 \\
RNO91						    & Narrow & \phantom{$-$}0.03 & \phantom{$-$}0.50 &			18.7	& \phantom{$-$1}6.8 & \phantom{1}$+$5.0 \\
RCrA-IRS5A					    & Medium & \phantom{$-$}0.12 & \phantom{$-$}1.93 &			15.2	& \phantom{$-$1}7.4 & \phantom{1}$+$5.7 \\
							    & Narrow & \phantom{$-$}0.36 & \phantom{$-$}0.96 &	\phantom{1}2.5	& \phantom{$-$1}6.0 \\
							    & Narrow &           $-$0.28 &           $-$0.20 &	\phantom{1}0.7	& \phantom{$-$1}5.7 \\
HH100-IRS					    & Medium & \phantom{$-$}0.05 & \phantom{$-$}1.06 &			19.1	& \phantom{$-$1}1.7 & \phantom{1}$+$5.6 \\
							    & Narrow & \phantom{$-$}0.20 & \phantom{$-$}0.95 &	\phantom{1}4.5	& \phantom{$-$1}5.4 \\
							    & Narrow & \phantom{$-$}0.35 & \phantom{$-$}0.31 &	\phantom{1}0.8	& \phantom{$-$1}6.3 \\
\hline\\
\end{tabular}
\tablefoot{
	All values are obtained from (multiple) Gaussian fits to each line profile. Non-saturated absorption components have negative values of $T_{\rm MB}^{\rm peak}$ and $\int T_{\rm MB}$ d$\varv$. Saturated absorption features are not included in this table. See Table \ref{tab:features} for an overview of what components are present in which source.
	\tablefoottext{a}{$\Delta\varv$ corresponds to the FWHM of the component.}
	\tablefoottext{b}{Inverse P-Cygni source; component parameters not shown.}
	\tablefoottext{c}{Regular P-Cygni source; component parameters not shown.}
		}
\label{tab:intensity_all}
\end{table*}

\clearpage

\section{CO supplementary material}\label{app:co3-2}

\begin{figure*}
\begin{center}
\includegraphics[width = 17cm]{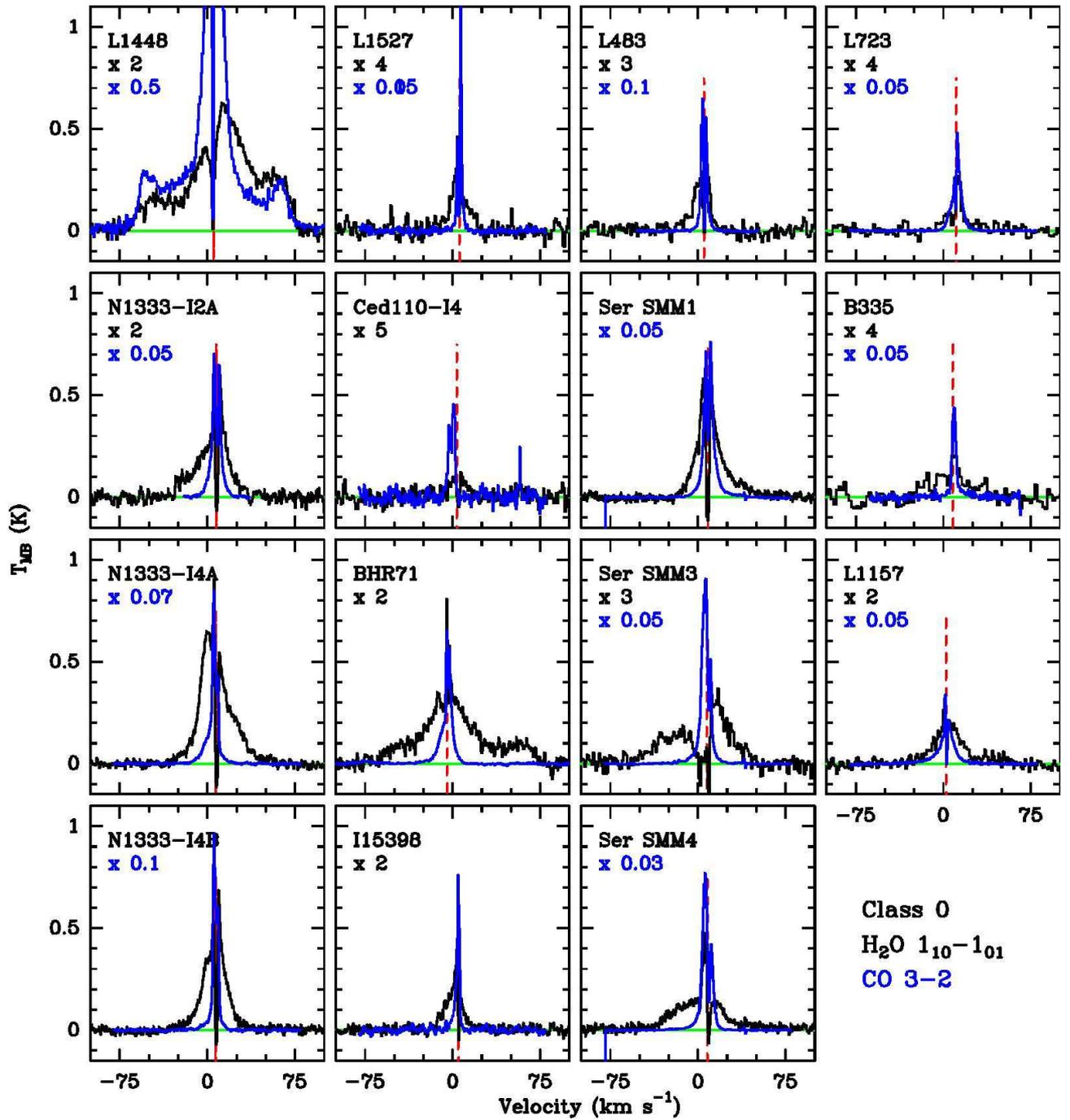}
\end{center}
\caption{Continuum-subtracted H$_2$O 1$_{10}$--1$_{01}$ spectra of the observed Class 0 sources as presented in Fig. \ref{fig:class0} with CO 3--2 spectra scaled and overplotted in blue. The CO 3--2 scaling factor is written in blue for each spectrum.}
\label{fig:class0co}
\end{figure*}

\begin{figure*}
\begin{center}
\includegraphics[width = 17cm]{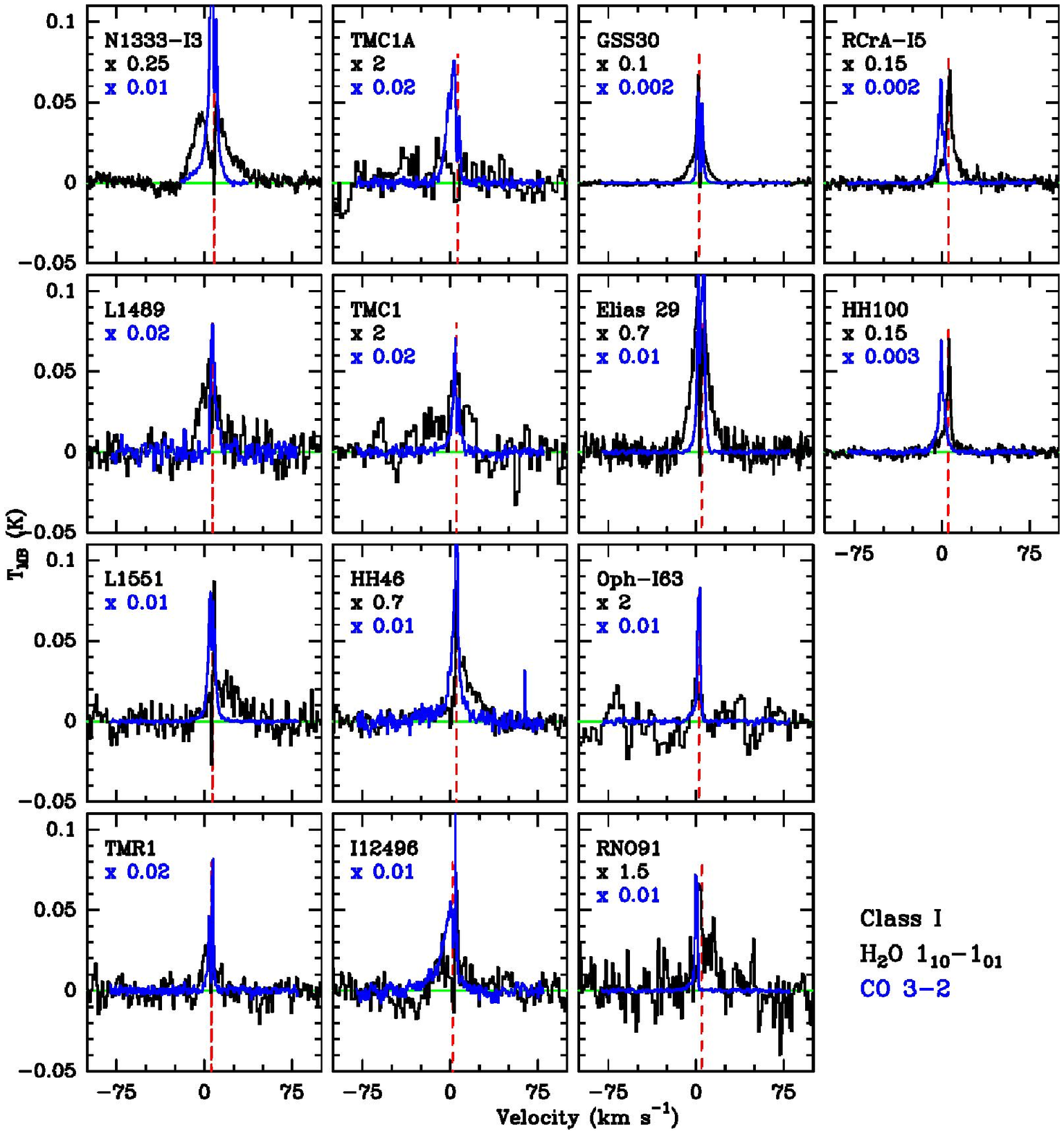}
\end{center}
\caption{Same as Fig. \ref{fig:class0co} but for Class I sources.}
\label{fig:class1co}
\end{figure*}

\begin{figure*}
\begin{center}
\includegraphics[width=17cm]{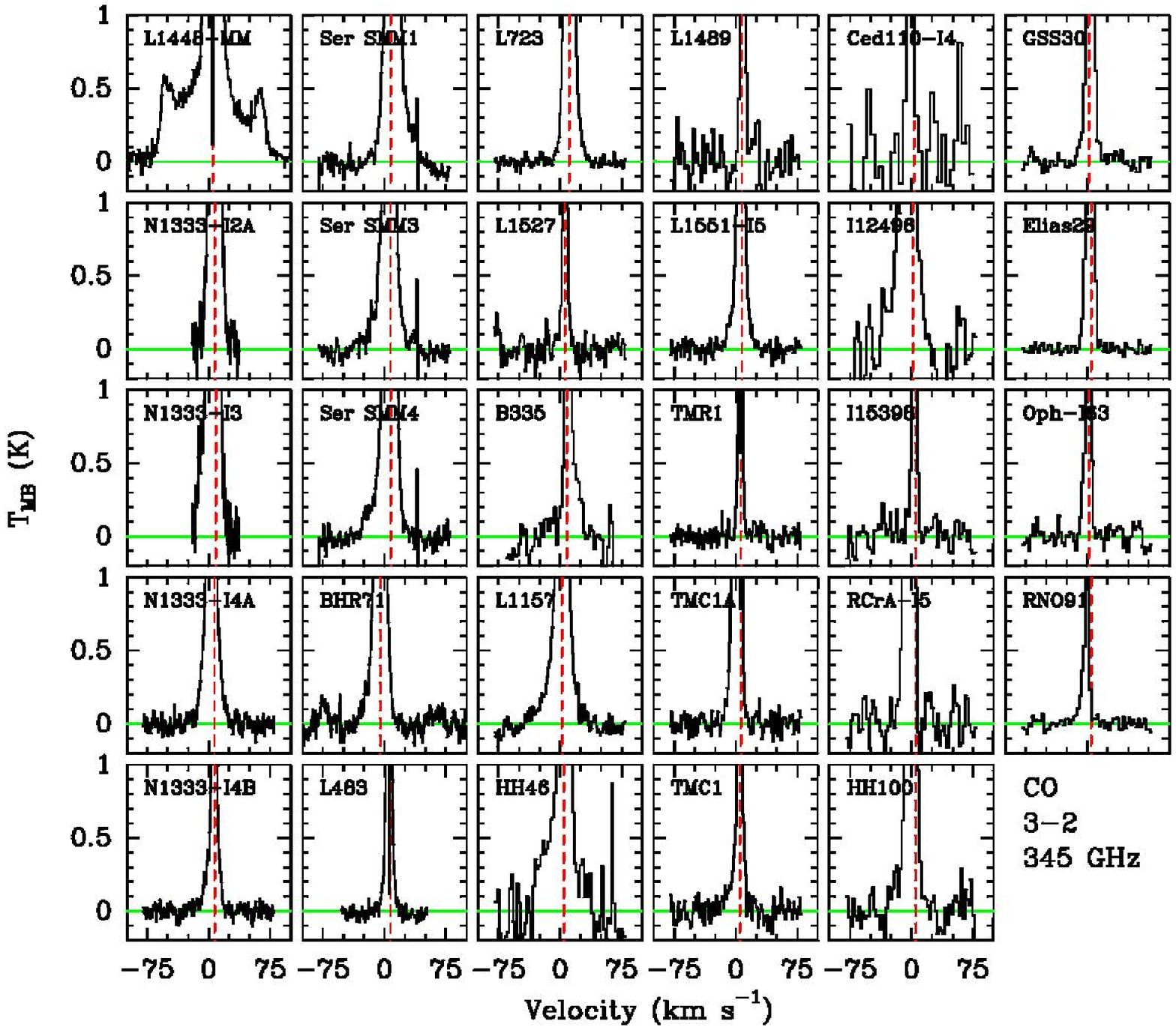}
\end{center}
\caption{CO 3--2 data for each source as obtained with the JCMT or APEX. The velocity and intensity scales are the same as in Figs. \ref{fig:class0} and \ref{fig:class1}.}
\label{fig:co3-2}
\end{figure*}

The CO 3--2 data are presented here. This includes the spectra overplotted on the H$_2$O spectra (Figs. \ref{fig:class0co} and \ref{fig:class1co}). Magnifications to highlight the CO emission in the line wings are shown in Fig. \ref{fig:co3-2} where the spectra are shown on the same absolute scale as the H$_2$O data in Figs. \ref{fig:class0} and \ref{fig:class1}. The integrated intensities are provided in Table \ref{tab:intensity_co}.

\begin{table*}[t]
\caption{Observed integrated CO 3--2 intensity in a 40\arcsec\ beam from the JCMT and APEX, and CO outflow force, $F_{\rm CO}$.}
\centering
\begin{tabular}{l c c c c c c}
\hline \hline
Source & rms\tablefootmark{a}  & $T_{\rm MB}^{\rm peak}$ & $\int T_{\rm MB}~{\rm d}\varv$ & $\Delta\varv_{\rm max}$\tablefootmark{b} & $F_{\rm CO}$\tablefootmark{c} & Reference \\
       & (mK) & (K) & (K\,km\,s$^{-1}$) & (km s$^{-1}$) & ($M_\odot$ yr$^{-1}$ km s$^{-1}$) & \\
\hline
L1448-MM				&	0.07	&	\phantom{1}6.2	&				120	&				134	&	9.6($-$5)	&	1,2,3	\\
NGC1333-IRAS2A			&	0.11	&			14.3	&				106	&	\phantom{1}26	&	2.4($-$4)	&	3	\\
NGC1333-IRAS4A			&	0.06	&			12.7	&	\phantom{1}69	&	\phantom{1}34	&	2.3($-$4)	&	3	\\
NGC1333-IRAS4B			&	0.04	&	\phantom{1}9.9	&	\phantom{1}46	&	\phantom{1}24	&	1.2($-$4)	&	3	\\
L1527					&	0.17	&			12.4	&	\phantom{1}29	&	\phantom{15}7	&	1.6($-$4)	&	1,4	\\
Ced110-IRS4				&	0.77	&	\phantom{1}9.8	&	\phantom{1}50	&	\phantom{15}7	&	9.6($-$5)	&	5	\\
BHR71					&	0.05	&			13.2	&	\phantom{1}85	&				144	&	2.6($-$3)	&	6	\\
IRAS15398				&	0.15	&	\phantom{1}8.8	&	\phantom{1}25	&	\phantom{1}10	&	3.9($-$5)	&	5	\\
L483\tablefootmark{d}	&	0.05	&	\phantom{1}7.7	&	\phantom{1}33	&	\phantom{1}17	&	1.5($-$5)	&	1	\\
Ser SMM1				&	0.10	&			15.5	&				133	&	\phantom{1}46	&	1.1($-$4)	&	7	\\
Ser SMM4				&	0.09	&			25.8	&				142	&	\phantom{1}43	&	1.1($-$4)	&	7	\\
Ser SMM3				&	0.09	&			18.3	&				191	&	\phantom{1}39	&	3.8($-$5)	&	7	\\
L723					&	0.07	&	\phantom{1}9.8	&	\phantom{1}64	&	\phantom{1}29	&	6.0($-$5)	&	1	\\
B335\tablefootmark{d}	&	0.38	&	\phantom{1}9.1	&	\phantom{1}44	&	\phantom{1}10	&	2.4($-$5)	&	1	\\
L1157					&	0.08	&	\phantom{1}8.0	&	\phantom{1}58	&	\phantom{1}43	&	1.5($-$4)	&	8	\\ \hline
NGC1333-IRAS3			&	0.09	&			17.7	&				116	&	\phantom{1}36	&	6.2($-$4)	&	2,3	\\
L1489					&	0.54	&	\phantom{1}5.7	&	\phantom{1}21	&	\phantom{15}5	&	1.4($-$5)	&	4	\\
L1551-IRS5				&	0.08	&	\phantom{1}8.2	&	\phantom{1}56	&	\phantom{1}26	&	2.2($-$3)	&	2,4	\\
TMR1					&	0.12	&	\phantom{1}6.2	&	\phantom{1}13	&	\phantom{15}7	&	6.6($-$6)	&	1,4	\\
TMC1A					&	0.13	&	\phantom{1}3.8	&	\phantom{1}35	&	\phantom{1}17	&	8.7($-$6)	&	1,4	\\
TMC1					&	0.14	&	\phantom{1}3.7	&	\phantom{1}19	&	\phantom{1}14	&	2.6($-$6)	&	1,4	\\
HH46-IRS				&	0.55	&			19.0	&	\phantom{1}90	&	\phantom{1}17	&	1.7($-$4)	&	5	\\
IRAS12496				&	0.42	&			12.5	&	\phantom{1}85	&	\phantom{1}24	&	\tablefootmark{e}	&		\\
GSS30-IRS1				&	0.10	&			29.3	&				120	&	\phantom{1}17	&	\tablefootmark{e}	&		\\
Elias29					&	0.09	&			19.8	&	\phantom{1}78	&	\phantom{1}17	&	2.1($-$6)	&	1,9	\\
IRS63					&	0.13	&			11.5	&	\phantom{1}31	&	\phantom{1}12	&	6.4($-$7)	&	1,9	\\
RNO91					&	0.09	&	\phantom{1}8.1	&	\phantom{1}23	&	\phantom{1}12	&	4.2($-$6)	&	9	\\
RCrA-IRS5A				&	0.35	&			34.9	&				174	&	\phantom{1}17	&	\tablefootmark{e}	&		\\
HH100-IRS				&	0.39	&			23.4	&				120	&	\phantom{1}17	&	4.4($-$7)	&	5	\\
\hline\\
\end{tabular}
\tablefoot{
	\tablefoottext{a}{Measured in 0.3 km\,s$^{-1}$ bins.}
	\tablefoottext{b}{Line width measured at the 4-$\sigma$ level.}
	\tablefoottext{c}{CO outflow force obtained from the literature. All values have been scaled to the distances used here. When multiple values were available, the average is used.}
	\tablefoottext{d}{Single-pointing data from the JCMT (beam-size is 15\arcsec).}
	\tablefoottext{e}{No values available in the literature. IRAS12496 has an outflow seen nearly face-on, and so any value for $F_{\rm CO}$ is uncertain at best; GSS30-IRS1 is located in a crowded region of Ophiuchus, and it is not possible to determine which outflow the source may be driving in single-dish CO 3--2 maps (van der Marel et al. in prep.).}
		}\tablebib{
	(1) \citet{bontemps96}; (2) \citet{cabrit92}; (3) \citet{curtis10}; (4) \citet{hogerheijde98}; (5) \citet{vankempen09b}; (6) \citet{bourke97}; (7) \citet{dionatos10}; (8) \citet{bachiller01}; (9) van der Marel et al. (in prep.).
	}
\label{tab:intensity_co}
\end{table*}


\section{DUSTY modelling}\label{app:dusty}

All envelopes are modelled using the 1D spherically symmetric dust radiative transfer code DUSTY \citep{ivezic97}. The procedure for modelling the envelopes is the same as outlined previously in \citet{schoier02} and \citet{jorgensen02}, this work being an extension and update of the latter. In this setup, the density profile is assumed to be a power-law with $n \propto r^{-p}$ and the inner boundary, $r_{\rm in}$, of the envelope is set to be where the dust temperature is equal to 250 K. Three parameters are varied: the slope of the density profile, $p$, the size of the envelope, $Y=r_{\rm out} / r_{\rm in}$, and the dust optical depth at 100 $\mu$m, $\tau_{100}$. The dust opacity is taken from \citet{ossenkopf94}, Table 1, column 5, the so-called OH5, corresponding to dust grains with thin ice mantles. The radial extent and slope of emission as observed at 450 and 850 $\mu$m is modelled first, whereby $p$ is determined. Secondly the spectral energy distribution (SED) is fitted with the remaining free parameters, $Y$ and $\tau_{100}$. The envelope mass is directly related to $\tau_{100}$ and is best constrained by the SCUBA 850 $\mu$m flux.

The emission profiles were obtained from the SCUBA Legacy Archive \citep{difrancesco08} or from LABOCA \citep[e.g., for the case of HH46-IRS;][]{vankempen09a}. For sources with no available sub-mm continuum maps the emission was assumed to fall off with a slope of 1.5, and the derived envelope mass is therefore more uncertain due to poor sampling of the SED at long wavelengths (BHR71 and Ced110-IRS4, and HH46-IRS at 450 $\mu$m). Fixing the slope is most relevant for southern sources, for which little or no continuum data at 450 and 850 $\mu$m exist. The SEDs were assembled from the literature and consist of data points between 60 and 1300 $\mu$m. The long wavelength SED points ($\lambda > 200$ $\mu$m) are obtained from SCUBA/LABOCA, Bolocam and SEST. The short-wavelength points now include the \textit{Spitzer}-MIPS points not available previously \citep{evans09} and new \textit{Herschel}-PACS data (Karska et al. in prep., Green et al. in prep.). Model images were convolved to the appropriate observing beam for comparison. To ensure that the mass of the envelope corresponds to that measured from the sub-mm fluxes \citep[e.g.,][]{shirley00}, the 850 and 450 $\mu$m fluxes were given high weight in the fit.

\begin{figure*}
\begin{center}
\includegraphics[width=14.5cm]{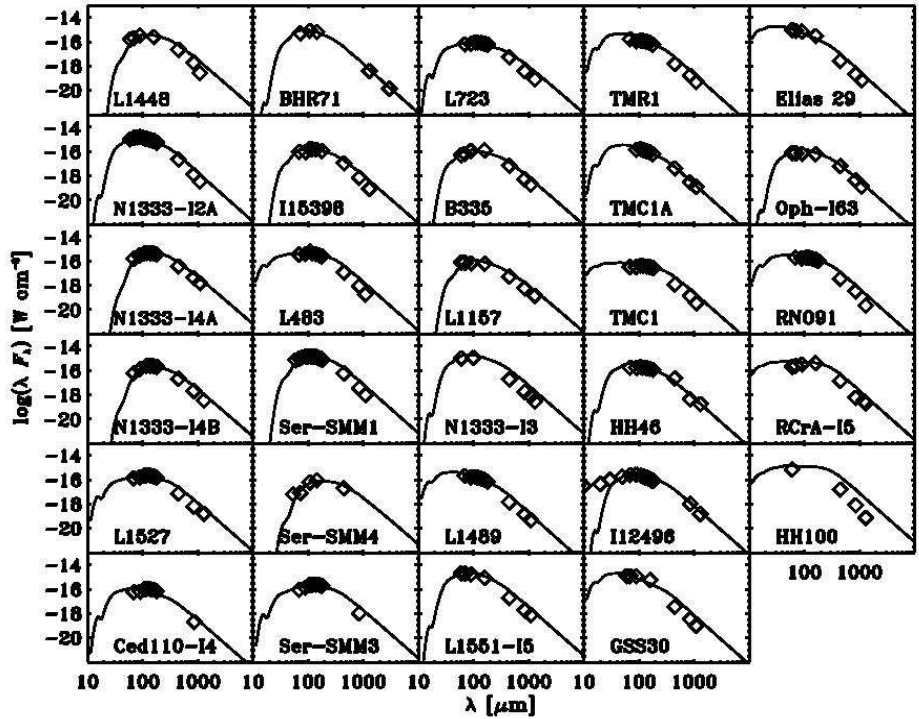}
\end{center}
\caption{SEDs for each source (diamonds). The full line shows the best-fit SED.}
\label{fig:sed}
\end{figure*}

\begin{figure*}
\begin{center}
\includegraphics[width=14.5cm]{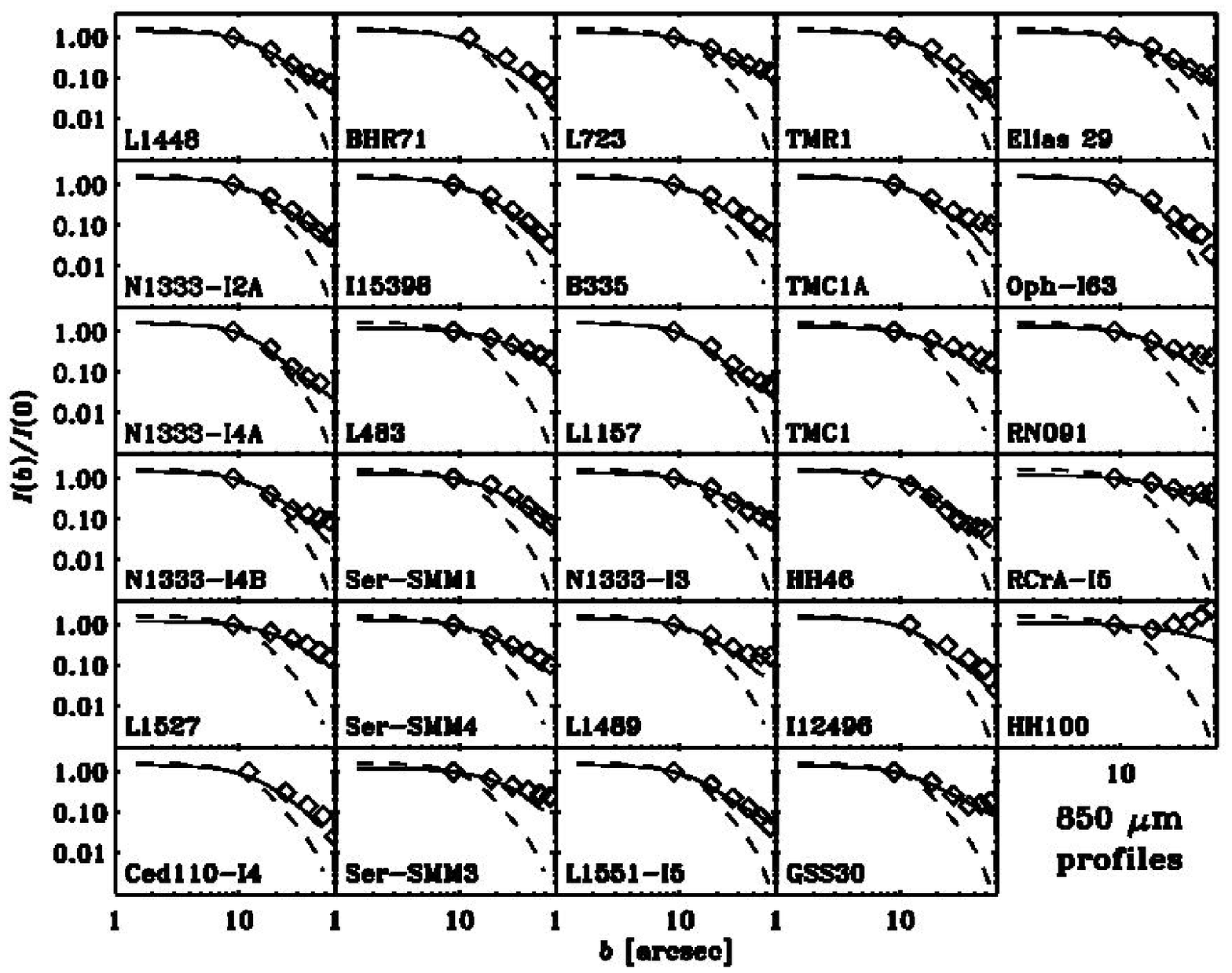}
\end{center}
\caption{850 $\mu$m radial profile for each source (diamonds). The dashed line shows the beam profile, and the full line shows the best-fit radial profile.}
\label{fig:850prof}
\end{figure*}

\begin{figure*}
\begin{center}
\includegraphics[width=14.5cm]{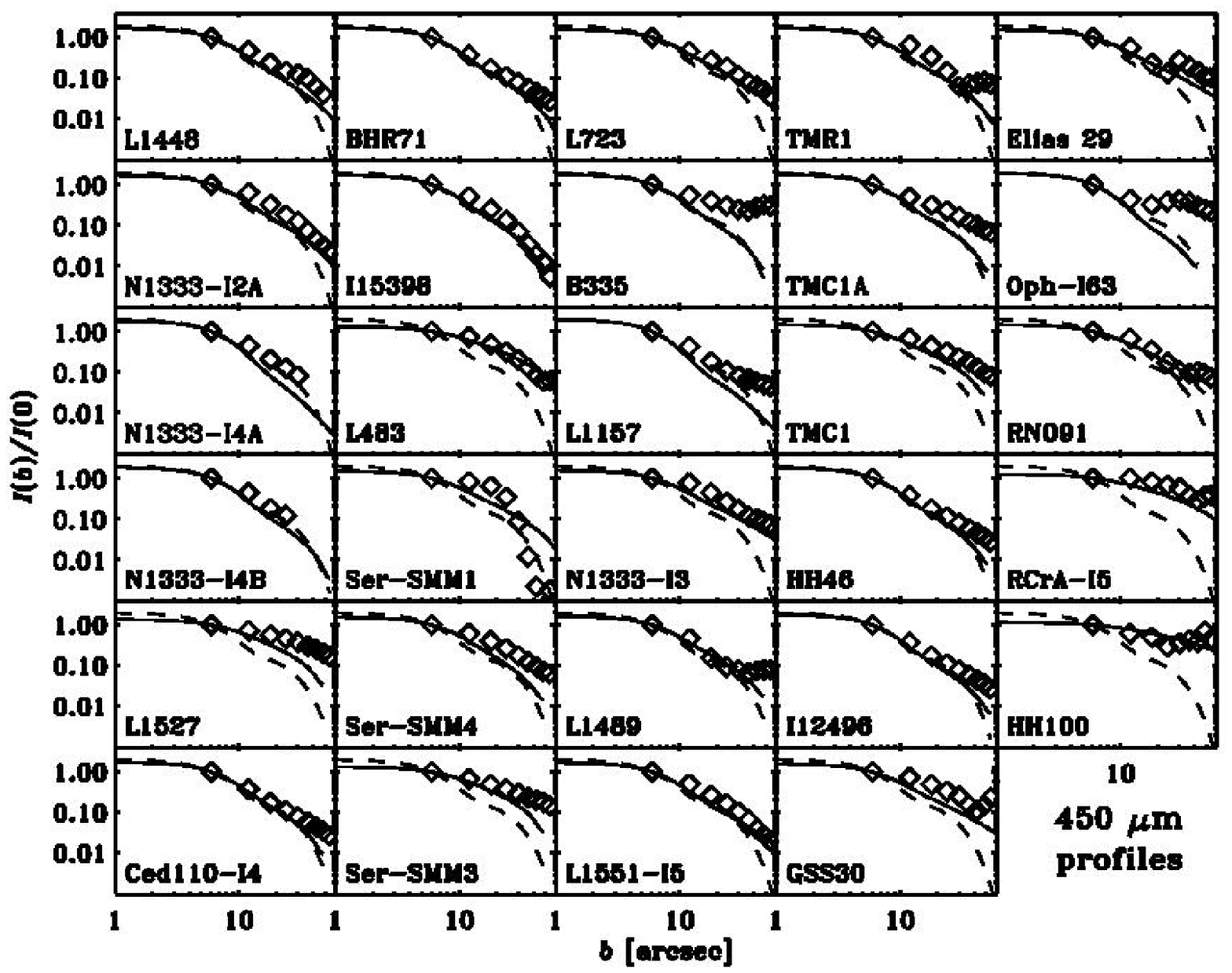}
\end{center}
\caption{450 $\mu$m radial profile for each source (diamonds). The dashed line shows the beam profile, and the full line shows the best-fit radial profile.}
\label{fig:450prof}
\end{figure*}

The best-fit parameters along with the model results are listed in Table \ref{tab:dustyparam}. An overview of the fitted data and results are shown in Figs. \ref{fig:sed}-\ref{fig:450prof}. Sources which have previously been analysed by \citet{jorgensen02} were re-analysed, and in general there is good agreement between the two sets of results. The uncertainty on $p$ and $\tau_{100}$ is of the order of 0.2--0.3, while it is somewhat higher for $Y$, of the order of $\sim$500.

The envelope mass is measured either at the $T_{\rm dust}$ = 10 K radius or at the $n$ = 10$^4$ cm$^{-3}$ radius, depending on which is smaller. The envelope masses range from 0.04 $M_\odot$ (Elias 29) to $\sim$16 $M_\odot$ (Ser SMM1). These masses are in good agreement with masses derived from the sub-mm flux alone, e.g., through scaling of the 850 $\mu$m flux assuming a constant dust temperature \citep{shirley00}. The mass estimates presented here fall well within the range of masses derived for these sources using other methods \citep{bontemps96, hogerheijde97, hogerheijde98, chen08, enoch09, jorgensen09}.

In this context it is important to note, that the modelling of the inner envelope is not realistic. Specifically, embedded disks are not taken into account and the model is expected to fail on scales smaller than a few 100 AU. In the Class I phase, disks can contribute significantly to the sub-mm fluxes \citep[e.g.,][]{jorgensen07, lommen08}, and their contribution has not been subtracted here. Therefore, the inferred envelope masses are upper limits, particularly for the Class I sources. The effect of an embedded disk is to add a point-source contribution to the flux at the center of the system and thus artificially increase the power-law density slope of the envelope. If the point source is subtracted, the $p-$value would decrease as well as the overall envelope mass. Thus, the values of $p$ and $M_{\rm env}$ presented here for Class I sources are overestimates.

\begin{table*}
\caption{Best-fit DUSTY parameters and predictions.}
\small
\begin{center}
\begin{tabular}{l r r r @{} c c c c c c c}
\hline\hline
       & \multicolumn{3}{c}{Fit} & & \multicolumn{6}{c}{Physical parameters} \vspace{3pt} \\  \cline{2-4} \cline{6-11}
Source & $p$ & $Y$ & $\tau_{100}$ & & $r_{\rm in}$ & $r$(10 K) & $M_{\rm env}$ & $n_{\rm in}$ & $n$(1000 AU) & $N$(H$_2$) \\ 
       &     &     &              & & (AU)     & (AU)      & ($M_\odot$)         & (cm$^{-3}$) & (cm$^{-3}$) & (cm$^{-2}$)\\ \hline
L1448-MM        & 1.5 & \phantom{1}900 & 3.2 & &           20.7 & 6.1(3) & \phantom{1}3.9\phantom{0} & 1.3(9) & 3.9(6) & 7.7(23) \\
NGC1333-IRAS2A  & 1.7 & \phantom{1}500 & 1.5 & &           35.9 & 1.7(4) & \phantom{1}5.1\phantom{0} & 4.9(8) & 1.7(6) & 3.7(23) \\
NGC1333-IRAS4A  & 1.8 &           1000 & 7.7 & &           33.5 & 6.9(3) & \phantom{1}5.6\phantom{0} & 3.1(9) & 6.7(6) & 1.9(24) \\
NGC1333-IRAS4B  & 1.4 & \phantom{1}800 & 4.3 & &           15.0 & 3.8(3) & \phantom{1}3.0\phantom{0} & 2.0(9) & 5.7(6) & 1.0(24) \\
L1527           & 0.9 &           1200 & 0.3 & & \phantom{1}5.4 & 4.6(3) & \phantom{1}0.9\phantom{0} & 8.9(7) & 8.1(5) & 6.9(22) \\
Ced110-IRS4     & 1.4 &           1400 & 0.5 & & \phantom{1}4.1 & 3.4(3) & \phantom{1}0.2\phantom{0} & 8.5(8) & 3.9(5) & 1.2(23) \\
BHR71           & 1.7 & \phantom{1}500 & 2.0 & &           24.8 & 9.9(3) & \phantom{1}2.7\phantom{0} & 9.4(8) & 1.8(6) & 4.9(23) \\
IRAS15398       & 1.4 &           1000 & 1.7 & & \phantom{1}6.2 & 2.7(3) & \phantom{1}0.5\phantom{0} & 1.9(9) & 1.6(6) & 4.1(23) \\	
L483            & 0.9 &           1000 & 0.2 & &           12.5 & 1.2(4) & \phantom{1}4.4\phantom{0} & 2.7(7) & 5.1(5) & 4.9(22) \\
Ser SMM1        & 1.3 & \phantom{1}500 & 2.0 & &           31.0 & 1.2(4) &           16.1\phantom{0} & 3.8(8) & 4.1(6) & 4.9(23) \\
Ser SMM4        & 1.0 &           1600 & 2.4 & & \phantom{1}6.8 & 2.8(3) & \phantom{1}2.1\phantom{0} & 7.9(8) & 5.4(6) & 4.8(23) \\
Ser SMM3        & 0.8 &           1200 & 0.4 & & \phantom{1}8.9 & 6.7(3) & \phantom{1}3.2\phantom{0} & 4.7(7) & 1.1(6) & 8.7(22) \\
L723            & 1.2 &           2800 & 0.5 & & \phantom{1}8.4 & 6.6(3) & \phantom{1}1.3\phantom{0} & 2.5(8) & 8.0(5) & 1.1(23) \\
B335            & 1.4 &           1200 & 1.4 & & \phantom{1}9.8 & 4.9(3) & \phantom{1}1.2\phantom{0} & 1.0(9) & 1.5(6) & 3.4(23) \\
L1157           & 1.6 &           2200 & 2.5 & &           14.3 & 5.4(3) & \phantom{1}1.5\phantom{0} & 1.7(9) & 2.0(6) & 6.1(23) \\ \hline
NGC1333-IRAS3   & 1.5 &           1000 & 1.2 & &           39.2 & 1.8(4) & \phantom{1}9.5\phantom{0} & 2.9(8) & 1.9(6) & 3.0(23) \\
L1489           & 1.5 & \phantom{1}800 & 0.2 & & \phantom{1}8.4 & 6.7(3) & \phantom{1}0.2\phantom{0} & 2.5(8) & 1.9(5) & 5.9(22) \\
L1551-IRS5      & 1.8 & \phantom{1}900 & 1.5 & &           28.9 & 1.4(4) & \phantom{1}2.3\phantom{0} & 6.9(8) & 1.2(6) & 3.7(23) \\
TMR1            & 1.6 & \phantom{1}900 & 0.3 & & \phantom{1}8.8 & 7.9(3) & \phantom{1}0.2\phantom{0} & 4.1(8) & 2.1(5) & 8.9(22) \\
TMC1A           & 1.6 & \phantom{1}900 & 0.4 & & \phantom{1}7.7 & 6.7(3) & \phantom{1}0.2\phantom{0} & 5.2(8) & 2.2(5) & 9.9(22) \\
TMC1            & 1.1 &           1800 & 0.1 & & \phantom{1}3.7 & 5.0(3) & \phantom{1}0.2\phantom{0} & 8.5(7) & 1.8(5) & 2.4(22) \\
HH46-IRS        & 1.6 & \phantom{1}800 & 1.0 & &           28.5 & 1.7(4) & \phantom{1}4.4\phantom{0} & 3.5(8) & 1.2(6) & 2.5(23) \\
IRAS12496       & 1.6 & \phantom{1}800 & 1.3 & &           12.0 & 6.1(3) & \phantom{1}0.8\phantom{0} & 1.1(9) & 9.2(5) & 3.2(23) \\
GSS30-IRS1      & 1.6 &           1000 & 0.2 & &           16.2 & 1.6(4) & \phantom{1}0.1\phantom{0} & 1.2(8) & 1.7(5) & 4.9(22) \\
Elias 29        & 1.6 &           1000 & 0.1 & &           15.5 & 1.6(4) &           \phantom{1}0.04 & 6.5(7) & 8.3(4) & 2.5(22) \\
Oph-IRS63       & 1.6 & \phantom{1}900 & 1.4 & & \phantom{1}6.6 & 3.2(3) & \phantom{1}0.3\phantom{0} & 2.2(9) & 6.9(5) & 3.4(23) \\
RNO91           & 1.2 & \phantom{1}900 & 0.2 & & \phantom{1}6.6 & 6.0(3) & \phantom{1}0.5\phantom{0} & 1.3(8) & 3.3(5) & 4.9(22) \\
RCrA-IRS5A      & 0.8 &           1000 & 0.1 & &           10.1 & 1.0(4) & \phantom{1}2.0\phantom{0} & 1.1(7) & 2.8(5) & 2.5(22) \\
HH100-IRS       & 0.5 &           1000 & 0.1 & &           15.5 & 1.6(4) & \phantom{1}8.1\phantom{0} & 1.7(6) & 2.2(5) & 2.5(22) \\
\hline
\end{tabular}
\end{center}
\label{tab:dustyparam}
\end{table*}

\end{document}